\documentclass[11pt]{amsart}
 
\usepackage{amsmath,amstext,amsopn,amsfonts,eucal,amssymb,amsthm,array}
\usepackage{graphicx,verbatim,url}
\usepackage[bf,textfont={small}]{caption}

\usepackage{multirow}

\usepackage{wrapfig}

\newcommand\N{{\mathbb N}}
\newcommand\Z{{\mathbb Z}}

\def\lp{\left(}
\def\rp{\right)}
\def\St{\mathcal{S}}
\def\A{\mathcal{A}}

\newtheorem{thm}{Theorem}[section]

\theoremstyle{definition}
\newtheorem{def.}[thm]{Definition}
\newtheorem{ex.}[thm]{Example}

\theoremstyle{remark}
\newtheorem*{rem}{Remark}
\newtheorem*{pf.}{Proof}

\newcolumntype{M}{>{$}c<{$}}

\begin{document}

\title{Active Tile Self-assembly, Self-similar Structures and Recursion}
\author{N. Jonoska$^{\dagger,1}$, D. Karpenko$^{*,\dagger,1}$}\thanks{$^{*}${\it The corresponding author}\\  \indent   $^\dagger${\bf Department of Mathematics and Statistics, University of South Florida}
\\ \indent$^1$ This work has been supported in part by NSF grants CCF-1117254 and DMS-0900671.}
\address{Department of Mathematics and Statistics, University of South Florida,}
\email{jonoska@math.usf.edu, dkarpenk@mail.usf.edu}
\date{}

\begin{abstract}
We present an active tile assembly model  which  extends Winfree's abstract tile assembly model \cite{winfree_phd}  to tiles that are capable of transmitting and receiving binding site activation signals. In addition, we introduce a mathematical framework to address self-similarity and recursion within the model. 
The model is applied to show a recursive assembly of an archetypal self-similar aperiodic tiling  known as the L-shape tiling.
\end{abstract}

\maketitle

\section{Introduction}
DNA self-assembly is based on an inherent property of its chemical structure: a given nucleic acid sequence binds to its perfect Watson-Crick complement. Partially complementary sequences may also bind but these bonds are weaker. Moreover, DNA can assume a variety of rigid forms other than the familiar linear double helix. Synthetic molecules have been designed and shown to assemble into branched species \cite{ned2,ned3}, and more complex species that entail the lateral fusion of DNA double helices \cite{ned4}, such as DNA double crossover (DX) molecules \cite{ned5}, triple crossover (TX) molecules \cite{TX} and paranemic crossover (PX) molecules. DX and TX molecules have been used as tiles and building blocks for large nanoscale arrays \cite{winfree_phd,Furong}.

DNA-based tile arrays can carry out computation: problems can be encoded in single-stranded DNA portions at the edges of the tiles (the ``sticky ends'') and the computation can be carried out by their self-assembly. Winfree introduced the abstract tile assembly model (aTAM)~\cite{winfree_phd} and showed that 2D self-assembled arrays made of DX or TX DNA tiles can simulate the dynamics of a bounded one dimensional cellular automaton and so are potentially capable of performing computations as a Universal Turing machine. Since then, several successful experiments have confirmed the possibility of computation by array-like DNA self-assembly: binary addition (simulation of XOR) using TX molecules as tiles \cite{mao}, Sierpinski triangle assembly \cite{Sierpinski,Origami-seed-CA-assembly}, and a binary counter \cite{DNAcount} by DX molecules. Transducer simulations with programmed inputs by TX DNA molecules have also been reported \cite{New-NACO,Banani-submitted}.

DNA-based tiles also provide a way to obtain new materials both directly, as the array is a physical structure in itself, and indirectly, by allowing the tiles to contain markers or scaffolding which could direct the assembly of other molecular structures on top. Two-dimensional periodic tilings form crystal lattices and have been studied the most to date (see \cite{DNAcrystalarrays}). It is the aperiodic tilings which can carry out computation. In this paper we show a model for assembling self-similar aperiodic motifs: these, if materialized, could yield quasi-crystallographic structures, which are quite rare in nature and in general quite difficult to design and assemble. Our model is an extension of the abstract tile assembly model to tiles that are capable of transmitting and receiving binding site activation signals. One of the well-known aperiodic set constructions consists of modified square tiles due to Robinson\cite{Robinson}. A tile set of just four ``active'' tiles utilizing the idea of signaling has been proposed in \cite{DNARobinson} for a set of Robinson tiles which are capable of hierarchically assembling a self-similar square pattern. In this paper, we propose a theoretical model for active tile assembly and a formal definition of self-similar and recursive assembly. We apply our model to another well-known aperiodic self-similar tiling, the L-shape tiling (also known as the chair tiling \cite{Goodman}). 

But why augment the existing model with signaling and how? In what has become the standard tile assembly model, the ``sticky ends’’ of the tile are formed by single stranded DNA containing a sequence of unpaired bases which can bind to a strand of complementary bases (on another tile). One important problem in the assembly process is error control: how to prevent ``tiles'' which have partial base matches from binding in the ``wrong'' place in the lattice. At high enough temperatures, weaker bonds are not stable; but, unfortunately, they can last long enough for the tile to form new and stable bonds in an incorrect position. One possible way to control the errors is by staging the assembly: by only allowing certain tiles in solution at a certain point, then adding other tiles in stages after a sufficient amount of time has passed for most correct bonds to form. A theoretical analysis of staged assembly is given in Demaine, et al. \cite{staging}. 

Signaling is another possible solution, proposed in different variants in \cite{Reif} and \cite{DNARobinson}. The idea is to keep certain binding sites dormant until a ``signal'' is received to activate them, which allows indirect control over the assembly process. The existing DNA signaling mechanisms rely on strand displacement techniques (see \cite{Yurke}) and have already been applied in a variety of ways. For example, devices whose activity is controlled by DNA strands have been produced in \cite{Yurke,Yan}: these utilized DNA ``fuel'' strands, which were also recently used to construct a large number of enzyme-free logic gates \cite{DNAlogiccircuits,DNAgatecircuits,Erik-Lulu-Science}. Additionally, structures that can perform simple ``walking" (based on strand displacement) on an arranged platform have been reported in \cite{ned-walker,hao-walker,niles-walker}. The three most recent walking devices \cite{Ned-cargo-walker,DNAsignals,Stojanovic-spider} showed a significant robustness, allowing the directions and the actions of the walker to be guided by a sequence of strand displacements incorporated within the walking platform.

These mechanisms allow one to construct DNA based tiles with sticky ends that begin in an inactive state - bound to a protective strand which can later be displaced by a signal initiated 
\begin{wrapfigure}{r}{.5\textwidth}
\begin{center}\vskip -3mm
\includegraphics[scale=.5]{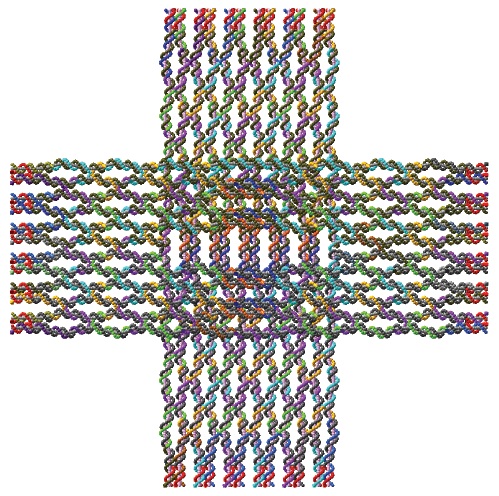}
\end{center} \vskip -3mm
\caption{{\small A DNA origami tile~\cite{DNAcrystalarrays}}.}
\label{fig:DNAtile}
\end{wrapfigure} 
by some other tile, resulting in activation. A particular DNA structure that can be used as a tile, an origami cross shape~\cite{DNAcrystalarrays}, 
     is shown in Figure~\ref{fig:DNAtile} (DNA origami is a method of creating arbitrary shapes from DNA by folding a single strand into the 
desired shape and using shorter ``helper'' strands to bind it rigidly together \cite{DNAorigami}). The size of this tile is 
about $100\times 100$~nm, which allows enough space to accommodate signaling, unlike the DX and TX molecules 
which may and have been used for computation~\cite{winfree_phd} but are not large enough to incorporate the existing 
signaling mechanisms. 

We present the formal active assembly model in the next section, using a generalization of Winfree's aTAM known as the two-handed assembly model (2HAM) \cite{2HAM} (a part of what the authors of \cite{2HAM} call a ``multiple tile model''). In it, the tile types are unique squares defined as 4-tuples $(\sigma_{n},\sigma_{e},\sigma_{s},\sigma_{w})$ representing the ``glues'' at the north, east, south, and west edges, each glue having an associated strength in the form of a non-negative integer. Tile assemblies are produced when smaller tile assemblies bind together. The ``two'' in 2HAM refers to the number of tile assemblies which may bind at a time. A bond is formed between adjacent tiles of two assemblies if their adjacent edges have matching glue labels and has the corresponding strength. Every tile assembly is a mapping of the integer lattice into a set of tile types and the bonds between all tiles are represented by a binding graph (the vertices are the tiles and edges are the connections among them).

The formal model proposed in this paper is that of an active programmable self-assembly which is able to build up arrays hierarchically. In such a system, the signaling capabilities of the tiles allow them to assume multiple identities throughout the assembly process. Moreover, a tile is allowed to have more than one type of ``glue'' on any given edge and these may be initially inactive (unable to bind) until they receive an activation signal. This approach allows for step by step assembly in a potentially controlled and robust way without actually staging the process. 

We note that a formal description of the same process of ``active assembly'' allowing tile glues to be deactivated and tile assemblies to break apart was introduced in \cite{Jensignals}. Although also based on the 2HAM, it differs in notation from our model and includes an added notion of ``pending actions.'' The latter allows signal transmission to be delayed by an arbitrary amount of time in order to better approximate the physical interactions among actual DNA tiles. Both deactivation and ``pending actions'' are compatible with the model presented here, and we discuss the specifics of their incorporation into our model in the concluding chapter.

The work presented here suggests a potentially feasible model for recursion in molecular self-assembly; and, through the use of signaling techniques, it has the potential to build self-similar structures hierarchically. In Section 2, we formally define active tiles and describe the hierarchical assembly in terms of a tile modification function. We also formally define the notions of recursive assembly and self-similarity. In Section 3, we apply the model to an archetypal example of a self-similar aperiodic tiling - the L-shape tiling \cite{Goodman} - and prove that the designed tiling system recursively assembles a non-periodic self-similar L-shape based array, providing a tiling of the plane. We end the paper with some concluding remarks about the presented model and future avenues of research.

\section{Model Description}\label{basicdef}

In this section, we give the formal description of our Active Tile Assembly Model. In the first subsection, we define active tiles while in the second, we present binding graphs and tile assemblies. The dynamics of signaling within the assembly process are represented through the tile modification function which we describe in the third subsection. In the last two subsections we use the tile modification function to construct hierarchical tile assembly sets and to define an active tile assembly system as well as the properties of recursion and self-similarity.

\subsection{Basic Definitions}\label{basicdefs}
Let $\Sigma^{+}$ be a finite set and define the complementary set $\Sigma^{-}=\{-c\vert\ c\in\Sigma^{+}\}$.
We call $\Sigma=\Sigma^{+}\cup\Sigma^{-}$ the set of all \emph{labels}. We use $|c|$ to denote $c$ if $c\in\Sigma^{+}$ and to denote $-c$ if $c\in\Sigma^{-}$. We also have $-(-c)=c$. We assume that $\Sigma$ is fixed throughout our discussion.

\begin{def.}
The \emph{strength function} is defined to be
\[s:\Sigma^{+}\rightarrow\Z^{+} \textrm{ where } \Z^{+}=\left\{ x\in\Z\vert\ x \geq 0 \right\},\]
and for $c\in\Sigma$, $s(|c|)$ is called the \emph{strength} of $c$. 
\label{def:strength}\end{def.}

\noindent
The strength of a label refers to the strength of the bond between it and its complementary label.

\begin{def.} Define the set of \emph{signals}:
\[{\hat{\Sigma}}=\left\{ c_{i}^{j}|\ c\in\Sigma^{+};\ i,j\in\left\{ +y, +x, -y, -x, 0 \right\},\ j\neq 0 \right\},\]
where $c_{i}^{j}$ for $i\neq 0$ means that $c$ can be \emph{transmitted} from the direction $i$ in the direction $j$, whereas $c_{0}^{j}$ means that $c$ can be initiated in the $j$ direction and we refer to it as an \emph{initiation signal.}
\label{def:signals}\end{def.}

The direction notation in this paper follows the axial directions of the xy-plane. Although it is a departure from the conventional notation of North, East, South, and West~\cite{2HAM}, we believe that this notation gives a more natural correspondence between the parts of a tile and its coordinate position in the plane and simplifies the description of signal transmission between adjacent tiles (e.g., we can say that a signal $c_{i}^{j}$ will ``connect'' to the signal of the form $c_{-j}^{k}$; here, the notation $j$ and $-j$ indicates a more direct relationship between the corresponding sides than, say, $N$ and $S$).

\begin{def.} A \emph{tile} $t$ is a 4-tuple of tile sides, $t=(t_{+y}, t_{+x}, t_{-y}, t_{-x})$, 
such that each tile \emph{side}~$t_{i}$, $i\in\{+y,+x,-y,-x\}$, is an ordered pair of sets of labels:
\[t_{i}=(A,I)\in\mathcal{P}(\Sigma)\times\mathcal{P}(\Sigma)\]
where $\mathcal{P}(\Sigma)$ denotes the power set of $\Sigma$. We refer to $A$ as the set of \emph{active labels} and to $I$ as the set of \emph{inactive labels}.
\end{def.}
\begin{figure}[htb]
\begin{center}\vskip -.7cm
\includegraphics[scale=.8]{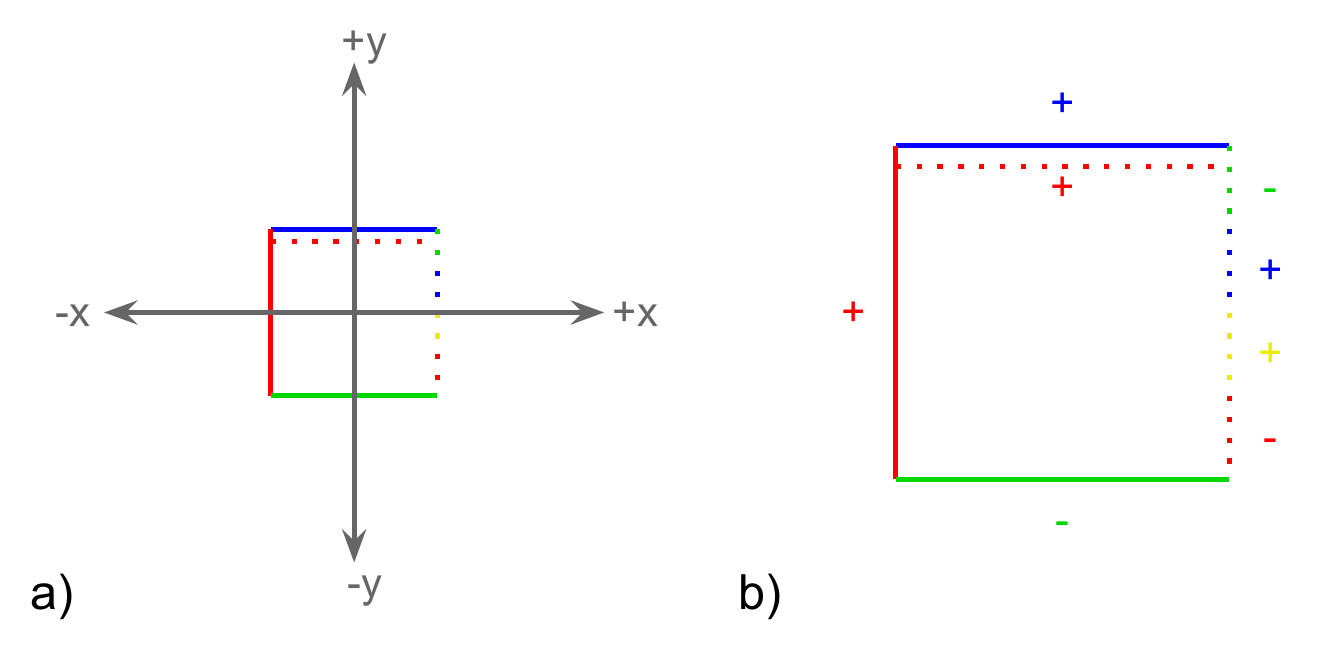}
\end{center}\vskip -.7cm
\caption{{\small a) A tile $t=(t_{+y}, t_{+x}, t_{-y}, t_{-x})$ placed on the coordinate plane. Solid lines represent active labels and dashed lines represent inactive labels. Subscripts on tile sides indicate directions normal to those tile sides. b) A tile with complementary labels marked by $+$/$-$. See Example \ref{ex:SimpleTile}}.} 
\vskip -.7cm
\label{fig:SimpleTile}
\end{figure}
\begin{ex.}
\label{ex:SimpleTile}
Refer to Figure \ref{fig:SimpleTile} for an example tile $t$. Let $\Sigma^{+}=\{r,b,y,g\}$, corresponding to the red, blue, yellow, and green labels. Then $\Sigma^{-}=\{-r,-b,-y,-g\}$, so $\Sigma=\{r,b,y,g,-r,-b,-y,-g\}$, and
\begin{align*}
t_{+y} &= \left(\left\{b\right\},\left\{r\right\}\right) &
t_{+x} &= \left(\emptyset, \left\{-r,y,b,-g\right\}\right) &
t_{-y} &= \left(\left\{-g\right\},\emptyset\right) &
t_{-x} &= \left(\left\{r\right\},\emptyset\right)
\end{align*}
are the tile sides for $t=(t_{+y}, t_{+x}, t_{-y}, t_{-x}). \hfill\blacksquare$
\end{ex.}

\begin{def.}\label{def:activetile}
An \emph{active tile} is an ordered triple $\tau=\left(t,\mathcal{A},\mathcal{S}\right)$ where $t$ is a tile and $\mathcal{A,S}\subseteq\hat{\Sigma}$ are the sets of \emph{activation signals} and \emph{transmission signals} respectively, such that
\begin{enumerate}
\item for $t_{a}=(A,I)$ 
\begin{itemize}
\item[] a) if $c\in{A}$, then $-c\notin A$ and $c,-c\notin I$ 
\item[] b) if $d\in I$, then $-d\notin I$.
\end{itemize}
\item $c_{0}^{a}\notin\mathcal{A}$ for any $a\in\{+y,+x,-y,-x\}$
\end{enumerate}
Condition (1) states that a tile side cannot contain redundant or complementary labels; condition (2) requires that the set of activation signals does not contain initiation signals.
\end{def.}

\begin{figure}[hbt]
\begin{center}
\includegraphics[scale=.75]{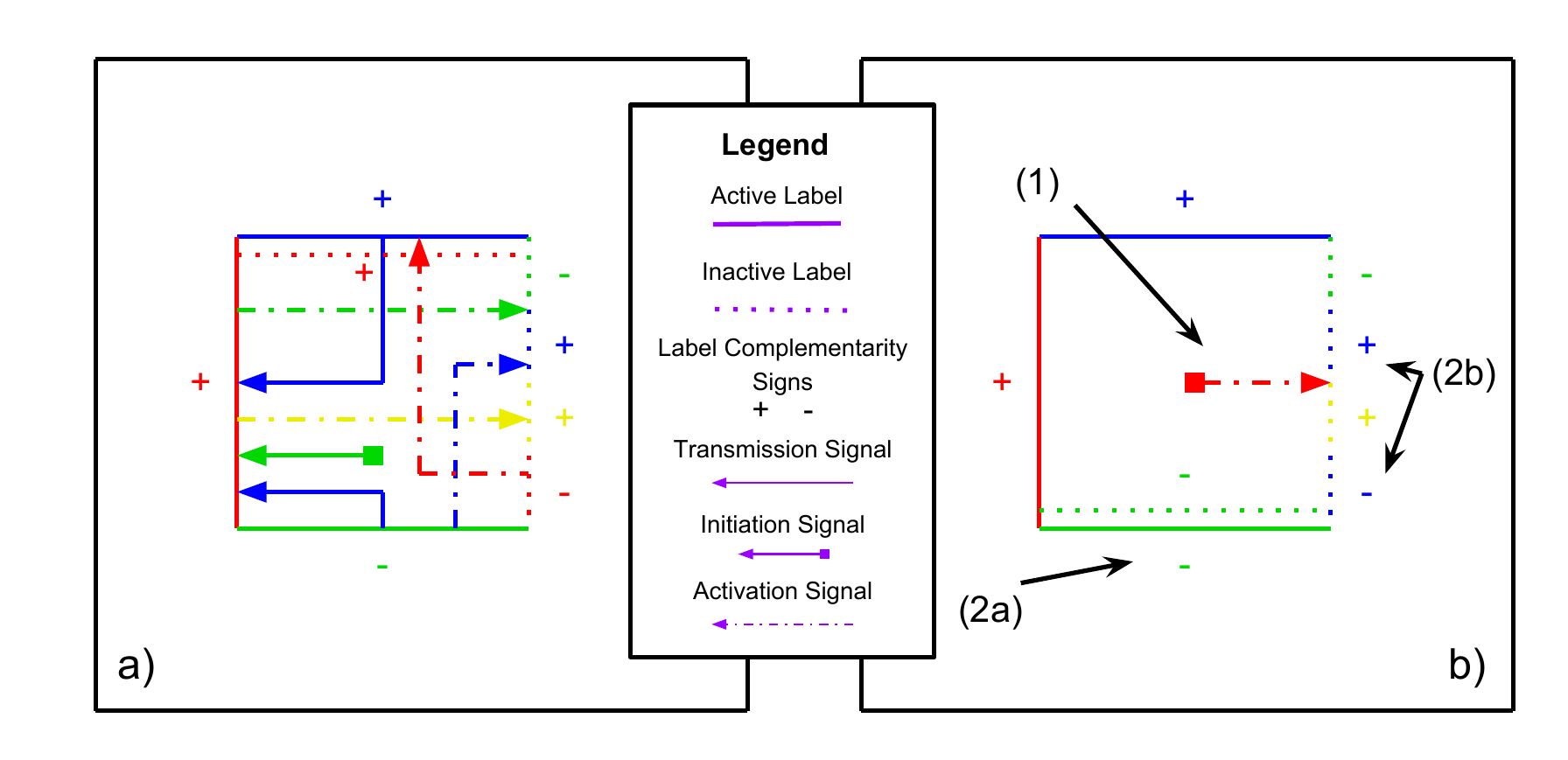}
\end{center}
\caption{a) An active tile obtained by adding signaling to tile $t$. See Example \ref{ex:ActiveTile}. b) Example of failed conditions (1), (2a) and (2b) in the active tile definition (Def.~\ref{def:activetile}). This figure establishes a convention for depicting active tiles in the rest of the paper.}
\label{fig:ActiveTile} 
\vskip -.7cm
\end{figure}

Condition (1) comes from practical considerations of actual DNA dynamics: if a tile side contains active complementary ends, they may bind to each other instead of binding to complementary strands on another tile. Since we suppose that a signal for a given label would also activate the corresponding complementary label, we forbid complementary labels from co-existing even in the inactive state. Condition (2) eliminates superfluous signaling (i.e. a tile which activates one of its own labels is equivalent to a tile with that label already active). See Fig.\ref{fig:ActiveTile}b for an example of a tile which violates these conditions.
\begin{ex.}
\label{ex:ActiveTile}
Refer to Figures \ref{fig:SimpleTile} and \ref{fig:ActiveTile}a) for the sample tile with $\Sigma=\{r,b,y,g,-r,-b,-y,-g\}$ and $t=(t_{+y}, t_{+x}, t_{-y}, t_{-x})$ as in Example \ref{ex:SimpleTile}. The signaling shown in Figure \ref{fig:ActiveTile} is:
\[\mathcal{A}=\left\{r_{+x}^{+y},\ b_{-y}^{+x},\ g_{-x}^{+x},\ y_{-x}^{+x} \right\}\]
\[\mathcal{S}=\left\{b_{+y}^{-x},\ b_{-y}^{-x},\ g_{0}^{-x} \right\}\]
and the corresponding active tile is $\tau=(t,\mathcal{A,S}).\hfill\blacksquare$
\end{ex.}

\subsection{Tile Assemblies and the Associated Binding Graphs}\label{sec:tileassembly}

Let $\mathcal{T}$ be any set of active tiles over an alphabet $\Sigma$ with a strength function $s$ and consider a partial mapping:
\[\alpha:\Z^{2}\rightarrow\mathcal{T}.\]
We call such a mapping a \emph{configuration} and define its associated \emph{binding graph} to be the weighted graph $G_{\alpha}=(V,E)$, with $V\subset\Z^{2}$ such that $(i,j)\in V$ if and only if $\alpha((i,j))$ is defined, and $E\subset {V\choose 2}$ such that
\[E=E_h\cup E_v\]
\[E_h=\left\{ \left\{ \left(i,j\right),\left(i+1,j\right) \right\} \vert\ (i,j), (i+1,j)\in V \right\}\]
\[E_v=\left\{ \left\{ \left(i,j\right),\left(i,j+1\right) \right\} \vert\ (i,j), (i,j+1)\in V\right\}.\] 

The weight function for $G$ is defined in the following way.
Let $e=\{v,w\}=\{(i,j),(i,j+1)\}$ be a vertical edge, and let 
\[\alpha(v)=\tau_{v}=(t^{v},\mathcal{A}^{v},\mathcal{S}^{v}),\quad \alpha(w)=\tau_{w}=(t^{w},\mathcal{A}^{w},\mathcal{S}^{w})\]
with 
\[t^v =(t^v_{+y}, t^v_{+x}, t^v_{-y}, t^v_{-x}),\quad t^w =(t^w_{+y}, t^w_{+x}, t^w_{-y}, t^w_{-x})\] 
(where $t^v_{+y}=(A^v_{+y},I^v_{+y})$ and similarly for the rest). We consider the intersection 
\[C_{e}=A^v_{+y}\cap -A^w_{-y}\] 
where $-A^w_{-y}=\{-c\vert\ c\in A^w_{-y}\}$. Then we define the \emph{weight} of the (vertical) edge $e$ to be  
\[w(e)=\left\{ \begin{array}{ll}
		 \sum_{c\in C_{e}} s(|c|) & \quad \text{if}\ C_{e}\neq\emptyset\\
		 0 & \quad \text{otherwise}.\\
\end{array} \right.\]
The definition for the weight of horizontal edges is analogous.

In other words, the vertices of the binding graph are formed by points in the integer lattice which we presume are occupied by tiles. There are horizontal and vertical edges indicating horizontal and vertical adjacencies. Each edge is assigned a weight corresponding to the sum of the strengths of the complementary active labels of the adjacent tile sides. We say that adjacent tiles whose adjacent sides contain complementary labels $c$ and $-c$ contain a bond of strength $s(|c|)$. 

With this definition of a binding graph in hand, we proceed to make the following definitions. 
\begin{def.}
Given a set of active tiles $\mathcal{T}$ and an integer $\theta\geq 0$ called the \emph{temperature parameter}, a \emph{tile assembly instance} for $\theta$ is a configuration $\alpha$ where either 
\begin{itemize}
\item[(a)] $\alpha$ is defined for only a single point in $\Z^{2}$ or 
\item[(b)] its associated graph $G_{\alpha}=(V,E)$ satisfies the following two conditions: 
	\begin{enumerate}
	\item[(i)] $G$ is connected, and 
	\item[(ii)] the sum of the weights of the edges in any edge cut\footnote{An \emph{edge cut} of a graph $G=(V,E)$ is a set of all edges with one vertex in $U$ and the other in $V\backslash U$ for some $U\subseteq V$.} of $G$ is greater than or equal to $\theta$. 
	\end{enumerate}
\end{itemize}
\end{def.}	

Observe that condition (b) in the definition, part (ii) of which is generally referred to as \emph{$\theta$-stability}, forces the tile assembly instance to be composed of at least two tiles; we refer to $\alpha$ in the special case of (a) as a \emph{unit tile instance}. In literature, the temperature parameter $\theta$ is usually set equal to 2.

We use the terms \emph{tile assembly} and \emph{unit tile} to denote the equivalence classes $[\alpha]$ of the respective instances under translation\footnote{We do not allow rotation explicitly in the model. To account for rotation, we later require that each tile set contain ``rotated'' copies of every tile.}:
\[\alpha\sim\beta\ \text{if and only if}\ \exists k,l\in\Z\ \text{such that}\ \alpha(i,j)=\beta(i+k,j+l)\ \forall i,j\in\Z.\]

We adopt the convention that the \emph{class representative} (with respect to translation) of a tile assembly $[\alpha]$ is the tile assembly instance $\alpha'\in[\alpha]$ such that 
\[\min\left\{i\ \vert\  \alpha'(i,k)\text{ is defined for some } k\in\Z\right\}=0\]
and
\[\min\left\{j\ \vert\  \alpha'(k,j)\text{ is defined for some } k\in\Z\right\}=0;\]
that is, the tile assembly instance is positioned with the left-most tile on the $y$-axis and the bottom-most tile on the $x$-axis.

\subsection{The Tile Modification Function}\label{tilemodfunction}
\label{sec:tilemodf}
\noindent In this section we define the tile modification function $f$ which simulates signal transmission and binding site activation for each tile assembly. This function allows each tile within the assembly to change its identity according to the signals it ``sends to'' and ``receives from'' its neighbors. 

Let $\alpha$ be a tile assembly instance and denote $\alpha(z)$, the tile associated with $z$, by $\alpha_{z}$. For $z\in\Z^{2}$ and  $\alpha_{z}=\tau=(t,\mathcal{A,S})\in\mathcal{T}$, let
\begin{align*}
\tau_{+y} &=\alpha_{z+(0,1)}\\
\tau_{+x} &=\alpha_{z+(1,0)}\\
\tau_{-y} &=\alpha_{z+(0,-1)}\\
\tau_{-x} &=\alpha_{z+(-1,0)}
\end{align*}
define the neighbors of $\tau$ on the respective sides. If $\tau_{i}$ is not defined for some $i\in\{+y,+x,-y,-x\}$ write $\tau_{i}=\varepsilon$, otherwise write $\tau_{+y}~=~(t^{+y}, \mathcal{A}^{+y}, \mathcal{S}^{+y})$ (similarly for the other three neighbors).

Let $N=\{(0,1), (0,-1), (1,0), (-1,0)\}$. A function $f:\mathcal{T}^{5}\rightarrow\mathcal{T}$ is a \emph{tile modification function} if
\begin{align*}
f(\alpha_{z+N}) &= f(\tau, \tau_{+y}, \tau_{+x}, \tau_{-y}, \tau_{-x})\\
 &= \alpha'_{z} = (t',\mathcal{A}',\mathcal{S}')
\end{align*} 
transforms the active tile $\tau=\alpha_z$ into another active tile $\alpha_{z}'$ with the following three rules.

Let $t=(t_{+y},t_{+x},t_{-y},t_{-x})$ and $t'=(t'_{+y},t'_{+x},t'_{-y},t'_{-x})$.
\begin{enumerate}
\item \emph{Tile Side Modification} \\
If $t_{i}=(A,I)$, $i\in\{+y,+x,-y,-x\}$, then
	\[t_{i}'=\left(A\cup C,\ I\setminus \left(C\cup D\right)\right),\] 
	where
		\[C=\left\{ c\in I \vert\ \exists |c|_{j}^{i}\in \mathcal{A}\ \text{and}\ |c|_{0}^{-j}\in \mathcal{S}^{j}\ \text{for some}\ j\in \left\{+y,+x,-y,-x\right\} \right\}\]
		\[D=\left\{ c\in I \vert\ \nexists |c|_{j}^{i}\in \mathcal{A}\ \text{for any}\ j\in \left\{+y,+x,-y,-x\right\} \right\}\]
		Informally, if the tile side contains an inactive label $c$ with a corresponding activation signal and an adjacent tile contains a corresponding initiation signal, then the label $c$ becomes active. Inactive labels which can never be activated are removed from the tile.
\item \emph{Activation Signal Modification} \\
For $i,j\in\{+y,+x,-y,-x\}$,
\[\mathcal{A}' = \mathcal{A} \setminus \mathcal{A}_{\text{removed}}\]
\begin{align*}
\mathcal{A}_{\text{removed}} &= \left\{ c_i^j\in\mathcal{A} \vert\ \tau_i \neq \varepsilon\ \text{and either}\ \exists c_0^{-i}\in S^i\ \text{or}\ \nexists c_k^{-i}\in\mathcal{S}^i\ \text{for any}\ k\in\{+y,+x,-y,-x\} \right\}.
\end{align*}
	Informally, when the tile receives a signal, its corresponding activation signal is used and cannot be used again, so it is removed. Activation signals that can never be activated due to being adjacent to a tile which has no possibility of sending the corresponding signal are also removed.
\item \emph{Transmission Signal Modification} \\
For $i,j\in\{+y,+x,-y,-x\}$,
\[\mathcal{S}' = \left(\mathcal{S} \cup \mathcal{S}_{\text{added}} \right) \setminus \mathcal{S}_{\text{removed}}\]
\[\mathcal{S}_{\text{added}} = \left\{ c_{0}^{j} \vert\ \exists k\in\{+y,+x,-y,-x\}\ \text{such that}\ c_k^j \in \mathcal{S}\ \text{and}\ c_0^{-k}\in\mathcal{S}^{k} \right\} \]
\begin{align*}
\mathcal{S}_{\text{removed}} &= S_{1}\cup S_{2}\cup S_{3}\\
S_{1}&= \left\{ c_i^j\in\mathcal{S} \vert\ \tau_i \neq \varepsilon\ \text{and either}\ \exists c_0^{-i}\in S^i\ \text{or}\ \nexists c_k^{-i}\in\mathcal{S}^i\ \text{for any}\ k\in\{+y,+x,-y,-x\} \right\} \\
S_{2}&=\left\{ c_i^j\in\mathcal{S} \vert\ \tau_j \neq \varepsilon\ \text{and}\ \nexists c_{-j}^{k}\in\mathcal{S}^j \cup \mathcal{A}^j\ \text{for any}\ k\in\{+y,+x,-y,-x\}\right\}\\
S_{3}&=\left\{ c_0^j\in\mathcal{S}\vert\ \tau_j\neq\varepsilon \right\}.
\end{align*}
	Informally, when the tile receives a signal, its transmission signal is replaced with an initiation signal. Transmission signals which do not have matching signals in adjacent tiles, i.e., transmission signals that cannot be received or transmitted, are removed since they cannot perform any function. Once a tile is adjacent to another tile, all initiation signals in that direction are removed, since they are either immediately transmitted or can never be transmitted.
\end{enumerate}

We can extend $f$ to a tile assembly instance by applying $f$ simultaneously to every active tile in the tile assembly instance:
\[f(\alpha)=\alpha'\quad \text{if and only if}\quad \alpha'_{z}=f(\alpha_{z+N})\quad\forall z\in\Z^{2}.\]
We write $f([\alpha])=[\alpha']=[f(\alpha)]$ to denote the image of the tile assembly under $f$. 

\begin{figure}[htb]
\begin{center}
\includegraphics[scale=.75]{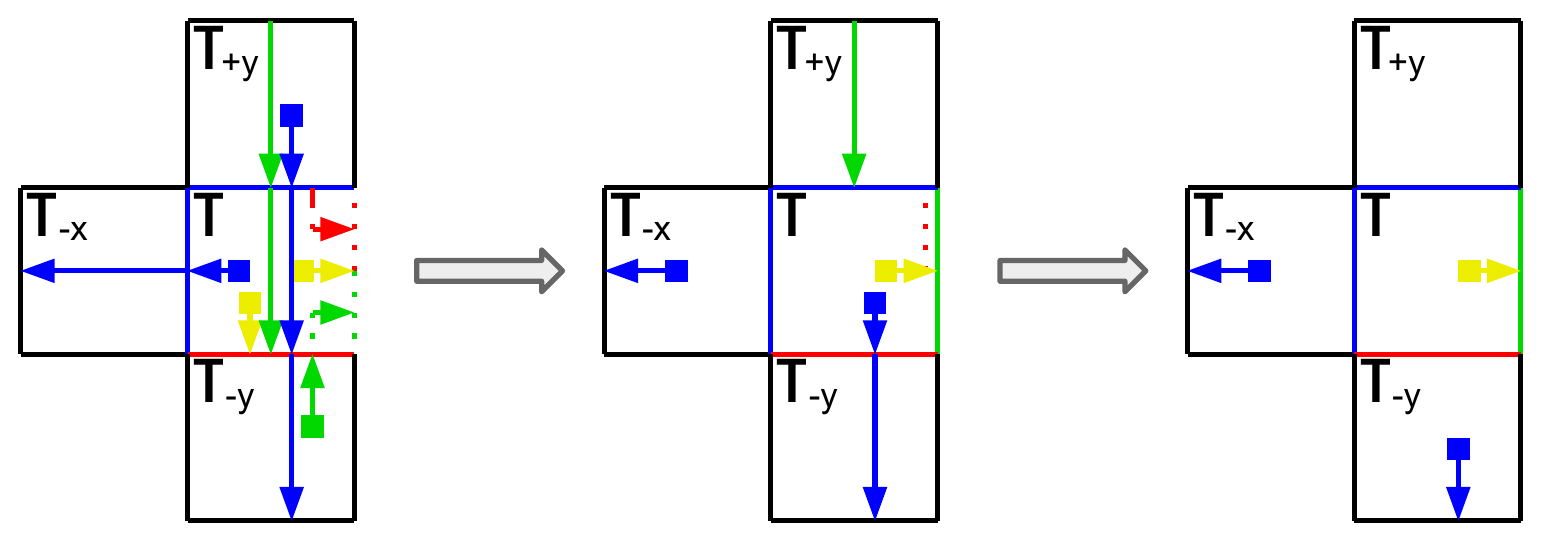}
\end{center}
\caption{The tile modification function acting on a tile assembly. Two iterations of the tile modification function complete the modification of the assembly. Refer to Example~\ref{ex:modfunction}.}
\label{fig:tilemod1}
\vskip -.7cm
\end{figure}

\begin{ex.}\label{ex:modfunction}
An example of $f$ acting on an assembly of four tiles is shown in Figure~\ref{fig:tilemod1}. We will focus on the tile $T=((t_{+y},t_{+x},t_{-y},t_{-x}),\mathcal{A},\mathcal{S})$ in the center of the assembly. Let $r,b,g,y$ correspond to the red, blue, green, and yellow labels, respectively, in the figure. For side modification, only the labels in $t_{+x}=(\emptyset,\{r,g\})$ are affected. In the first iteration, we have $C=\{g\}$ and $D=\emptyset$, because $g_{-y}^{+x}\in\mathcal{A}$ is an activation signal for the green label and the tile $T_{-y}$ contains an initiation signal $g_{0}^{+y}$, so we end up with $t_{+x}'=(\{g\},\{r\})$. For the activation signals, we begin with $\mathcal{A}=\{r_{+y}^{+x},g_{-y}^{+x}\}=\mathcal{A}_{\text{removed}}$ because $g_{-y}^{+x}$ was ``activated'' by $g_{0}^{+y}$ in the bottom tile and $r_{+y}^{+x}$ does not have a corresponding transmission signal in the top tile $T_{+y}$. So, $\mathcal{A}'=\emptyset$. Finally, in the transmission signal modification, $\mathcal{S}=\{b_{0}^{-x}, y_{0}^{+x}, y_{0}^{-y}, g_{+y}^{-y}, b_{+y}^{-y}\}$. $\mathcal{S}_{\text{added}}=\{b_{0}^{-y}\}$ since $b_{+y}^{-y}$ is adjacent to $b_{0}^{-y}$ in $T_{+y}$. For the same reason $b_{+y}^{-y}\in S_{1}$. $S_{1}$ does not contain $g_{+y}^{-y}$ because it is adjacent to $g_{+y}^{-y}$ in the $T_{+y}$; however, $g_{+y}^{-y}\in S_{2}$ because there is no corresponding transmission or activation signal in $T_{-y}$; and, $S_{3}=\{b_{0}^{-x}, y_{0}^{-y}\}$ because $T$ has neighbors on the $-x,-y$ sides but not on the $+x$ side, so $\mathcal{S}'=\{y_{0}^{+x}, b_{0}^{-y}\}$.

In the second iteration, $C=\emptyset$ and $D=\{r\}$ because there is no longer an activation signal for $r$ in $\mathcal{A}'$, so  $t_{+x}''=(\{g\},\emptyset)$. $\mathcal{A}'=\emptyset=\mathcal{A}''$ since activation signals can only be removed, never added, and $\mathcal{S}''=\{y_{0}^{+x}\}$ because $T$ has a neighbor on the $-y$ side and therefore $b_{0}^{-y}$ is removed.\hfill $\blacksquare$
\end{ex.}

\subsection{Hierarchical Tile Assembly Sets and the Active Tile Assembly System}
We use the term ``complete tile assembly'' to refer to a tile assembly which has reached a quiescent state under iterated application of the tile modification function. 
\begin{def.}
A \emph{complete tile assembly instance} for $\alpha$ is a tile assembly instance denoted $\hat{\alpha}$, such that
\[\hat{\alpha}=f^{n}(\alpha)\text{ for some }n\text{ and }f(\hat{\alpha})=\hat{\alpha}.\]
As noted, $[\hat{\alpha}]$ is the equivalence class of $\hat{\alpha}$ under translation, so we define the \emph{complete tile assembly} obtained from $[\alpha]$ as $\hat{[\alpha]}$.
\label{def:complete}\end{def.}
In figure~\ref{fig:tilemod1}, the tile assembly on the far right is complete.

A \emph{seed set} of unit tile representatives, denoted $\mathcal{T}_{0}$, is a set of unit tiles that is closed under $90^\circ$ rotations: for every $\alpha\in\mathcal{T}_{0}$, if $\alpha'(0,0)$ can be obtained from $\alpha(0,0)$ through a cyclic permutation of the direction coordinates $(+y, +x, -y, -x)$ in the tile description, then $\alpha'\in\mathcal{T}_{0}$.

Fix a temperature $\theta$. We define $\mathcal{T}_{i}$, $i\geq 1$, recursively as follows: $\alpha\in\mathcal{T}_i$, if and only if $\alpha\in\mathcal{T}_{i-1}$ or $\alpha=\hat{\alpha}'$ where $\alpha'$ is a tile assembly representative composed of two tile assembly instances, $\beta_{1}$ and $\beta_{2}$:
\[\alpha'(z)=\left\{\begin{array}{ll}
		    \beta_{1}(z) & \quad \text{if}\ z\in\text{dom}(\beta_{1})\\
		    \beta_{2}(z) & \quad \text{if}\ z\in\text{dom}(\beta_{2})\\
				\end{array} \right. \]
such that the domains of $\beta_{1}$ and $\beta_{2}$ are disjoint and such that there exist $\beta_{1}',\beta_{2}'\in\mathcal{T}_{i-1}$ with $[\beta_{1}]=[\beta_{1}']$ and $[\beta_{2}]=[\beta_{2}']$. In other words, $\alpha$ is in $\mathcal{T}_{i}$ if it is the completed instance of a composition of two tile assembly instances whose class representatives are in $\mathcal{T}_{i-1}$. 

We call $\mathcal{T}_{n}$ an \emph{active supertile set at stage} $n$. Observe that $\mathcal{T}_{i-1}\subseteq \mathcal{T}_{i}$.

\begin{def.}
An \emph{Active Tile Assembly System} (ATAS) is an ordered triple of a seed set, a strength function, and a temperature parameter: $(\mathcal{T}_{0},s,\theta)$, $\theta\in\Z^{+}$.
\end{def.}

\subsection{Recursive Assembly}\label{sec:recursive}
Now we proceed to define the recursive assembly of tile types and provide a bridge between the tiles of tiling theory and the unit squares of the TAM. We capitalize ``Tiles'' in what follows to emphasize that these are not necessarily unit squares. Intuitively, a Tile type could be thought of as a shape that consists of a fixed number of regions, each of which consists of subregions that all ``look'' the same (i.e., a region is a collection of instances of one kind of tile assembly). The recursive assembly allows the number of subregions in each region to increase without changing the region's ``kind,'' allowing the Tiles to grow into larger and larger versions of themselves.

Continuing formally, let $\mathcal{T}_{\infty}$ denote the set of all tile assembly representatives producible by an ATAS, i.e., $\displaystyle{\mathcal{T}_{\infty}=\bigcup_{n=0}^{\infty} \mathcal{T}_{n}}$. Then a \emph{Tile type} is a function
\[T:\Z^{+}\rightarrow\mathcal{T}_{\infty}\]
and $T(\ell)$ is a Tile type $T$ at level $\ell$. Furthermore, a Tile type $T$ is \emph{self-similar} if $\text{dom} (T(\ell+1))$ is geometrically similar to $\text{dom} (T(\ell))$ for all $\ell\geq 0$.

A finite set $\Theta=\{T_{1},\ldots,T_{n}\}$ of Tile types is said to be \emph{recursive} if for each $\ell>0$, $T_{i}(\ell)$ satisfies the following conditions\footnote{By union we mean the piecewise joining of the indicated configurations (recall that a configuration is a mapping from $\Z^{2}$ to the set of possible active tiles).}:
\[T_{i}(\ell)=\bigcup_{j=1}^{\rho(i)} R_{i,j}(\ell)\]
where $\rho(i)$ is the number of regions in the $i$th Tile type, and $R_{i,j}(\ell)$ is a configuration (but not necessarily a tile assembly instance), referred to as the $j$th \emph{region} in the $i$th Tile type (at level $\ell$. We note that for each Tile type $i$, the number of regions $\rho(i)$ does not depend on the level $\ell$.

We require that the regions are unions of tile assembly instances whose number increases with the level. The regions $R_{i,j}$ therefore satisfy
\[\text{dom}\left(R_{i,j_{1}}\left(\ell\right)\right) \cap \text{dom}\left(R_{i,j_{2}}\left(\ell\right)\right)=\emptyset\text{ for }j_{1}\neq j_{2}\] 
and
\[R_{i,j}(\ell)=\bigcup_{k=1}^{\eta_{i,j}(\ell)}r_{i,j}^{k}(\ell)\]
where $\eta_{i,j}:\N\rightarrow\N$ is a non-decreasing function with $\eta_{i,j}(\ell)$ specifying the number of subregions $r_{i,j}^{k}(\ell)$ in the $j$th region of the $i$th Tile type at level $\ell$. 

The \emph{subregions} $r_{i,j}^{k}(\ell)$, $r_{i,j}^{k'}(\ell)$ must be assembly instances with non-overlapping domains for $k\neq k'$ and further, $\forall k,k'\geq 0$,
\begin{enumerate}
\item $\forall\ell\ [r_{i,j}^{k}(\ell)]=[r_{i,j}^{k'}(\ell)]$
\item (a) $\forall\ell, m\ [r_{i,j}^{k}(\ell)]=[r_{i,j}^{k'}(m)]$ or \\(b) $\exists\ell_{0},q$ such that $\forall\ell\ [r_{i,j}^{k}(\ell)]=[T_{q}(\ell_{0}+\ell)]$.
\end{enumerate}
That is, all of the subregions in a given region must be instances of the same tile assembly for a fixed $\ell$ (condition 1) and also either be instances of the same tile assembly across all $\ell$ or be Tile types (conditions 2 (a) and (b) respectively). Note that $q$ and $\ell_{0}$ are specific to the $j$th region of the $i$th Tile type.

\begin{rem}
Observe that condition 2(b) allows a Tile type $i$ at level $\ell$ to contain a Tile type $j\neq i$ at level $\ell'$ as a subregion even if $\ell'\geq \ell$, since $T_{i}(\ell)$ need not appear before $T_{j}(\ell')$ does. One or more levels of some Tile types may assemble before the assembly of the first level of other Tile types begins.
\end{rem}

Thus, a Tile type in a recursive set is a sequence of tile assembly representatives indexed by level, each of which is the union of a number of configurations (regions) that is fixed for the Tile type and independent of level. Each of these regions, in turn, is the union of tile assembly instances that all belong to the same class but whose number is allowed to increase as the level increases. The class of instances corresponding to a given region, on the other hand, is either fixed and independent of the level or else encompasses a single Tile type. If the region $R$ corresponds to a single Tile type $T$, then $R$ at level 0, $R(0)$, consists of translated copies of $T(\ell')$ for some $\ell'\geq 0$ and, in general, $R(\ell)$ consists of translated copies of $T(\ell'+\ell)$.

\begin{def.} A subset $\Theta'=\{T_{i_{1}},\ldots,T_{i_{m}}\}$ of a recursive set of Tile types $\Theta=\{T_{1},\ldots,T_{n}\}$ where each $T_{i_{k}}$ is self-similar is called \emph{strongly self-similar} if for every region $R_{i_{k},j}$, $k=1,\ldots,m$, there exists a Tile type $T_{i_{k'}}\in\Theta'$ such that for all $\ell$, dom$\left(R_{i_{k},j}\left(\ell\right)\right)$ is geometrically similar to dom$\left(T_{i_{k'}}\right)$.
\end{def.}

In other words, a subset $\Theta'$ of a recursive set $\Theta$ is strongly self-similar if every region of each Tile type in $\Theta'$ is geometrically similar to some Tile type in $\Theta'$ (specific to the region).

\begin{rem}
Note the distinction between \emph{self-similar} and \emph{strongly self-similar}: the former is a property of the overall shape of a single Tile type, the latter is a property of the shapes of individual regions in a set of Tile types.
\end{rem}

\section{L-shape}

\begin{figure}[hbt]
\begin{center}
\includegraphics[scale=.5]{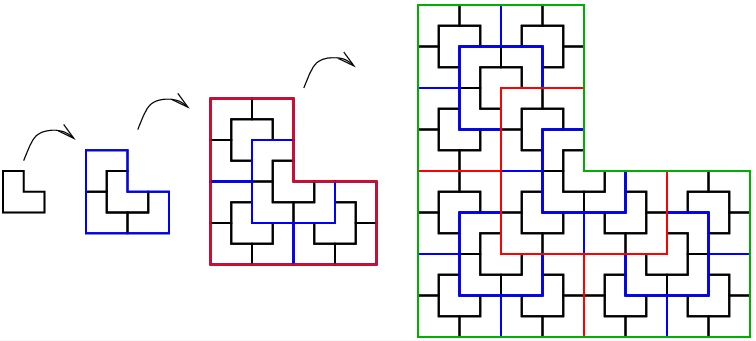}
\end{center}
\caption{Self-Similar L-Shape Tiling}
\label{fig:L-Shape}
\vskip -.7cm
\end{figure}

In the following section we present an active tile assembly system based on a well-known aperiodic
tiling of the plane obtained by a substitution tiling of an L-shaped tile (Fig.\ref{fig:L-Shape}). The substitution process consists of repeatedly ``inflating'' the L-shaped tile to obtain a larger L-shape. The larger L-shape is in fact produced by joining four smaller L-shaped tiles together: every subsequent tiling level is obtained from the previous one by substituting each L-shaped tile with four newly assembled L-shaped tiles. This is indicated in Fig.\ref{fig:L-Shape} where each level of the substitution is outlined with a different color. The tiling of the plane generated in this process is aperiodic\cite{Goodman}.

We provide a recursive set of L-shape Tile types and an ATAS which generates them. Then, we discuss the recursive properties of the system and prove that it yields an aperiodic tiling of the plane.

\subsection{Active Tile Assembly System for the L-Shape Tiling}
\begin{figure}[htb]
\begin{center}
\includegraphics[scale=.75]{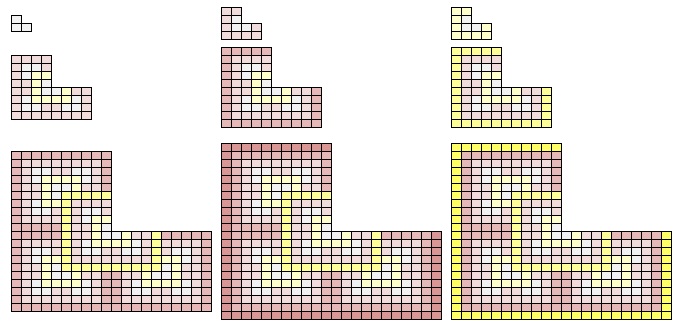}
\end{center}
\caption{Outline of three iterations of the L-shape unit tile construction. A yellow border represents the center role and a red one represents the outside role. Each successive level unbordered shape (left column) is composed of three red-bordered shapes (center column) and one yellow-bordered shape (right column) of the preceding level.}
\label{fig:L-ShapeTiling}
\vskip -.7cm
\end{figure}

The premise in the unit tile construction that we present here is illustrated in Fig.~\ref{fig:L-ShapeTiling}. We begin by joining three unit tiles together to obtain the smallest (level 0) L-shape. Then, one of two types of borders is built around each L-shape (yellow or red in the figure) and depending on the type of border, the L-shape assumes either the center role (yellow) or an outside role (red) in the next level of the assembly process. Each subsequent level of the assembly is formed by three outside L-shapes and one center L-shape. To proceed to the next level, a border is built once again around each of the assembled L-shapes and the process is repeated. We treat the two bordered versions of the smallest L-shape as two distinct Tile types in addition to the unbordered type. Fig.~\ref{fig:L-ShapeTiling} outlines three iterations of the process yielding levels 0, 1 and 2 of each Tile type.

We show that the active tile assembly system which produces the desired effect is $(\mathcal{T}_{0},s,2)$, with $\mathcal{T}_{0}$ and $s$ given in Fig.~\ref{fig:Time0}. Formally, the tiles are given in Tables~\ref{tab:L-sides} and \ref{tab:L-signals} in the Appendix. 

All of the tiles in the figure (and the two tables) are presented modulo rotation, so, for example, the tile G3, given by:
\begin{align*}
t_{+y} &= \lp\left\{ 1 \right\}, \emptyset\rp\\
t_{+x} &= \lp\left\{ 2 \right\}, \emptyset\rp\\
t_{-y} &= \lp\left\{ -66 \right\}, \emptyset\rp\\
t_{-x} &= \lp\emptyset, \left\{ -4 \right\}\rp\\
\A	   &= \{ 4_{+y}^{-x} \}\\
\St	   &= \{ 2_{0}^{-y}, 55_{-x}^{+y}, 55_{-y}^{-x} \}
\end{align*}
stands to also represent the tiles given by:
\begin{align*}
t_{+y} &= \lp\left\{ 2 \right\}, \emptyset\rp & t_{+y} &= \lp\left\{ -66 \right\}, \emptyset\rp & t_{+y} &= \lp\emptyset, \left\{ -4 \right\}\rp\\
t_{+x} &= \lp\left\{ -66 \right\}, \emptyset\rp & t_{+x} &= \lp\emptyset, \left\{ -4 \right\}\rp & t_{+x} &= \lp\left\{ 1 \right\}, \emptyset\rp\\
t_{-y} &= \lp\emptyset, \left\{ -4 \right\}\rp & t_{-y} &= \lp\left\{ 1 \right\}, \emptyset\rp & t_{-y} &= \lp\left\{ 2 \right\}, \emptyset\rp \\
t_{-x} &= \lp\left\{ 1 \right\}, \emptyset\rp & t_{-x} &= \lp\left\{ 2 \right\}, \emptyset\rp & t_{-x} &= \lp\left\{ -66 \right\}, \emptyset\rp\\
\A	   &= \{ 4_{-x}^{-y} \} & \A &= \{ 4_{-y}^{+x} \} & \A &= \{ 4_{+x}^{+y} \}\\
\St	   &= \{ 2_{0}^{+x}, 55_{-y}^{-x}, 55_{+x}^{-y} \} & \St &= \{ 2_{0}^{+y}, 55_{+x}^{+y}, 55_{+y}^{+x} \} & \St &= \{ 2_{0}^{-x}, 55_{+y}^{+x}, 55_{-x}^{+y} \}.
\end{align*}

For the remainder of this paper, we adopt the following notational convention: if $\tau$ is a unit tile in Figure~\ref{fig:Time0}, then $\tau_{N}$ refers to $\tau$ exactly as shown and $\tau_{E},\tau_{S},\tau_{W}$ refer respectively to 1, 2, and 3 counter-clockwise $90^\circ$ rotations of $\tau_{N}$. Thus, for example, G3$_N$ is G3 as in the first case listed and the three rotations are G3$_{E}$, G3$_{S}$, and G3$_{W}$ respectively. Thus, there are 28, up to rotation, equivalence classes encompassing a total of 112 unit tiles required for this tiling (only 28 DNA tiles in application, however, since physical DNA tiles are free to rotate in the plane).

\begin{figure}[hbt]
\begin{center}
\includegraphics[scale=.38]{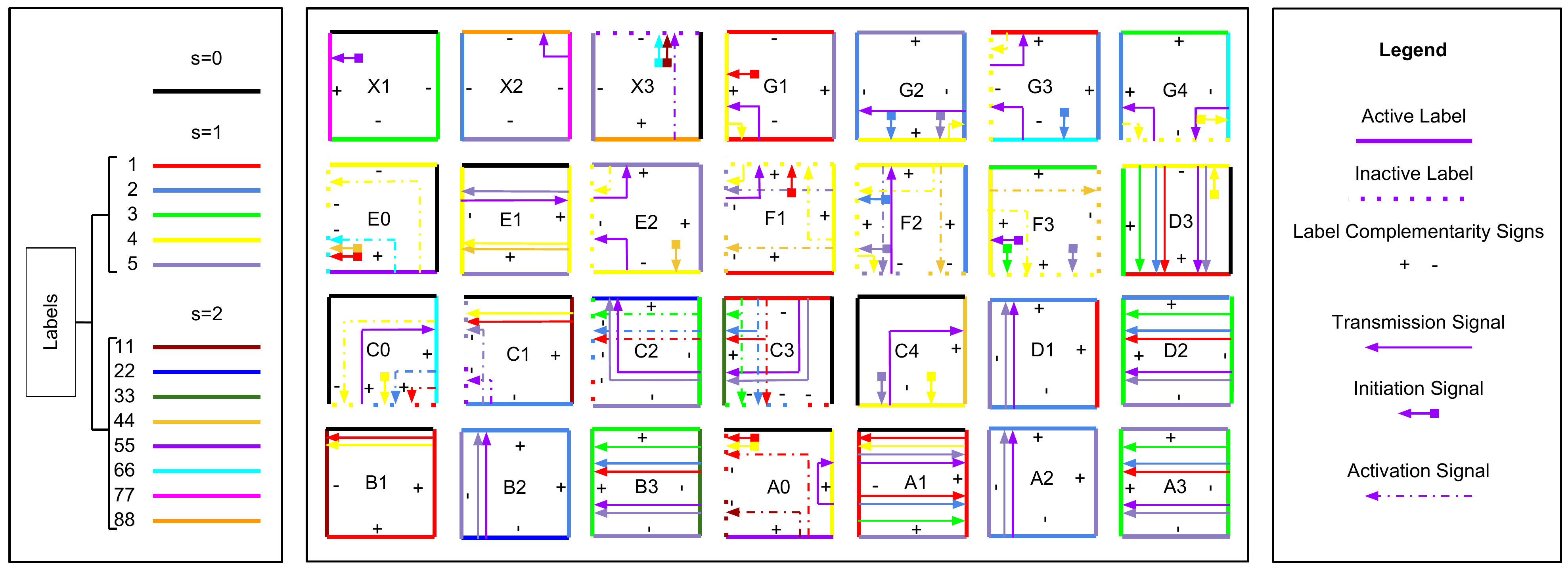}
\end{center}
\caption{The 28 rotational representatives for the set $\mathcal{T}_{0}$ of unit tiles for the L-shape tiling. The strength function $s$ is shown on the left: $s(c)=1$ for $c=1,2,3,4,5$ and $s(c)=2$ for $c=11,22,33,44,55,66,77,88$. Note that a black edge ($s=0$) represents $(A,I)=(\emptyset,\emptyset)$, i.e., the absence of any labels, and is not a label in itself.}
\vskip -.7cm
\label{fig:Time0}
\end{figure}

As shown in Fig.~\ref{fig:Time0}, labels 1, 2, 3, 4, and 5 are assigned a strength of 1, and labels 11, 22, 33, 44, 55, 66, 77, and 88 are assigned a strength of 2 (see Def.~\ref{def:strength}). Note that these labels are to be interpreted only as strings of symbols and do not hold any significance as numbers. 

Furthermore, the temperature parameter in $(\mathcal{T}_{0},s,2)$ set to ``2'' means that the sum of the strengths of all bonds which would have to be cut in order to separate a tile assembly into two parts would have to be at least 2. A consequence of this is that at the initial stage, only unit tiles with complementary labels of strength 2 can bind. 

\begin{figure}[hbt]
\begin{center}
\includegraphics[scale=.4]{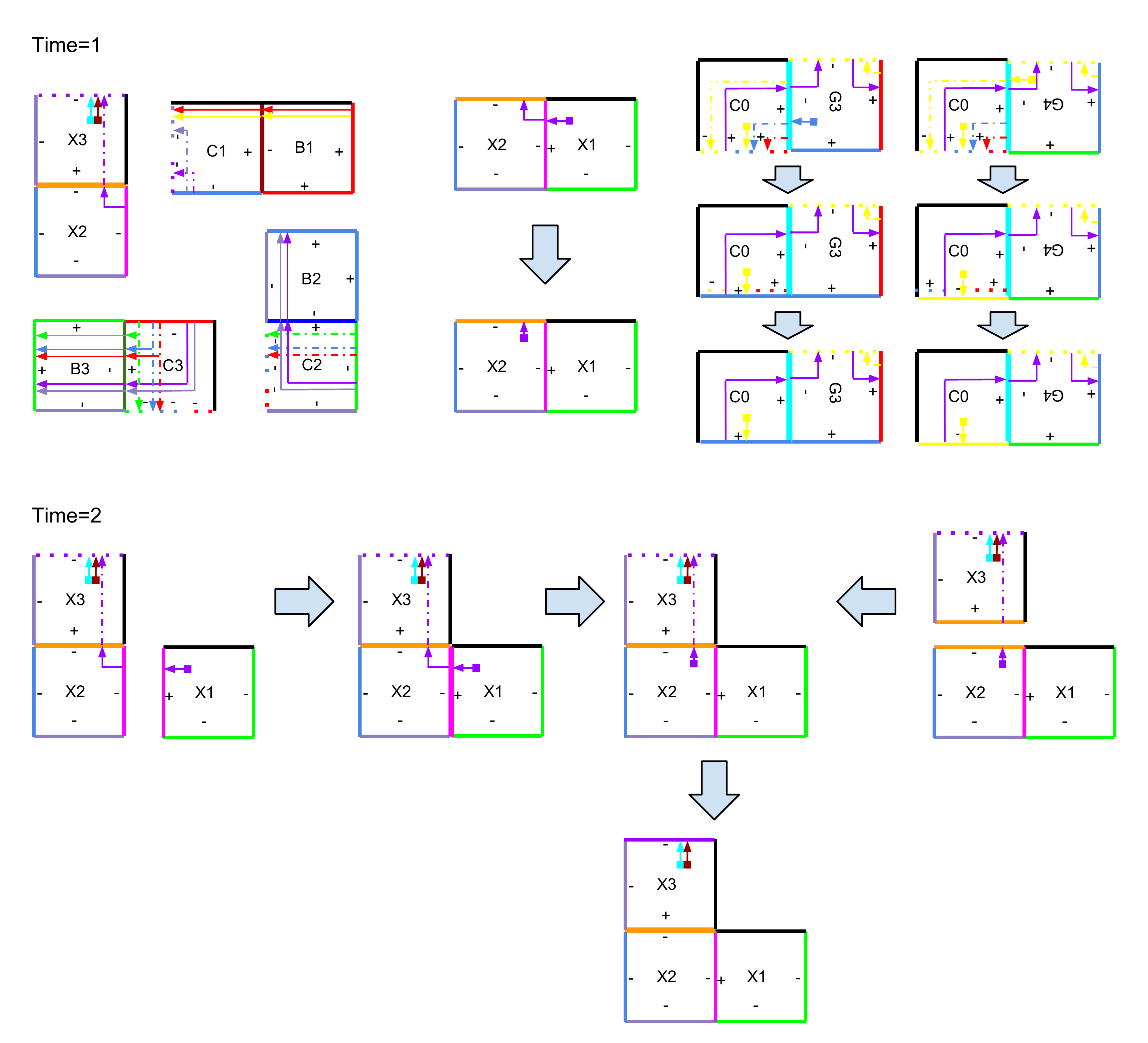}
\end{center}
\caption{Time 1 shows the assemblies added to $\mathcal{T}_{1}$. Time 2 shows the assemblies added to $\mathcal{T}_{2}$, the first L-shape assemblies in the construction.}
\label{fig:Time1-2}
\vskip -.7cm
\end{figure}

The initial assembly is shown in Fig.~\ref{fig:Time1-2}, under Time 1. The four assemblies on the left are complete (see Def.~\ref{def:complete}) assemblies as soon as their unit tiles bind because there is no signaling to be transmitted; the assembly X2-X1 requires one iteration of the tile modification function, as shown, to become a complete assembly; and the remaining two assemblies become complete after two\footnote{The first iteration activates labels 2 and 4 for assemblies C0-G3 and C0-G4 respectively and removes ineffectual signal pathways. The second iteration removes the ineffectual labels.} iterations of $f$. Thus, $\mathcal{T}_{1}$ is the set $\mathcal{T}_{0}$ plus the seven (modulo rotation) two-tile assemblies added at Time 1. 
\begin{rem}
Note that the word ``Time'' here counts the assembly stages. The assembly figures depict only the new tile assemblies added at each stage. The tile assemblies which can interact at a given stage $i$ (or ``Time'' $i$) are the tile assemblies in $\mathcal{T}_{i}$ and are given by the union of all tile assemblies shown in the preceding stages.
\end{rem}
At the next stage, Time 2 (Fig.~\ref{fig:Time1-2}), the only new bonds that can form are between X1 and the tile assembly X2-X3 or between X3 and the tile assembly X2-X1. The completed assembly in either case is the first L-shape: X1-X2-X3 (in the first case, it takes two iterations of $f$, in the second only one). It can be easily verified that no other additions can occur at this stage: that would require side by side labels on one of the two-tile assemblies to be matched with those of another two-tile assembly and none of the seven assemblies can be paired in that way.

Further stage by stage assembly is given in detail in Section~\ref{sec:self-assembly}. For now, we state the following key properties about the system.

\begin{thm}\label{thm:recset}
The active tile assembly system $(\mathcal{T}_{0},s,2)$ as given in Fig.~\ref{fig:Time0} yields a recursive set $\Theta$ of self-similar L-shape Tile types with a strongly self-similar subset. \end{thm}
We give the construction of a recursive set for the system in the next section (Section~\ref{sec:recset}) and show that it is, in fact, produced by the system in Section~\ref{sec:self-assembly}.

\begin{thm}\label{thm:unique}
All of the assemblies produced by the L-shape ATAS are components of the L-shape Tile types in set $\Theta$ from Theorem~\ref{thm:recset}. \end{thm}
The above theorem claims that the system produces only structures that take part in the assembly of some level of a Tile type in $\Theta$. This is proved in Section~\ref{sec:self-assembly}.

\begin{thm}\label{thm:aperiodicity}
The L-shape ATAS induces an aperiodic tiling of the plane.
\end{thm}
We prove this in Section~\ref{sec:aperiodicitypf}.

\subsection{A recursive set of L-shape Tile types}\label{sec:recset}
In the following discussion we write $\tau_{1}\tau_{2}$ to represent the only possible assembly formed by the two tiles in the ATAS $\left(\mathcal{T}_{0},s,2\right)$: e.g., C1$_{N}$B1$_{N}$ refers to the assembly formed by C1 and B1 in Figure~\ref{fig:Time1-2} and X2$_{N}$X3$_{N}$ to the assembly formed by X2 and X3 in the same figure (note that since there is only one way for these unit tiles to form an assembly, there is no ambiguity in notation). Also, for any assembly $T$, $T+(u,v)$ will mean a translation of the class representative of $T$ by the vector $(u,v)$.

We now proceed to define $\Theta=\{T_{1},\ldots,T_{12}\}$, a set of 12 (L-shaped) Tile types for $\left(\mathcal{T}_{0},s,2\right)$, to prove Theorem~\ref{thm:recset}.

 \begin{wraptable}{r}{.5\textwidth}
\begin{center}
\renewcommand{\arraystretch}{1.2}
\begin{tabular}{|c|r|c|c|c|}
\hline
\multicolumn{5}{|c|}{$\eta_{i,j}(\ell)$} \\
\hline\hline
\multicolumn{2}{|c|}{$i\mod 3$} & 1 & 2 & 0\\ 
\hline\hline
\multirow{17}{*}{$j$} & 1 & 1 & 1 & 1 \\
\cline{2-5}
 & 2 & 1 & 1 & 1 \\
\cline{2-5}
 & 3 & 1 & $3\cdot 2^{\ell}-4$ & $3\cdot 2^{\ell}-4$ \\
\cline{2-5} 
& 4 & 1 & 1 & 1 \\
\cline{2-5}
 & 5 & - & $3\cdot 2^{\ell+1}-6$ & 1 \\
\cline{2-5}
 & 6 & - & 1 & $3\cdot 2^{\ell+1}-7$ \\
\cline{2-5}
 & 7 & - & 1 & 1 \\
\cline{2-5}
 & 8 & - & $3\cdot 2^{\ell+1}-6$ & 1 \\
\cline{2-5}
 & 9 & - & 1 & 1 \\
\cline{2-5}
 & 10 & - & 1 & 1 \\
\cline{2-5}
 & 11 & - & $3\cdot 2^{\ell}-3$ & $3\cdot 2^{\ell+1}-7$ \\
\cline{2-5}
 & 12 & - & 1 & 1 \\
\cline{2-5}
 & 13 & - & - & 1 \\
\cline{2-5}
 & 14 & - & - & 1 \\
\cline{2-5}
 & 15 & - & - & 1 \\
\cline{2-5}
 & 16 & - & - & $3\cdot 2^{\ell}-3$ \\
\cline{2-5}
 & 17 & \phantom{$3\cdot 2^{\ell}$}-\phantom{$3\cdot 2^{\ell}$} & - & 1 \\
\hline
\end{tabular}\vspace{10pt}
\caption{Table of values for $\eta_{i,j}(\ell)$.}
\label{tab:subregions}
\end{center}
\end{wraptable}
We let the four $T_{i}$ for $i = 1\ \text{mod}\ 3$ represent the four possible rotations of the ``unbordered'' Tile types, for $i = 2\ \text{mod}\ 3$ - the ``A0-type bordered'' (or edge) Tile types, and for $i = 0\ \text{mod}\ 3$ - the ``E0-type bordered'' (or center) Tile types. The ``A0'' and ``E0'' in the name refer to the unit tiles A0 and E0 in Figure~\ref{fig:Time0} which initiate the construction of the corresponding border in each case, discussed in Section~\ref{sec:self-assembly}. In the following sections, we discuss only $T_{1}, T_{2}, T_{3}$ in full detail as the other Tile types are equivalent up to rotation.

The number of regions for all $T_{i}$ is specified as:
$$
\hspace{-8.5cm}
\rho(i)=\left\{\begin{array}{rrr}
		    4, & \quad i = 1\ \text{mod}\ 3\\
		    12, & \quad i = 2\ \text{mod}\ 3\\
   		    17, & \quad i = 0\ \text{mod}\ 3 
				\end{array} \right. 
$$
Thus, $T_{1}$ consists of 4 regions, $T_{2}$ consists of 12, $T_{3}$ of 17, and so on. Recall that the $j$th region of $T_{i}$, denoted $R_{i,j}$, a configuration, is defined as the union of assembly instances $r_{i,j}^{k}$, the subregions. We give the number of a region's subregions in terms of level $\ell$ in Table \ref{tab:subregions} as $\eta_{i,j}(\ell)$.

The type-specific descriptions are grouped into three sections: in each, we give $T_{i}(0)$ explicitly for all values of $i$ and also give the region and subregion descriptions, accompanied by a diagram, for a rotational representative of the four types.

\subsubsection{The unbordered Tile types}

\begin{table} [h!]
\begin{center}
\begin{tabular}{c c|c c}
\multicolumn{4}{l}{$T_{1}(0)$} \\
\multirow{2}{*}{$y$} & 1 & X3$_{N}$ & \\
 & 0 & X2$_{N}$ & X1$_{N}$\\
\cline{2-4}
 & & 0 & 1\\
 & \multicolumn{1}{c}{} & \multicolumn{2}{c}{$x$}\\
\end{tabular}
\begin{tabular}{c c|c c}
\multicolumn{4}{l}{$T_{4}(0)$} \\
\multirow{2}{*}{$y$} & 1 &  & X1$_{E}$ \\
 & 0 & X3$_{E}$ & X2$_{E}$\\
\cline{2-4}
 & & 0 & 1\\
 & \multicolumn{1}{c}{} & \multicolumn{2}{c}{$x$}\\
\end{tabular}
\begin{tabular}{c c|c c}
\multicolumn{4}{l}{$T_{7}(0)$} \\
\multirow{2}{*}{$y$} & 1 & X1$_{S}$ & X2$_{S}$ \\
 & 0 &  & X3$_{S}$\\
\cline{2-4}
 & & 0 & 1\\
 & \multicolumn{1}{c}{} & \multicolumn{2}{c}{$x$}\\
\end{tabular}
\begin{tabular}{c c|c c}
\multicolumn{4}{l}{$T_{10}(0)$} \\
\multirow{2}{*}{$y$} & 1 & X2$_{W}$ & X3$_{W}$ \\
 & 0 & X1$_{W}$ & \\
\cline{2-4}
 & & 0 & 1\\
 & \multicolumn{1}{c}{} & \multicolumn{2}{c}{$x$}\\
\end{tabular}
\caption{Table of $\ell=0$ assemblies for the unbordered Tile types.}
\label{tab:unborderedL0}
\end{center}
\end{table}
 In this section, we look at $T_{1},T_{4},T_{7},$ and $T_{10}$. The initial - level 0 - configurations $T_{1}(0),T_{4}(0),T_{7}
 (0),T_{10}(0)$ are given in Table \ref{tab:unborderedL0}. Recall that a Tile type at a given level is an assembly 
 representative (see end of Section~\ref{sec:tileassembly}), a map from the integer lattice into the set of active tiles. As 
 shown in Fig.~\ref{fig:Time1-2}, these initial assemblies are obtained in $\mathcal{T}_{2}$, the second stage of the ATAS 
 $(\mathcal{T}_{0},s,2)$. Refer to Figure~\ref{fig:Lvlnshape} for a diagram of the four regions of $T_{1}(\ell),T_{4}
 (\ell),T_{7}(\ell),T_{10}(\ell)$ for $\ell\geq 1$. These regions for $T_{1}$ are given as follows:

 \begin{figure}[h]
\begin{minipage}{.35\textwidth}
\begin{center}{\ } \vskip 1.5cm
\begin{center}
\includegraphics[width=2.5in]{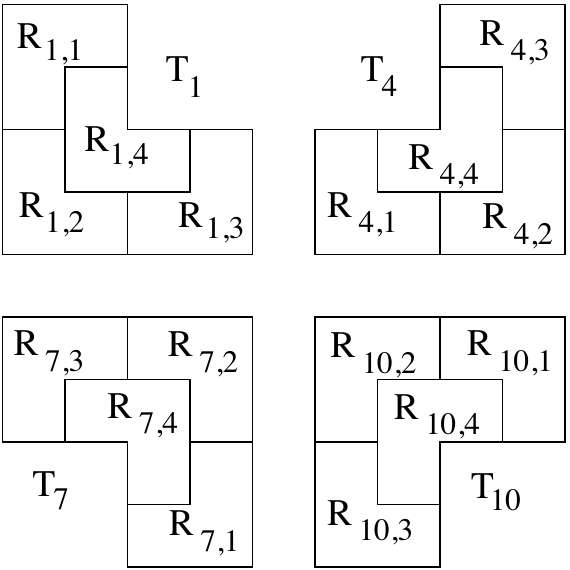}\\
\end{center}\vskip 1cm
(a)
\end{center}
\end{minipage}
\hfill
\begin{minipage}{.6\textwidth}
   \begin{center}
  \begin{center}
\includegraphics[scale=.45]{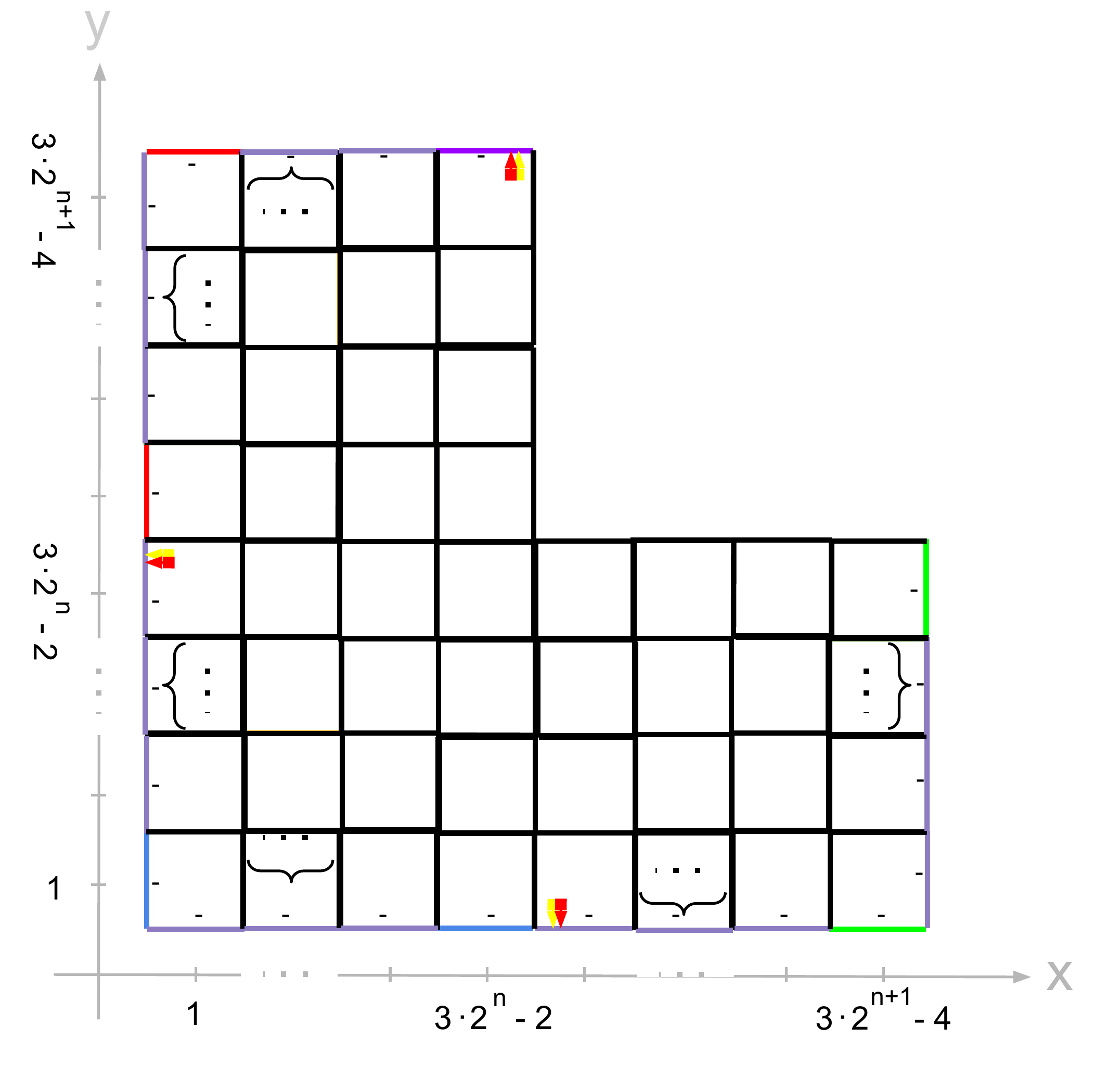}\\
\end{center}
(b)
   \end{center}
\end{minipage}
\caption{ (a) The regions of $T_{1}(\ell),T_{4}(\ell),T_{7}(\ell),$ and $T_{10}(\ell)$ for $\ell\geq 1$.  (b) {\small Level $n$ L-Shape Tile $T_{1}(n)+(1,1)$ in $\Z^{2}$ as a complete tile assembly. Tiles marked with ``...'' represent suppressed rows and columns consisting of identical tiles. The only signaling present is on the three tiles as indicated by the red and yellow arrows, i.e., signals of the form $1_{0}^{i}$, $4_{0}^{i}$ (see Section~\ref{sec:self-assembly}). The strength 1 labels -1,-2,-3,-5 are marked with red, blue, green, and light purple edges respectively. The only strength 2 label, -55 at $(3\cdot 2^{n}-2,3\cdot 2^{n+1}-4)$, is marked with a dark purple edge. No active or inactive labels other than the ones indicated are present.}}\label{fig:Lvlnshape}
\vskip -.7cm
\end{figure}

\begin{align*}
R_{1,1}(\ell) & = r_{1,1}^{1}(\ell) = T_{11}(\ell-1)+(0,3\cdot 2^{\ell}-2)\\ 
R_{1,2}(\ell) & = r_{1,2}^{1}(\ell) = T_{2}(\ell-1)+(0,0)\\ 
R_{1,3}(\ell) & = r_{1,3}^{1}(\ell) = T_{5}(\ell-1)+(3\cdot 2^{\ell}-2,0)\\ 
R_{1,4}(\ell) & = r_{1,4}^{1}(\ell) = T_{3}(\ell-1)+(3\cdot 2^{\ell -1}-1,3\cdot 2^{\ell -1}-1)
\end{align*} 
Recall that $T_{11},T_{2},T_{5}$ are A0-type bordered shapes and that $T_{3}$ is E0-type bordered (compare to the construction outlined in Figure~\ref{fig:L-ShapeTiling}, center and bottom shapes in the left column). Figure~\ref{fig:Lvlnshape} shows the outline of a general $T_{1}(n)$ Tile type. The Tile types $T_{4},T_{7},$ and $T_{10}$ are equivalent to $T_{1}$ up to rotation, so their region descriptions are analogous to the above and are thus omitted.
%
%
%
\subsubsection{A0-type bordered Tile types}
\begin{figure}[h]
\begin{center}
\includegraphics[scale=.4]{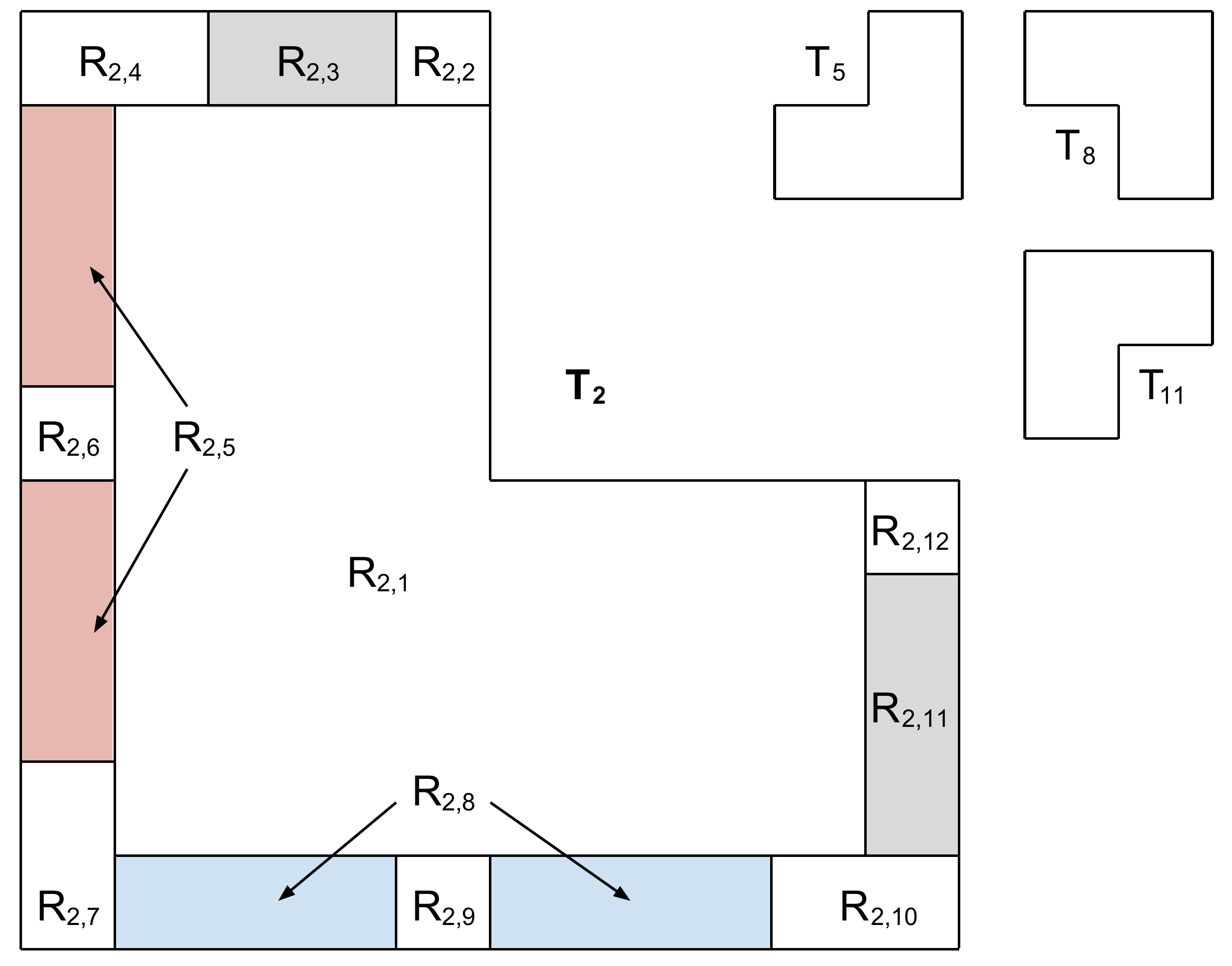}
\end{center}
\caption{The regions of $T_{2}(\ell)$ and outlines of $T_{5}(\ell),T_{8}(\ell)$ and $T_{11}(\ell)$ for $\ell\geq 1$. The regions marked in color consist of a number of subregions that increases with level. Note also that the two regions marked in red and blue ($R_{2,5}$ and $R_{2,8}$ respectively) each consist of two nonadjacent sets of subregions.}
\label{fig:A0TypeRegions}
\end{figure}

\begin{figure}[h!]
\begin{center}
\includegraphics[scale=.32]{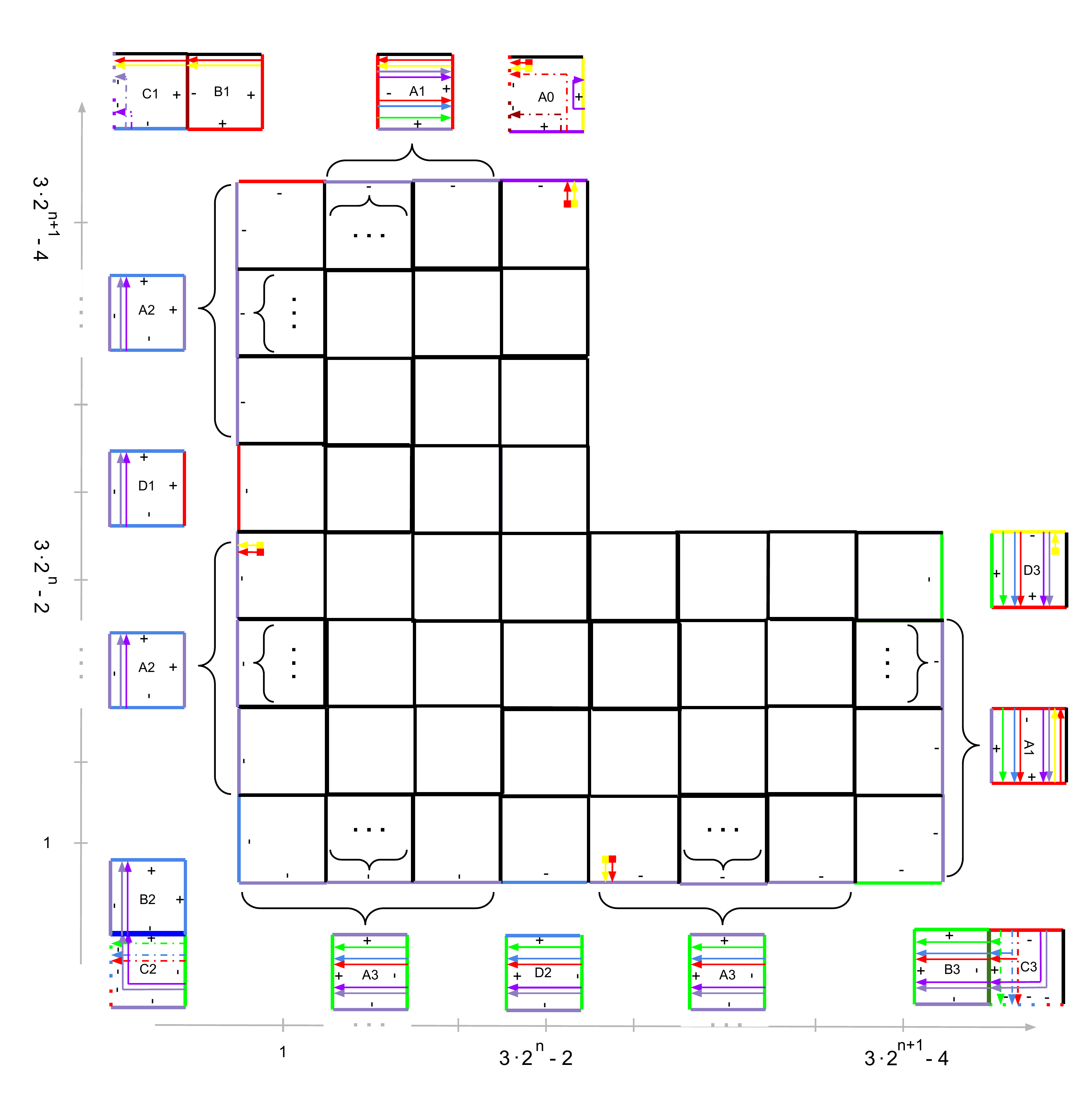}
\end{center}
\caption{Assembly of $T_{2}(n)$ Tile type: the A0-type border assembles around the Level $n$ L-Shape $T_{1}(n)+(1,1)$ (i.e., $T_{1}(n)$ positioned with bottom left corner tile at $(1,1)$).}
\label{fig:LvlnA0shape}
\end{figure}

We continue with $T_{2},T_{5},T_{8},$ and $T_{11}$. The level 0 instances $T_{2}(0),T_{5}(0),T_{8}(0),T_{11}(0)$ are given in Table \ref{tab:A0typeL0}. Refer to Figure \ref{fig:A0TypeRegions} for a diagram of the regions of $T_{2}(\ell)$ for $\ell\geq 1$. All regions of $T_{2}$ with the exception of $R_{2,1}$ consist of unit tile subregions forming a border around the $R_{2,1}$ region, which is the Tile type $T_{1}$ (see Figures~\ref{fig:A0TypeRegions}-\ref{fig:LvlnA0shape}). The regions are given as follows:
\begin{table} [t]
\begin{center}
\begin{tabular}{c c|c c c c}
\multicolumn{6}{l}{$T_{2}(0)$} \\
\multirow{4}{*}{$y$} & 3 & C1$_{N}$ & A0$_{N}$ & & \\
 & 2 & A2$_{N}$ & X3$_{N}$ & & \\
 & 1 & B2$_{N}$ & X2$_{N}$ & X1$_{N}$ & D3$_{N}$ \\
 & 0 & C2$_{N}$ & A3$_{N}$ & B3$_{N}$ & C3$_{N}$\\
\cline{2-6}
 & & 0 & 1 & 2 & 3\\
 & \multicolumn{1}{c}{} & \multicolumn{4}{c}{$x$}\\
\end{tabular}
\begin{tabular}{c c|c c c c}
\multicolumn{6}{l}{$T_{5}(0)$} \\
\multirow{4}{*}{$y$} & 3 & & & D3$_{E}$ & C3$_{E}$ \\
 & 2 & & & X1$_{E}$ & B3$_{E}$ \\
 & 1 & A0$_{E}$ & X3$_{E}$ & X2$_{E}$ & A3$_{E}$ \\
 & 0 & C1$_{E}$ & A2$_{E}$ & B2$_{E}$ & C2$_{E}$\\
\cline{2-6}
 & & 0 & 1 & 2 & 3\\
 & \multicolumn{1}{c}{} & \multicolumn{4}{c}{$x$}\\
\end{tabular}
\begin{tabular}{c c|c c c c}
\multicolumn{6}{l}{$T_{8}(0)$} \\
\multirow{4}{*}{$y$} & 3 & C3$_{S}$ & B3$_{S}$ & A3$_{S}$ & C2$_{S}$ \\
 & 2 & D3$_{S}$ & X1$_{S}$ & X2$_{S}$ & B2$_{S}$ \\
 & 1 & & & X3$_{S}$ & A2$_{S}$ \\
 & 0 & & & A0$_{S}$ & C1$_{S}$ \\
\cline{2-6}
 & & 0 & 1 & 2 & 3\\
 & \multicolumn{1}{c}{} & \multicolumn{4}{c}{$x$}\\
\end{tabular}
\begin{tabular}{c c|c c c c}
\multicolumn{6}{l}{$T_{11}(0)$} \\
\multirow{4}{*}{$y$} & 3 & C2$_{W}$ & B2$_{W}$ & A2$_{W}$ & C1$_{W}$ \\
 & 2 & A3$_{W}$ & X2$_{W}$ & X3$_{W}$ & A0$_{W}$ \\
 & 1 & B3$_{W}$ & X1$_{W}$ & & \\
 & 0 & C3$_{W}$ & D3$_{W}$ & & \\
\cline{2-6}
 & & 0 & 1 & 2 & 3\\
 & \multicolumn{1}{c}{} & \multicolumn{4}{c}{$x$}\\
\end{tabular}
\caption{Table of $\ell=0$ configurations for the A0-bordered Tile types.}
\label{tab:A0typeL0}
\end{center}
\end{table}
\begin{align*}
R_{2,1}(\ell) & = r_{2,1}^{1}(\ell) = T_{1}(\ell)+(1,1)\\ 
R_{2,2}(\ell) & = r_{2,2}^{1}(\ell) = A0_{N} + (3\cdot 2^{\ell}-2,3\cdot 2^{\ell +1}-3)\\ 
R_{2,4}(\ell) & = r_{2,4}^{1}(\ell) = C1_{N}B1_{N}+(0,3\cdot 2^{\ell +1}-3)\\ 
R_{2,6}(\ell) & = r_{2,6}^{1}(\ell) = D1_{N}+(0,3\cdot 2^{\ell}-1)\\
R_{2,7}(\ell) & = r_{2,7}^{1}(\ell) = B2_{N}C2_{N}+(0,0)\\  
R_{2,9}(\ell) & = r_{2,9}^{1}(\ell) = D2_{N}+(3\cdot 2^{\ell} -2,0)\\
R_{2,10}(\ell) & = r_{2,10}^{1}(\ell) = B3_{N}C3_{N}+(3\cdot 2^{\ell +1}-4,0)\\
R_{2,12}(\ell) & = r_{2,12}^{1}(\ell) = D3_{N}+(3\cdot 2^{\ell +1}-3,3\cdot 2^{\ell}-2)\\   
\end{align*}
with
\[R_{2,3}(\ell) = \bigcup_{k=1}^{3\cdot 2^{\ell}-4} r_{2,3}^{k}\]
where $r_{2,3}^{1}(\ell)=A1_{N}+(2,3\cdot 2^{\ell +1}-3)$ and  $r_{2,3}^{k+1}(\ell)=r_{2,3}^{k}(\ell)+(1,0)$;
\[R_{2,5}(\ell) = \bigcup_{k=1}^{3\cdot 2^{\ell +1}-6} r_{2,5}^{k}\]
where $r_{2,5}^{1}(\ell)=A2_{N}+(0,2),\quad r_{2,5}^{3\cdot 2^{\ell}-2}(\ell)=r_{2,5}^{3\cdot 2^{\ell}-3}(\ell)+(0,2)$ (this shift by 2 units reflects the skipping of the single unit tile region $R_{2,6}$, Fig.~\ref{fig:A0TypeRegions}), $\quad r_{2,5}^{k+1}(\ell)=r_{2,5}^{k}(\ell)+(0,1)$, $k\neq 3\cdot 2^{\ell}-3$;
\[R_{2,8}(\ell) = \bigcup_{k=1}^{3\cdot 2^{\ell +1}-6} r_{2,8}^{k}\]
where $r_{2,8}^{1}(\ell)=A3_{N}+(1,0),\quad r_{2,8}^{3\cdot 2^{\ell}-2}(\ell)=r_{2,8}^{3\cdot 2^{\ell}-3}(\ell)+(2,0)$ (this shift by 2 units reflects the skipping of the single unit tile region $R_{2,9}$, Fig.~\ref{fig:A0TypeRegions}), $\quad r_{2,8}^{k+1}(\ell)=r_{2,8}^{k}(\ell)+(1,0),\ k\neq 3\cdot 2^{\ell}-3$;
and
\[R_{2,11}(\ell) = \bigcup_{k=1}^{3\cdot 2^{\ell}-3} r_{2,11}^{k}\]
where $r_{2,11}^{1}(\ell)=A1_{W}+(3\cdot 2^{\ell +1}-3,1)$ and  $r_{2,11}^{k+1}(\ell)=r_{2,11}^{k}(\ell)+(0,1)$.
Refer to Figure~\ref{fig:LvlnA0shape} for an outline of the $T_{2}(n)$ type as a tile assembly. The constructions for $T_{5},T_{8},$ and $T_{11}$ are, once again, analogous to the above and are omitted here.

As we shall see in Section~\ref{sec:self-assembly}, both in the case of A0-type and E0-type bordered assemblies, the actual construction of the Tile type at a level ``starts'' with the unbordered assembly in region 1 binding to a unit tile (A0 or E0) in region 2, followed by the region 3 being filled in one tile at a time. In our case, the filling of the regions is sequential by design. But from the perspective of the recursive Tile type set, it does not matter how or in what order the regions are filled, only that it is possible to obtain structures with all regions filled in the indicated way.
\subsubsection{E0-type bordered Tile types}
\begin{table}[h!]
\begin{center}
\begin{tabular}{c c|c c c c}
\multicolumn{6}{l}{$T_{3}(0)$} \\
\multirow{4}{*}{$y$} & 3 & C0$_{N}$ & E0$_{N}$ & & \\
 & 2 & G1$_{N}$ & X3$_{N}$ & & \\
 & 1 & G3$_{N}$ & X2$_{N}$ & X1$_{N}$ & F3$_{E}$ \\
 & 0 & C0$_{E}$ & G2$_{N}$ & G4$_{N}$ & C0$_{S}$\\
\cline{2-6}
 & & 0 & 1 & 2 & 3\\
 & \multicolumn{1}{c}{} & \multicolumn{4}{c}{$x$}\\
\end{tabular}
\begin{tabular}{c c|c c c c}
\multicolumn{6}{l}{$T_{6}(0)$} \\
\multirow{4}{*}{$y$} & 3 & & & F3$_{S}$ & C0$_{W}$ \\
 & 2 & & & X1$_{E}$ & G4$_{E}$ \\
 & 1 & E0$_{E}$ & X3$_{E}$ & X2$_{E}$ & G2$_{E}$ \\
 & 0 & C0$_{E}$ & G1$_{E}$ & G3$_{E}$ & C0$_{S}$\\
\cline{2-6}
 & & 0 & 1 & 2 & 3\\
 & \multicolumn{1}{c}{} & \multicolumn{4}{c}{$x$}\\
\end{tabular}
\begin{tabular}{c c|c c c c}
\multicolumn{6}{l}{$T_{9}(0)$} \\
\multirow{4}{*}{$y$} & 3 & C0$_{N}$ & G4$_{S}$ & G2$_{S}$ & C0$_{W}$ \\
 & 2 & F3$_{W}$ & X1$_{S}$ & X2$_{S}$ & G3$_{S}$ \\
 & 1 & & & X3$_{S}$ & G1$_{S}$ \\
 & 0 & & & E0$_{S}$ & C0$_{S}$ \\
\cline{2-6}
 & & 0 & 1 & 2 & 3\\
 & \multicolumn{1}{c}{} & \multicolumn{4}{c}{$x$}\\
\end{tabular}
\begin{tabular}{c c|c c c c}
\multicolumn{6}{l}{$T_{12}(0)$} \\
\multirow{4}{*}{$y$} & 3 & C0$_{N}$ & G3$_{W}$ & G1$_{W}$ & C0$_{W}$ \\
 & 2 & G2$_{W}$ & X2$_{W}$ & X3$_{W}$ & E0$_{W}$ \\
 & 1 & G4$_{W}$ & X1$_{W}$ & & \\
 & 0 & C0$_{E}$ & F3$_{N}$ & & \\
\cline{2-6}
 & & 0 & 1 & 2 & 3\\
 & \multicolumn{1}{c}{} & \multicolumn{4}{c}{$x$}\\
\end{tabular}
\caption{Table of $\ell=0$ configurations for the E0-bordered Tile types.}
\label{tab:E0typeL0}
\end{center}
\end{table}

\begin{figure}[h]
\begin{center}
\includegraphics[scale=.35]{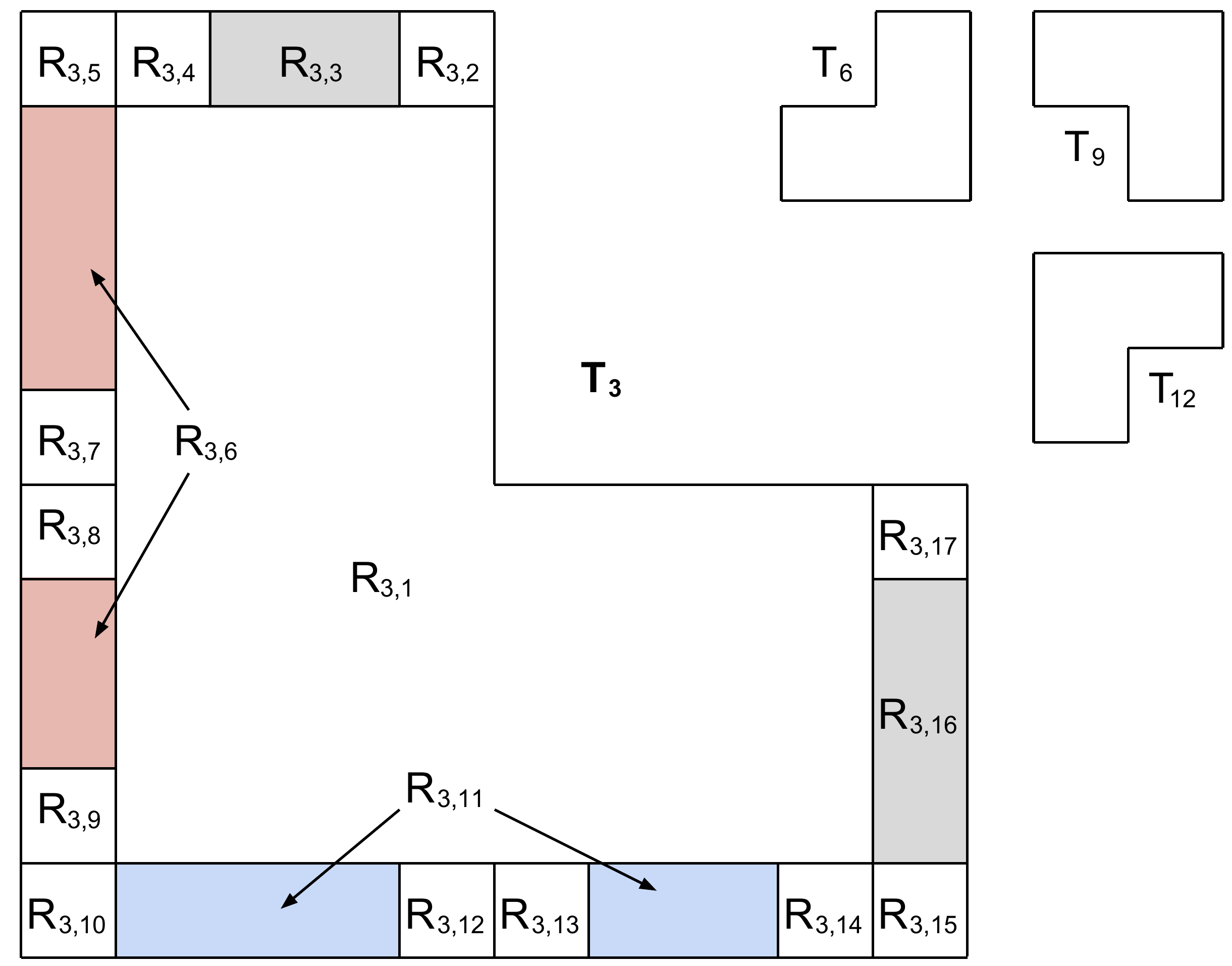}
\end{center}
\caption{The regions of $T_{3}(\ell)$ and outlines of $T_{6}(\ell),T_{9}(\ell)$ and $T_{12}(\ell)$ for $\ell\geq 1$. The regions marked in color consist of a number of subregions that increases with level. Just as in Figure \ref{fig:A0TypeRegions}, the two regions marked in red and blue ($R_{3,6}$ and $R_{3,11}$ respectively) each consist of two nonadjacent sets of subregions.}
\label{fig:E0TypeRegions}
\vskip -.7cm
\end{figure}
\begin{figure}[h!]
\begin{center}
\includegraphics[scale=.32]{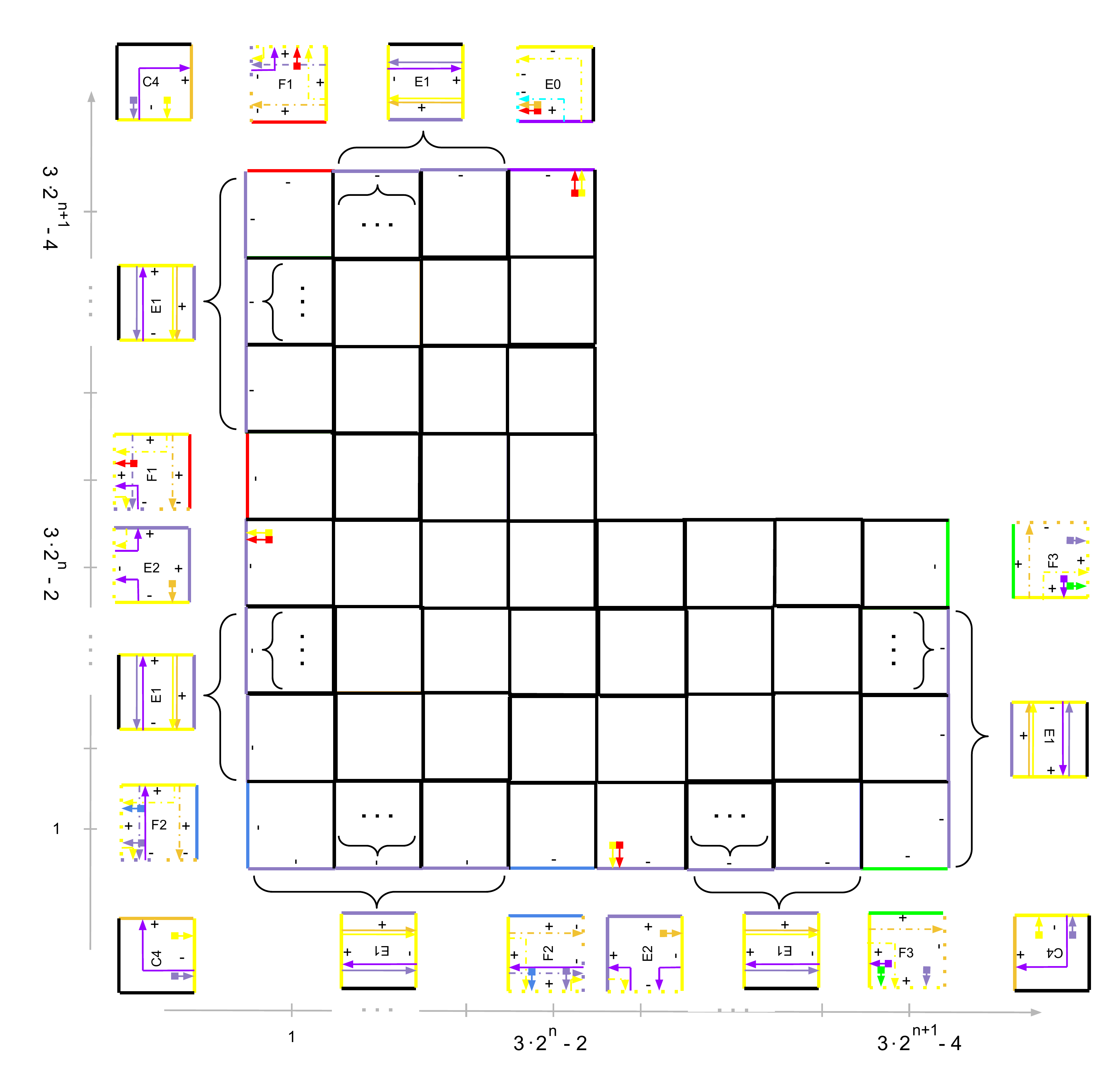}
\end{center}
\caption{Assembly of $T_{3}(n)$ Tile type: the E0-type border assembles around the Level $n$ L-Shape $T_{1}(n)+(1,1)$ (i.e., $T_{1}(n)$ positioned with bottom left corner tile at $(1,1)$).}
\vskip -.7cm
\label{fig:LvlnE0shape}
\end{figure}

We continue with $T_{3},T_{6},T_{9},$ and $T_{12}$. The construction is essentially identical to that of the Tile types $T_{2},T_{5},T_{8},$ and $T_{11}$, but using different unit tiles in the border. The initial structures $T_{3}(0),T_{6}(0),T_{9}(0),T_{12}(0)$ are given in Table \ref{tab:E0typeL0}. Refer to Figure \ref{fig:E0TypeRegions} for a diagram of the regions of $T_{3}(\ell)$ for $\ell\geq 1$. They are explicitly given as follows:
\begin{align*}
R_{3,1}(\ell) & = r_{3,1}^{1}(\ell) = T_{1}(\ell)+(1,1)\\ 
R_{3,2}(\ell) & = r_{3,2}^{1}(\ell) = E0_{N} + (3\cdot 2^{\ell}-2,3\cdot 2^{\ell +1}-3)\\ 
R_{3,4}(\ell) & = r_{3,4}^{1}(\ell) = F1_{N}+(1,3\cdot 2^{\ell +1}-3)\\ 
R_{3,5}(\ell) & = r_{3,5}^{1}(\ell) = C4_{N}+(0,3\cdot 2^{\ell +1}-3)\\ 
R_{3,7}(\ell) & = r_{3,7}^{1}(\ell) = F1_{E}+(0,3\cdot 2^{\ell}-1)\\
R_{3,8}(\ell) & = r_{3,8}^{1}(\ell) = E2_{N}+(0,3\cdot 2^{\ell}-2)\\
R_{3,9}(\ell) & = r_{3,9}^{1}(\ell) = F2_{N}+(0,1)\\
R_{3,10}(\ell) & = r_{3,10}^{1}(\ell) = C4_{E}+(0,0)\\  
R_{3,12}(\ell) & = r_{3,12}^{1}(\ell) = F2_{E}+(3\cdot 2^{\ell} -2,0)\\
R_{3,13}(\ell) & = r_{3,13}^{1}(\ell) = E2_{E}+(3\cdot 2^{\ell} -1,0)\\
R_{3,14}(\ell) & = r_{3,14}^{1}(\ell) = F3_{N}+(3\cdot 2^{\ell +1}-4,0)\\
R_{3,15}(\ell) & = r_{3,15}^{1}(\ell) = C4_{S}+(3\cdot 2^{\ell +1}-3,0)\\
R_{3,17}(\ell) & = r_{3,17}^{1}(\ell) = F3_{E}+(3\cdot 2^{\ell +1}-3,3\cdot 2^{\ell}-2)\\   
\end{align*}
with
\[R_{3,3}(\ell) = \bigcup_{k=1}^{3\cdot 2^{\ell}-4} r_{3,3}^{k}\]
where $r_{3,3}^{1}(\ell)=E1_{N}+(2,3\cdot 2^{\ell +1}-3)$ and  $r_{3,3}^{k+1}(\ell)=r_{3,3}^{k}(\ell)+(1,0)$;
\[R_{3,6}(\ell) = \bigcup_{k=1}^{3\cdot 2^{\ell +1}-7} r_{3,6}^{k}\]
where $r_{3,6}^{1}(\ell)=E1_{E}+(0,2),\quad r_{3,6}^{3\cdot 2^{\ell}-3}(\ell)=r_{3,6}^{3\cdot 2^{\ell}-4}(\ell)+(0,3)$ (this shift by 3 units reflects the skipping of the two single unit tile regions $R_{3,7}$ and $R_{3,8}$, Fig.~\ref{fig:E0TypeRegions}), $\quad r_{3,6}^{k+1}(\ell)=r_{3,6}^{k}(\ell)+(0,1), k\neq 3\cdot 2^{\ell}-4$;
\[R_{3,11}(\ell) = \bigcup_{k=1}^{3\cdot 2^{\ell +1}-7} r_{3,11}^{k}\]
where $r_{3,11}^{1}(\ell)=E1_{S}+(1,0),\quad r_{3,11}^{3\cdot 2^{\ell}-2}(\ell)=r_{3,11}^{3\cdot 2^{\ell}-3}(\ell)+(3,0)$ (this shift by 3 units reflects the skipping of the two single unit tile regions $R_{3,12}$ and $R_{3,13}$, Fig.~\ref{fig:E0TypeRegions}), $\quad r_{3,11}^{k+1}(\ell)=r_{3,11}^{k}(\ell)+(1,0), k\neq 3\cdot 2^{\ell}-3$;
and
\[R_{3,16}(\ell) = \bigcup_{k=1}^{3\cdot 2^{\ell}-3} r_{3,16}^{k}\]
where $r_{3,16}^{1}(\ell)=E1_{W}+(3\cdot 2^{\ell +1}-3,1)$ and  $r_{3,16}^{k+1}(\ell)=r_{3,16}^{k}(\ell)+(0,1)$.
Figure~\ref{fig:LvlnE0shape} shows the outline of the general $T_{3}(n)$ assembly. As with the previous cases, the construction of regions for $T_{6},T_{9},$ and $T_{12}$ follows analogously and is omitted here.

This concludes the construction of the recursive set of Tile types for Theorem~\ref{thm:recset}. What remains to be shown for the proof of this theorem is that these structures are, in fact, obtainable from the ATAS $(\mathcal{T}_{0},s,2)$. We do this in the next section.

We finish this section by pointing out that $T_{i}\in\Theta$ is self-similar for all $i$, since the geometrical shape of the union of all regions does not change for any $i$. Furthermore, the set containing any of the Tile types $T_{1},T_{4},T_{7},T_{10}$ is a strongly self-similar subset of $\Theta$: every region in one of these four Tile types consists of a single tile type in $\Theta$, but every Tile type in $\Theta$ is geometrically similar to every other type, so the conclusion follows.

We point out that the entire set $\Theta$ is not strongly self-similar because A0- and E0-type bordered Tile types contain regions that stretch, e.g., $R_{2,3}$ (see Figure~\ref{fig:A0TypeRegions}), and that aren't geometrically similar to any Tile type in $\Theta$ (refer to Section~\ref{sec:recursive} for the definitions of self-similarity).

\subsection{Self-Assembly}\label{sec:self-assembly}
Our goal now is to show that the structures we defined with Tile types do in fact assemble in the system $(\mathcal{T}_{0},s,2)$, finishing the proof of Theorem~\ref{thm:recset}. The construction will proceed by induction and show that structures that do not take part in the assembly of Tile types in $\Theta$ cannot be produced, proving Theorem~\ref{thm:unique}. The set of unit tiles is shown in Figure~\ref{fig:Time0} (also Tables~\ref{tab:L-sides} and \ref{tab:L-signals} in the Appendix). 
\begin{figure}[h!]
\begin{center}
\includegraphics[scale=.25]{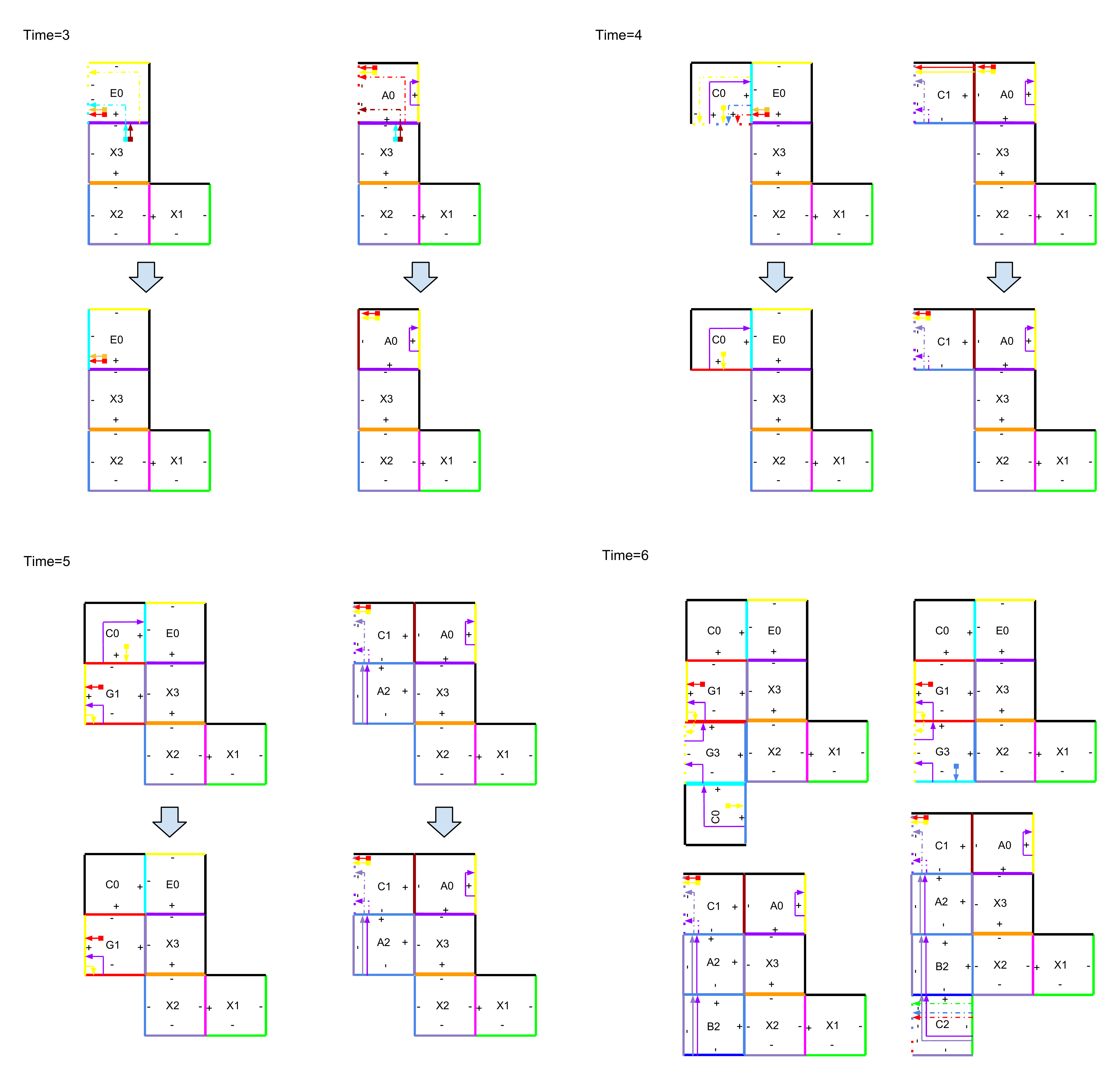}
\end{center}\vskip -.7cm
\caption{L-shape assembly, Time 3-6. The completed constructions depicted are the only constructions which can be added to $\mathcal{T}_{i}$ for corresponding $i$. Beginning with this figure, only the initial and completed assemblies are shown in all figures, unless indicated otherwise: intermediate steps of the tile modification function are suppressed.}
\vskip -.7cm
\label{fig:Time3-6}
\end{figure}

For the base case of the induction we need to obtain $T_{1}(1)$ from the level 0 Tile types. We have already seen that X1$_{N}$X2$_{N}$X3$_{N}$ assembles in Figure~\ref{fig:Time1-2} (Time 2), which is $T_{1}(0)$. We proceed to demonstrate the assembly of $T_{2}(0)$ and $T_{3}(0)$ (all other $T_{i}(0)$ are rotated copies of one of the three level 0 types mentioned and assemble in the same way).

We have already established that no assemblies other than the ones in Fig.~\ref{fig:Time1-2} could be formed from pairing unit tile and double assemblies. Now, observe that 
all active labels of $T_{1}(0)$ are strength 1 except for the label -55 at the top, which has strength 2. All of the strength 1 labels are ``negative'' labels. A ``negative'' label can only bind to a ``positive'' label, so two ``positive'' labels would have to be present on the same side of a two-tile assembly in order to bind to $T_{1}(0)$. The only positive labels appropriately positioned on the existing assemblies are the $+2$ labels on C0-G3, but they do not match the labels on $T_{1}(0)$. Therefore, the only possible binding that can occur is with the single strength 2 label, $-55$. 
\begin{figure}[h!]
\begin{center}
\includegraphics[scale=.25]{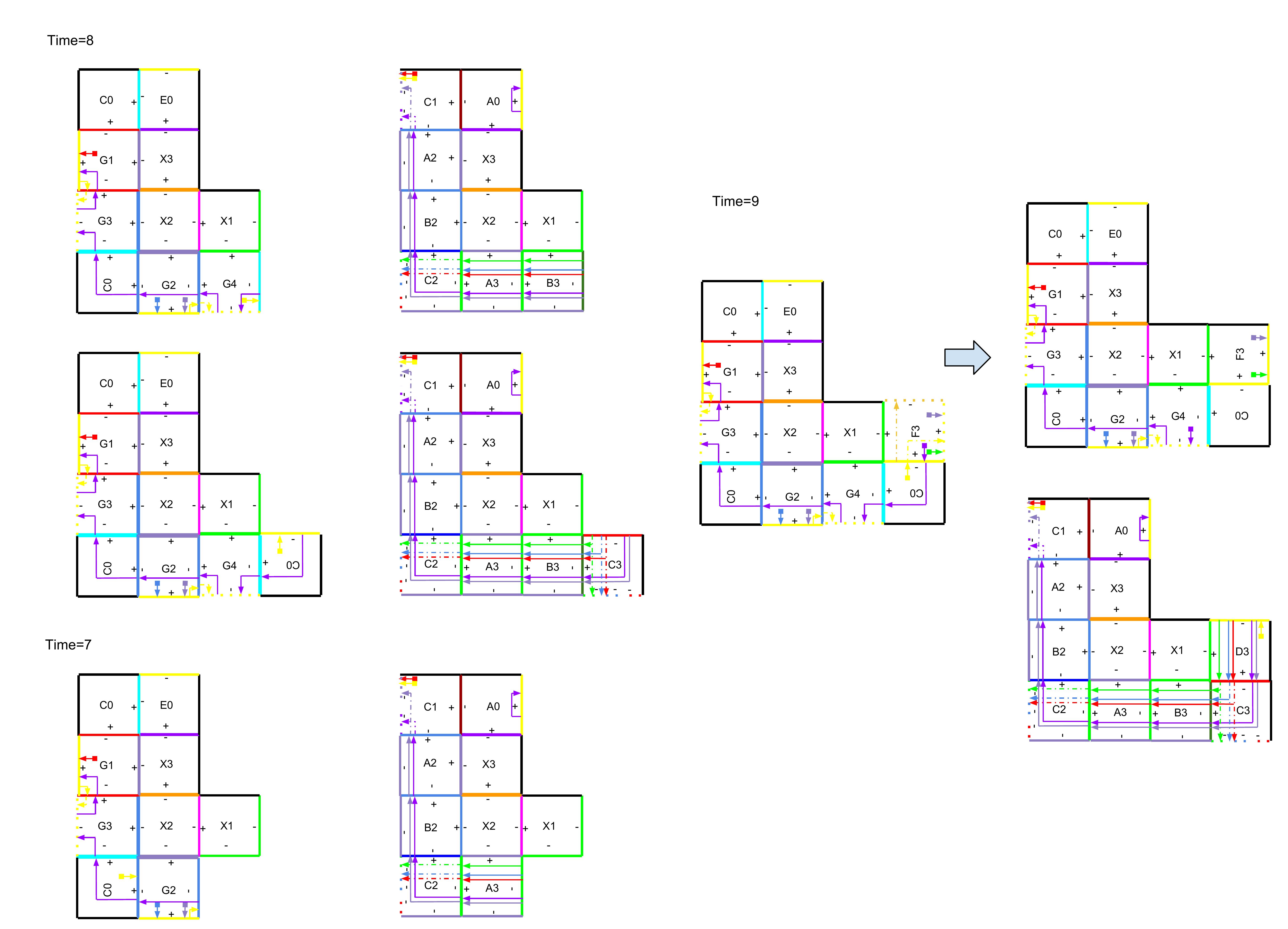}
\end{center}
\caption{L-shape assembly. Time 7 (bottom left), Time 8 (top left), Time 9 (right). Again, these are the only new constructions that can be added to the active tile set.}
\vskip -.7cm
\label{fig:Time7-9}
\end{figure}

We can see that the complementary +55 label appears in both E0 and A0 unit tiles in Figure~\ref{fig:Time0}. The two different tiles allow the creation of two different borders around $T_{1}(0)$ depending on which one of E0 or A0 attaches to $T_{1}(0)$. This local non-determinism allows us to give the L-shapes one of two different roles (the center and edge roles, respectively) in the next assembly stage. Note that border initiation is the only case when two tile choices are available for binding at a single site. The assembly process is shown in Fig.~\ref{fig:Time3-6}: Time 3-5 shows the only possible binding events and the resulting complete tile assemblies. Note that signals are transferred from A0 to C1 in Time 4. No structures other than the ones shown can be constructed from earlier assemblies because there are no assemblies which contain two appropriately positioned strength 1 ``positive'' labels to match two of the ``negative'' ones, except in the newly formed corners. So, the construction can only proceed by matching labels in these corners (which may also contain a ``negative'' and a ``positive'' strength 1 label). An activated strength 2 label does not occur until the formation of the next level Tile type.

Throughout the rest of the L-shape assembly, we use the fact that the unbordered Tile types only have sequences of strength 1 ``negative'' labels along their sides except for the single strength 2 label -55 which initiates the creation of the A0 or E0 border. So, only this label of strength 2 or the new corners formed by two negative (or, sometimes, negative and positive) labels can serve as attachment points for other tiles.

Note that in Figure~\ref{fig:Time3-6}, Time 6, both the A0-type assembly and the E0-type assembly may be augmented by two different assemblies: either the unit tiles B2 and G3 respectively, or the two-tile assemblies formed at Time 1, B2 and C2 assembly or G3 and C0 assembly. However, in the former case, Time 7 would produce the latter case. Thus, the only new assemblies in Time 7 (Figure~\ref{fig:Time7-9}) are the ones from Time 6 assemblies that were created with the two-tile assemblies.

This process, when continued, yields the desired $T_{2}(0)$ and $T_{3}(0)$ assemblies (shown in Tables~\ref{tab:A0typeL0} and \ref{tab:E0typeL0}) at Time 9 in Figure~\ref{fig:Time7-9}. The rotated Tile types (not shown) are obtained simultaneously in a similar manner.

\begin{figure}[hbt]
\begin{center}
\includegraphics[scale=.35]{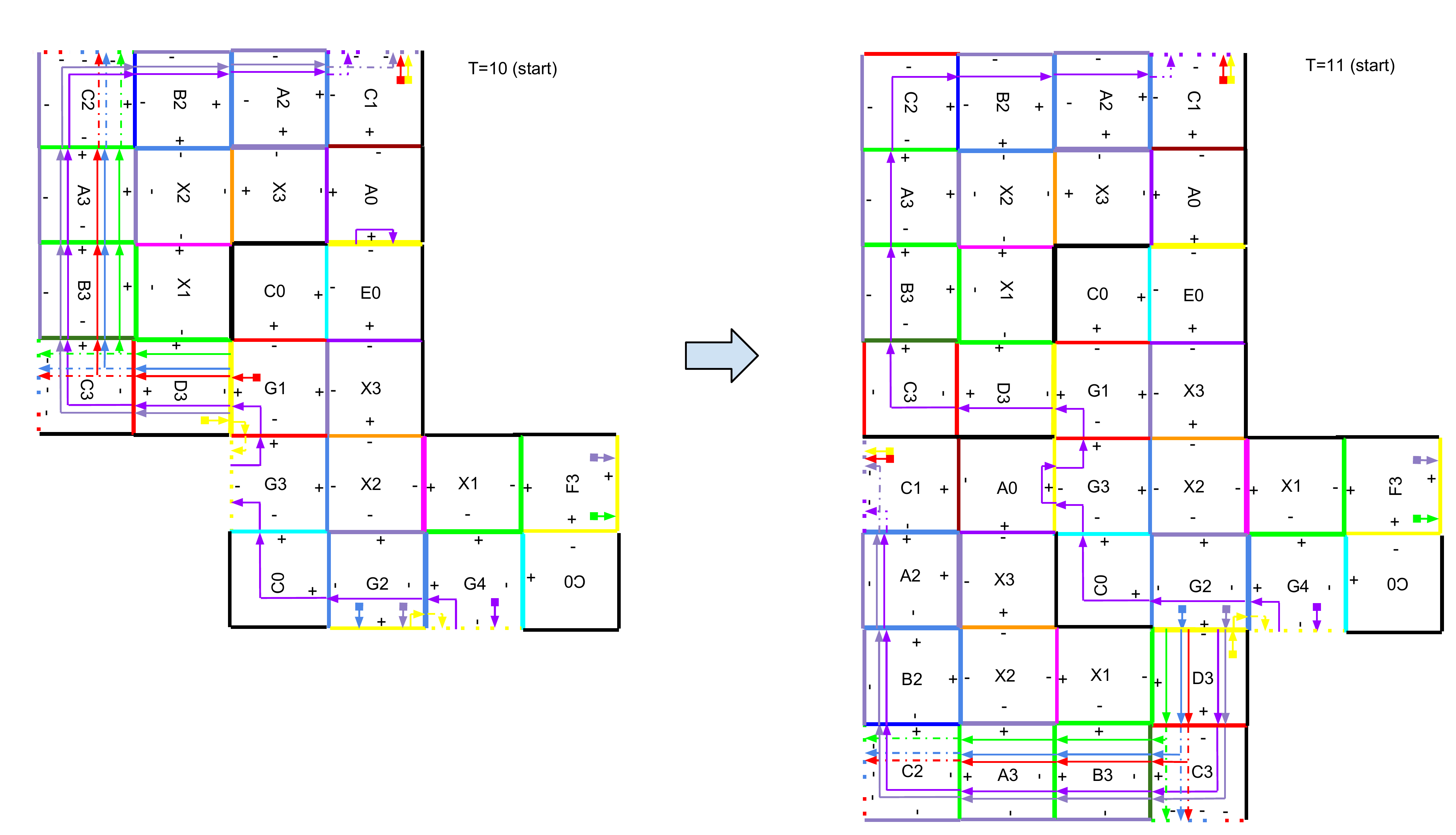}
\end{center}
\caption{Time 10-11, assembling the level 2 shape. The A0-type assembly ($T_{11}(0)$) binds via A0 and D3 to E0 and G1 of $T_{3}(0)$. The red signal from G1 is transmitted to C2 and C3, activating the corresponding labels. A (yellow) signal from D3 is transmitted through G1 and activates the -4 label in G3 which allows an A0-type assembly ($T_{2}(0)$) to bind at G3 and G2 sites. Broken transmission signals and disconnected inactive labels are cleaned up. In the figure on right, a process similar to the previous takes place, resulting in the activation of designated labels on the C1, C2, and C3 tiles in the A0-type and the G4 tile in the E0-type.}
\label{fig:Time10-11}
\vskip -.7cm
\end{figure}

\begin{figure}[ht]
\begin{center}
\includegraphics[scale=.3]{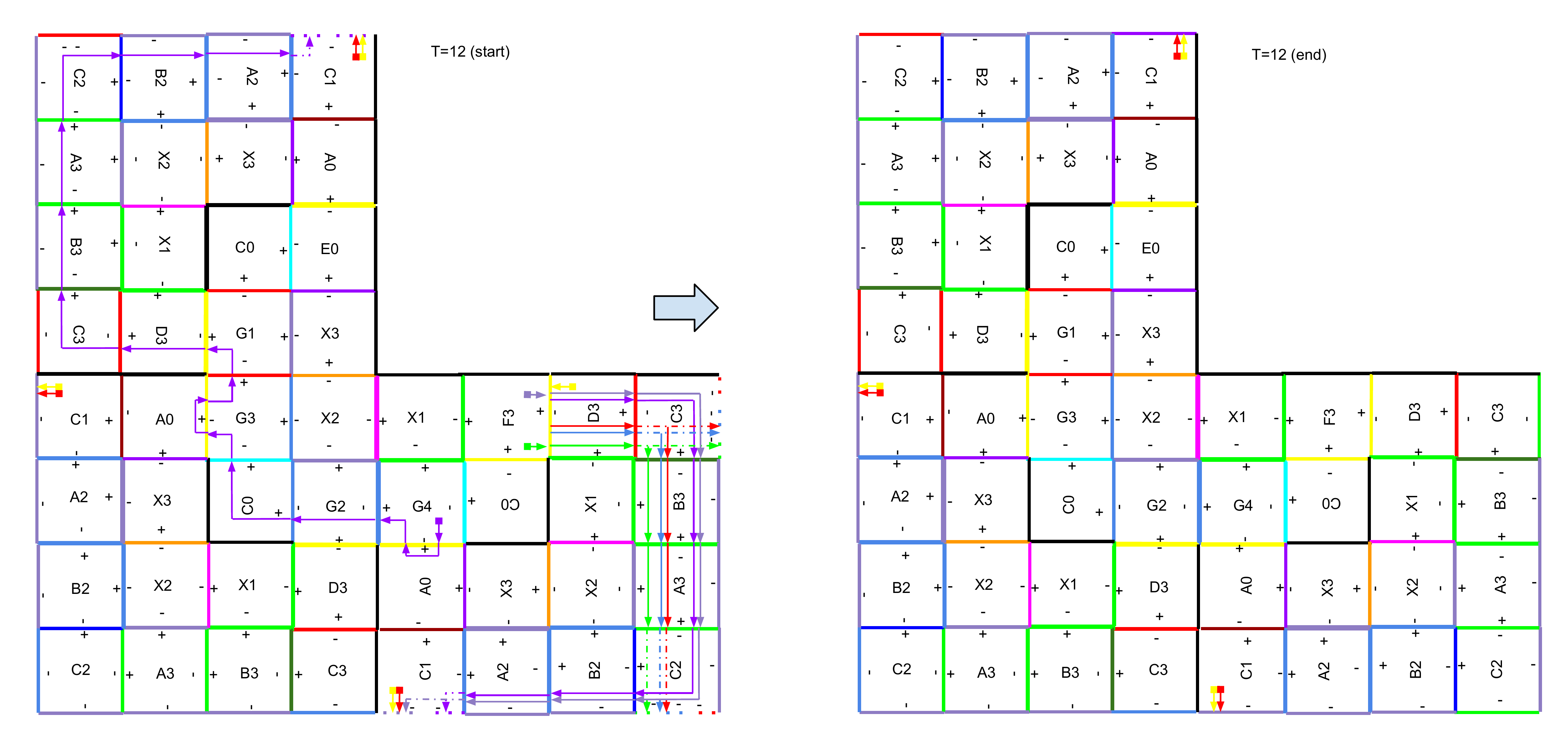}
\end{center}
\caption{Time 12 completes the assembly of $T_{1}(1)$ unbordered L-shape from four level 0 bordered shapes.}
\label{fig:Time12}
\vskip -.7cm
\end{figure}

We can now assemble $T_{1}(1)$. In the $T_{3}(0)$ assembly (and in the corresponding rotations) we see two perpendicular label 4's (yellow) (negative on the E0 tile and positive on the G1 tile), which can be matched by the $T_{11}(0)$ assembly (a rotation of $T_{2}(0)$), as it has a $+4$ on the A0 and a $-4$ on the D3 tile. When these attach (see Fig.\ref{fig:Time10-11}, Time 10), a signal from G1 activates the label $-1$ (red) on C2 and C3, and a (yellow) signal from D3 is sent through G1 to G3, activating the $-4$ label in the completed Time 10 (not shown). Now the $T_{3}(0)$ portion of the $T_{11}(0)T_{3}(0)$ assembly in Time 10 has active -4 and +4 labels in tiles G3 and G2 respectively, allowing $T_{2}(0)$ to attach at Time 11 (Fig.~\ref{fig:Time10-11}). A signal from G2 activates label $-2$ (blue) on C2 and C3, and label $-5$ on C1 in the completed Time 11 (not shown). The $-4$ label (yellow) on G4 is activated similarly to G3. Finally, in Fig.~\ref{fig:Time12} (left), the third A0 Tile type, $T_{5}(0)$, can attach to the -4 label in G4 and and +4 label in F3. A signal from F3 activates label $-3$ (green) on C2 and C3, and label $-5$ on C1. Also, a signal travels from G4 to A0, then back to G4 and eventually to C1 on the $T_{11}(0)$ (Fig.~\ref{fig:Time12}, top left), activating the $-55$ label (Fig.~\ref{fig:Time12}, right). In this way, we obtain $T_{1}(1)$ in Figure~\ref{fig:Time12}. 

Note that starting with the Time 10 stage, we also obtain incomplete L-shape constructions from completed A0-type assemblies binding to the E0-type assemblies with only a partially assembled border (for example, to the E0-type shape like in Time 7, Fig.\ref{fig:Time7-9}). These are not depicted as they ultimately lead to constructions that result in the desired L-shapes. An example of the partial border mixed assemblies can be seen in Figures \ref{fig:Time37}-\ref{fig:Time40-41} in the Appendix and also in \cite{mythesis}.

We can now proceed with the inductive step. Suppose we begin with a shape like in Figure~\ref{fig:Lvlnshape}, that is, with $T_{1}(n)$ for some $n\geq 1$ (clearly, $T_{1}(1)$ satisfies the diagram) and suppose we have $T_{2}(n-1)$ and $T_{3}(n-1)$ (and all corresponding rotations). We first show that from $T_{1}(n)$ we can obtain both $T_{2}(n)$ and $T_{3}(n)$. This is demonstrated in Figures~\ref{fig:LvlnA0shape} and \ref{fig:LvlnE0shape}. Note that $T_{1}(n)$ (Figure~\ref{fig:Lvlnshape}) contains only negative strength 1 labels around its borders except for the strength 2 label, -55, in the top right corner. So, $T_{1}(n)$ cannot interact with any assemblies except for the unit tiles A0 and E0, yielding two possible border initiations, shown in Figures~\ref{fig:LvlnA0shape} and \ref{fig:LvlnE0shape}. Upon close examination, it is evident that beginning with the A0 or E0 tile in the corner, it is possible to construct the borders one (or two, where possible) unit tile at a time exactly as indicated. 

Moreover, observe that the exposed edges of the growing assembly in either figure never contain ``$+$'' strength 1 labels except for the $+4$ (yellow) labels on A0 (Fig.~\ref{fig:LvlnA0shape}), and F1, F2, and F3 unit tiles (Fig.~\ref{fig:LvlnE0shape}. Note that although each of these three tiles appears twice, the +4 label is activated only the second instance of each, after receiving a signal from the C4 tiles). In Figure~\ref{fig:LvlnE0shape}, an exposed -4 complementary label first appears on E0, parallel to the L-shape outline. Additional active -4 labels on this assembly are exposed only one at a time and are perpendicular to the L-shape border while building the E0 border. Here we note that it is geometrically impossible for any assembly to bind simultaneously to a perpendicular positive or negative 4 label and a parallel positive or negative 4 label on another assembly. The reason is as follows: if the labels appear on the same side of the L-shape border, e.g., parallel +4 on F1 (or -4 on E0) and perpendicular -4 on E1 (during the border building process, just as one of these unit tile attaches), the matching assembly would have to have a parallel -4 (or +4) and a perpendicular +4 exposed, but no exposed perpendicular +4 labels ever appear on any assembly. If the labels appear on different sides, the matching assembly would have to round two or more corners, whereas the existing L-shape assemblies are only capable of rounding one corner. So, only pairs of parallel labels 4 in Figure~\ref{fig:LvlnE0shape} remain to be considered. Because none of the labels 4 appear on the same side, the E0-bordered assembly is geometrically prevented from binding to its own type as it assembles. On the other hand, Figure~\ref{fig:LvlnA0shape} shows that the A0-bordered figure has a perpendicular +4 label on A0 and acquires a perpendicular -4 label on D3 at the last stage of its assembly, allowing it to bind to the +4 and -4 labels on adjacent sides of the E0-bordered assembly, specifically, to the E0 and F1 (the only correctly positioned complements).

What this shows is that no level of either assembly (E0 or A0 bordered) can interact with another level except for an A0 Tile type to bind to an E0 Tile type of the same level: first at E0 and F1, then at E2 and F2, and finally at E2 and F3 (this sequence of binding events is enforced by signaling - E2 is only activated when the binding at E0 and F1 takes place). Observe that the next border cannot be initiated until the unbordered shape is completely assembled, producing $T_{1}(n+1)$. The reader is invited to compare the tile assemblies in Figures~\ref{fig:LvlnA0shape} and \ref{fig:LvlnE0shape} to the region outlines in Figures~\ref{fig:A0TypeRegions} and \ref{fig:E0TypeRegions} respectively. ($T_{2}(1)$ and $T_{3}(1)$ are shown as an example in Figure~\ref{fig:Lvl2Completed} in the Appendix).

The position coordinates for the tiles in the L-shapes are derived from the fact that $T_{1}(0)$, the assembly obtained at Time 2 (see Figure~\ref{fig:Time1-2}), measures $1\times 2$ tiles on the short and the long sides respectively, and if $T_{1}(\ell)$ measures $L\times 2L$ tiles, then the bordered versions $T_{2}(\ell)$ and $T_{3}(\ell)$ each measure $(L+1) \times (2L+2)$ tiles. 

Let $s_{\ell}$ denote the length of the short side of $T_{1}(\ell)$. Since $T_{1}(\ell+1)$ is obtained by combining four of the bordered shapes of the previous level as in Figure~\ref{fig:Lvlnshape}a), we get the following relation:
\begin{align*}
s_{\ell} &= 2s_{\ell-1}+2,\quad\text{for }\ell\geq 1\\
&= 3\cdot 2^{\ell}-2,\ \ \text{for }\ell\geq 0
\end{align*}
where the second equality can be easily verified using the first: $s_{0}=1=3\cdot 2^{0}-2$, so if $s_{\ell}=3\cdot 2^{\ell}-2$, $\ell\geq 0$, it follows that $s_{\ell+1} = 2(3\cdot 2^{\ell}-2)+2 = 3\cdot 2^{\ell+1}-4+2 = 3\cdot 2^{\ell+1}-2$
as required. 

Based on the above, we can also compute the precise stage at which a Tile type at a given level assembles. For example $T_{1}(0)$ assembles at stage 2 and for $\ell\geq 1$, $T_{1}(\ell)$ assembles at stage $9\cdot 2^{\ell+1}-6\ell-18$.

So far, we have shown that the Tile types of the previous section do result from the self-assembly process of this tiling system. We have also shown why this tiling system does not produce any assemblies that are not components of the Tile types in $\Theta$. This concludes the proof of Theorems~\ref{thm:recset} and \ref{thm:unique}.

\begin{figure}[ht]
\begin{center}
\includegraphics[scale=.25]{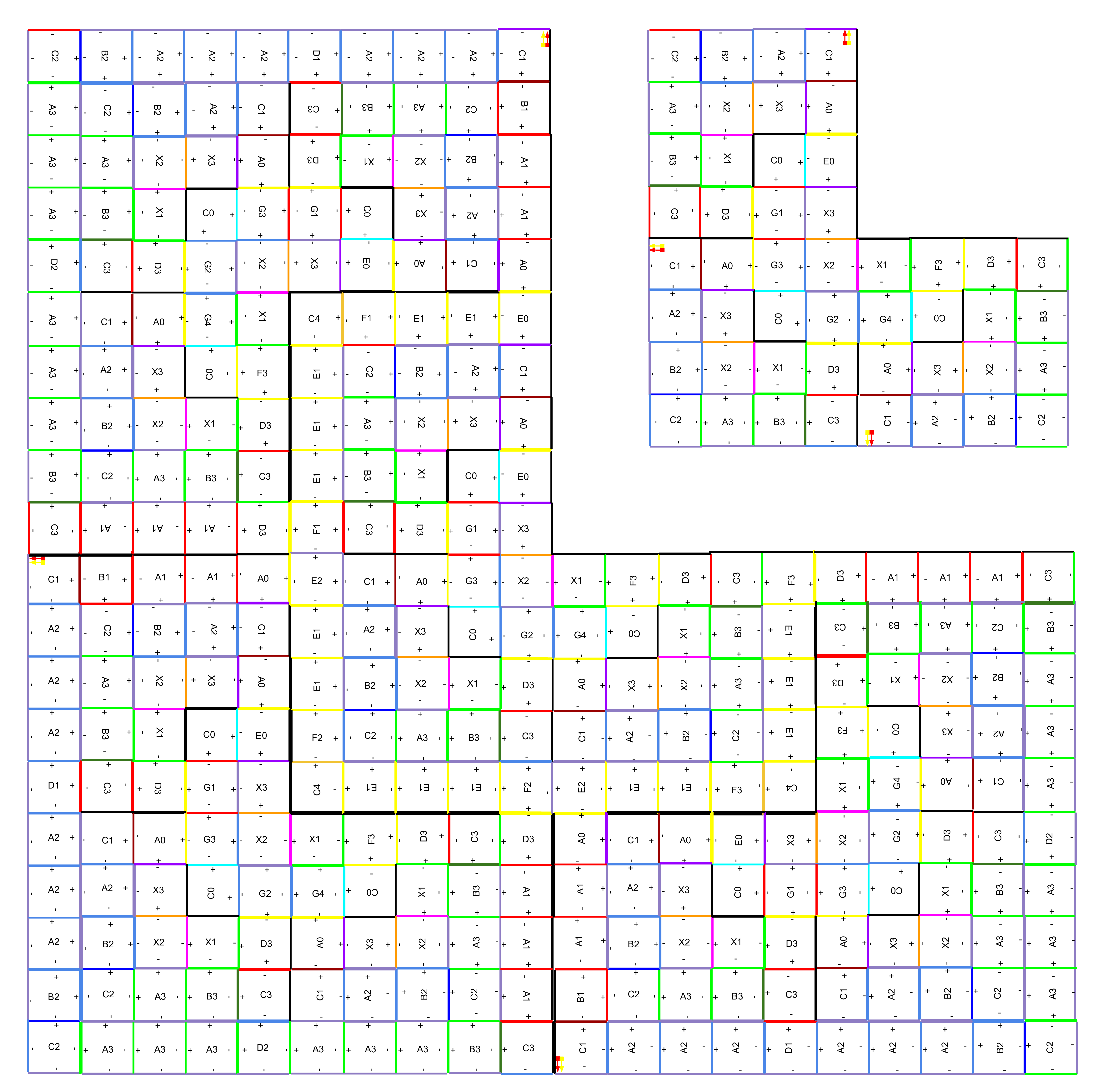}
\end{center}
\caption{$T_{1}(1)$ and $T_{1}(2)$ L-Shapes side by side. Note the positioning of corner markings (red labels -1, blue labels -2, green labels -3, and the purple -55 label) and observe that the rest of the outside tile edges are filled with the -5 label.}
\label{fig:Lvl2-3L-Shape}
\vskip -.7cm
\end{figure}

As a final note here, we include a side by side comparison of the assembled $T_{1}(1)$ and $T_{1}(2)$ structures in Figure~\ref{fig:Lvl2-3L-Shape}. The assembly process from $T_{1}(1)$ up to $T_{1}(2)$ (stage 42) is shown explicitly in the Appendix and can also be found in \cite{mythesis}.

\subsection{Aperiodicity}\label{sec:aperiodicitypf}
In this section we give the proof of Theorem~\ref{thm:aperiodicity}. First, we show that the L-shape Tile types provide a tiling of the plane by considering the $T_{1}$ (unbordered) Tile type. We note that for each $\ell$, the corresponding L-shape can be divided into three square regions, and that each of the square regions in an $\ell+1$ L-shape covers more than 4 times the area covered by a square region of the level $\ell$ L-shape. Thus, since these squares grow without bound, given any sized square portion of the plane, there exists some $m$ such that the level $m$ L-shape covers it. As $m\rightarrow\infty$, the L-shape assembly instances provide a tiling of the plane.\footnote{This is, in fact, a consequence of K\"onig's lemma.}

For the proof of aperiodicity, refer to Figure \ref{fig:Lvl2-3L-Shape} and Figures \ref{fig:LvlnA0shape} and \ref{fig:LvlnE0shape}. Notice that for a given level $\ell$ L-shape $T_{1}(\ell)$, there is a unique outermost C2 marked with a $-2$ label and a unique outermost C3 marked with a $-2$ label. These two unit tiles, furthermore, are separated by a region of A3's, where the exact number of the A3's indicates the level $\ell$. This uniqueness is guaranteed by construction: every level $\ell$ L-shape is formed by an E0-type bordered level $\ell-1$ L-shape, $T_{3}(\ell-1)$ in the center that is adjacent to three A0-type bordered level $\ell-1$ L-shapes, $T_{2}(\ell-1),T_{5}(\ell-1),T_{11}(\ell-1)$ (Fig.~\ref{fig:Lvlnshape}). Each of the A0-type shapes contains a single C2 and a single C3 as part of its border. The label counter-clockwise adjacent to the +22 (dark blue) label on C2 and to the +33 (dark green) label on C3 (see Fig.\ref{fig:Lvl2-3L-Shape}) is determined by the A0-type level $\ell-1$ shape's position relative to the level $\ell-1$ E0-type shape. The three possible orientations yield three possible label values (-1,-2, or -3). Consequently, the particular configuration of C2 and C3 carrying the $-2$ label is indeed unique within the level $\ell$ L-shape. 

Fix an L-shape tiling of the entire plane, $\xi$. We consider $T_{1}(\ell)$ and $T_{7}(\ell)$. Observe that $T_{7}(\ell)$ (refer to Fig.~\ref{fig:Lvlnshape}a)) appears inside $T_{1}(\ell+2)$ (it is nested inside $T_{8}(\ell)$ that is inside $T_{11}(\ell+1)$) by construction. The unique placement of C2 and C3 followed by the fixed number of A3's in $T_{1}(\ell)$ and $T_{7}(\ell)$ guarantees that no translation of an L-shape tiling of the plane is possible by a vector with either component less than $3\cdot 2^{\ell}-2$ (the length of the short side of a level $\ell$ L-shape $T_{1}$ or $T_{7}$) for any $\ell$ ($T_{1}$ and $T_{7}$ together cover the four possible orientations of the C2 C3 segment). Otherwise, supposing this is not the case for some $\ell\geq 2$, consider the plane tiling $\xi$ induced by the L-shape assembly instances. Then a C2 C3 border configuration as described above for the level $\ell$ L-shape would have to appear more than once inside that level $\ell$ L-shape $T_{1}$ (if at least one of the vector coordinates is positive) or $T_{7}$ (if both vector coordinates are negative) - and this, we have shown, cannot happen. Since this is true for every level $\ell$, it follows that the tiling $\xi$ is not invariant under any translation.

\section{Conclusion}
In this paper, we presented an abstract self-assembly model for active DNA-based tiles - tiles with capacity to receive and transmit binding site activation signals. We also gave a definition of recursive assembly and 
self-similarity based on this abstract model, which, to the best of our knowledge, is the first attempt to describe recursion in self-assembly. The aperiodic L-shape assembly system that we described illustrates not only these new concepts of active tiles and recursive tilings but also the power of signaling in the 
self-assembly process. It was the signaling capability of the tiles that ensured that no extraneous structures could arise and enforced the order of assembly without the necessity of staging.

\begin{figure}[h]
\begin{center}
\includegraphics[scale=.7]{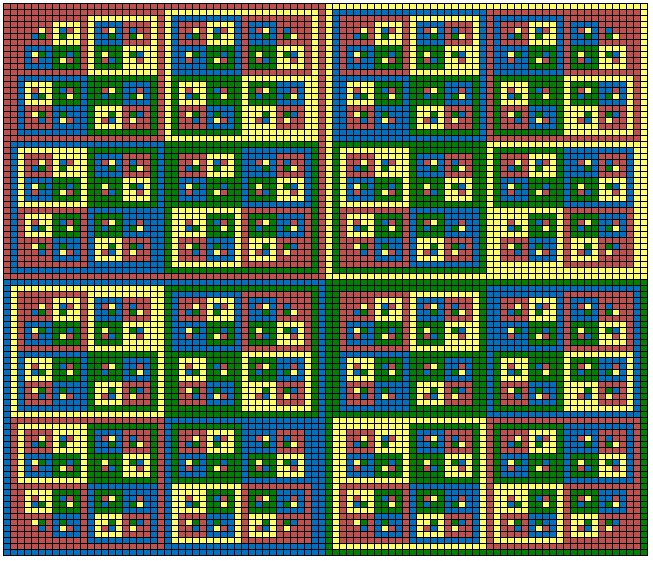}
\end{center}
\caption{A non-periodic square tiling utilizing the border technique; here, four different border types - red, yellow, green, and blue - determine a particular square's position in the larger shape.}
\label{fig:square}
\vskip -.7cm
\end{figure}

We point out  that the method used for the creation of the L-shape Tiles can be used to create a variety of aperiodic patterns: for example the square pattern in Figure \ref{fig:square}. The principle is to use local nondeterminism to create the desired hierarchy. At each level, each instance of the shape is assigned a role in the next level by obtaining a partial border, selected at random, which grows in length with proportion to the size of the shape. This growth in length is what guarantees non-periodicity as increasingly longer border lines consisting of one kind of tile types will be found further away from a given tile, disallowing translational symmetry. Essentially, the entire identity of the larger tile is contained in its outside border, so the borders mediate all binding and signaling which occurs among the shapes.

As mentioned in the introduction, this model can be adjusted to accommodate different assumptions to better approximate the physical self-assembly process. For example, the alternate description of an active tile model made in \cite{Jensignals} includes the ability to deactivate what we here have called ``edge labels'' and allowing tiles to break away from a constructed assembly. In fact, adding a deactivation capacity to the model we have presented is straightforward. If desired, the triple $(\tau,\mathcal{A},\mathcal{S})$ could be expanded to include a set of deactivation signals and a set of deactivation pathways, becoming a 5-tuple. The tile modification function would have to be expanded to remove deactivation signals when they are used and to remove labels that have been deactivated in a manner completely analogous to activation signal processing. However, after any deactivation, it would be necessary to check whether the assembly is still $\theta$-stable - if not, the simulation of the remaining signal transmission would need to be continued on separate assemblies (which, if not stable themselves, would need to be broken up as well). The definition of active tile sets (the stage hierarchy of sets) would also have to be modified to include assemblies which are obtained in this manner. We did not include deactivation in the model presented here in order to simplify the exposition in this first introduction to the model and to recursion in self-assembly.

Likewise, one could account for the variability in signal transmission rates, as the authors in \cite{Jensignals} do, by allowing tiles to ``wait'' before transmitting a signal. Our assembly model is rather simplified in the sense that we assume that all signal transmissions occur simultaneously and that all signaling across a single assembly is completed before it can bind to another one. In solution, all these processes of binding and signal transmission are occurring in parallel and at variable rates; thus, a more accurate model of physical reality would allow an assembly at any point in its signal transmission cycle to bind with other assemblies, which means for any assembly $\alpha$ we would have to add assemblies $f^{i}(\alpha)$ for every $i\geq 0$ (where $f$ is the tile modification function) to the active supertile set of the relevant stage. An even more accurate model could account for a different rates of signal transmission by writing $c_{x}^{i}$ instead of $c_{0}^{i}$ whenever a tile receives an activation signal from another tile; this could represent a pending signal. A tile containing such a signal could replace the $x$ with a $0$ at any iteration of the tile modification function, in which case it would act the same way as the already defined activation signals.

We note that neither of the possible extensions that we mentioned would affect the assembly process in the L-shape ATAS that we described in this paper since the design of the tiles does not allow two Tile types at a given level to join before both of them are completed (with respect to the tile modification function, i.e., signal transmission). Thus, in our case, the simpler model is sufficient for an accurate presentation of the given recursion.

\newpage

\section*{Appendix}

\begin{table}[h]
\begin{center}
{
\renewcommand{\arraystretch}{0.9}
\begin{tabular}{|M|*{4}{M|}}
\hline
\multicolumn{5}{|c|}{\textbf{Tile Index: Tile Sides with Active and Inactive Label Sets}}\\
\hline
\text{Tile} & t_{+y} & t_{+x} & t_{-y} & t_{-x}\\
\hline
X1 & \lp\emptyset,\emptyset\rp 
& \lp\left\{ -3 \right\}, \emptyset\rp 
& \lp\left\{ -3 \right\}, \emptyset\rp 
& \lp\left\{ 77 \right\}, \emptyset\rp 
\\ \hline
X2 & \lp\left\{ -88 \right\}, \emptyset\rp
& \lp\left\{ -77 \right\}, \emptyset\rp
& \lp\left\{ -5 \right\}, \emptyset\rp
& \lp\left\{ -2 \right\}, \emptyset\rp
\\ \hline
X3 & \lp\emptyset, \left\{ -55 \right\}\rp
& \lp\emptyset, \emptyset\rp
& \lp\left\{ -5 \right\}, \emptyset\rp
& \lp\left\{ 88 \right\}, \emptyset\rp
\\ \hline
G1 & \lp\left\{ -1 \right\}, \emptyset\rp
& \lp\left\{ 5 \right\}, \emptyset\rp
& \lp\left\{ -1 \right\}, \emptyset\rp
& \lp\left\{ 4 \right\}, \emptyset\rp
\\ \hline
G2 & \lp\left\{ 5 \right\}, \emptyset\rp
& \lp\left\{ -2 \right\}, \emptyset\rp
& \lp\left\{ 4 \right\}, \emptyset\rp
& \lp\left\{ -2 \right\}, \emptyset\rp
\\ \hline
G3 & \lp\left\{ 1 \right\}, \emptyset\rp
& \lp\left\{ 2 \right\}, \emptyset\rp
& \lp\left\{ -66 \right\}, \emptyset\rp
& \lp\emptyset, \left\{ -4 \right\}\rp
\\ \hline
G4 & \lp\left\{ 3 \right\}, \emptyset\rp
& \lp\left\{ -66 \right\}, \emptyset\rp
& \lp\emptyset, \left\{ -4 \right\}\rp
& \lp\left\{ 2 \right\}, \emptyset\rp
\\ \hline
F1 & \lp\emptyset, \left\{ 4 \right\}\rp
& \lp\left\{ 4 \right\}, \emptyset\rp
& \lp\left\{ 1 \right\}, \emptyset\rp
& \lp\emptyset, \left\{ -5, -44 \right\}\rp
\\ \hline
F2 & \lp\left\{ 4 \right\}, \emptyset\rp
& \lp\left\{ 2 \right\}, \emptyset\rp
& \lp\emptyset, \left\{ -5, -44 \right\}\rp
& \lp\emptyset, \left\{ 4 \right\}\rp
\\ \hline
F3 & \lp\left\{ 3 \right\}, \emptyset\rp
& \lp\emptyset, \left\{ -44 \right\}\rp
& \lp\emptyset, \left\{ 4 \right\}\rp
& \lp\left\{ 4 \right\}, \emptyset\rp
\\ \hline
E0 & \lp\left\{ -4 \right\}, \emptyset\rp
& \lp\emptyset, \emptyset\rp
& \lp\left\{ 55 \right\}, \emptyset\rp
& \lp\emptyset, \left\{ -4, -66 \right\}\rp
\\ \hline
E1 & \lp\emptyset, \emptyset\rp
& \lp\left\{ 4 \right\}, \emptyset\rp
& \lp\left\{ 5 \right\}, \emptyset\rp
& \lp\left\{ -4 \right\}, \emptyset\rp
\\ \hline
E2 & \lp\left\{ 5 \right\}, \emptyset\rp
& \lp\left\{ 5 \right\}, \emptyset\rp
& \lp\left\{ -4 \right\}, \emptyset\rp
& \lp\emptyset, \left\{ -4 \right\}\rp
\\ \hline
D1 & \lp\left\{ 2 \right\}, \emptyset\rp
& \lp\left\{ 1 \right\}, \emptyset\rp
& \lp\left\{ -2 \right\}, \emptyset\rp
& \lp\left\{ -5 \right\}, \emptyset\rp
\\ \hline
D2 & \lp\left\{ 2 \right\}, \emptyset\rp
& \lp\left\{ -3 \right\}, \emptyset\rp
& \lp\left\{ -5 \right\}, \emptyset\rp
& \lp\left\{ 3 \right\}, \emptyset\rp
\\ \hline
D3 & \lp\left\{ -4 \right\}, \emptyset\rp
& \lp\emptyset, \emptyset\rp
& \lp\left\{ 1 \right\}, \emptyset\rp
& \lp\left\{ 3 \right\}, \emptyset\rp
\\ \hline
C0 & \lp\emptyset, \emptyset\rp
& \lp\left\{ 66 \right\}, \emptyset\rp
& \lp\emptyset, \left\{ 1, 2, -4 \right\}\rp
& \lp\emptyset, \emptyset\rp
\\ \hline
C1 & \lp\emptyset, \emptyset\rp
& \lp\left\{ 11 \right\}, \emptyset\rp
& \lp\left\{ -2 \right\}, \emptyset\rp
& \lp\emptyset, \left\{ -5, -55 \right\}\rp
\\ \hline
C2 & \lp\left\{ 22 \right\}, \emptyset\rp
& \lp\left\{ -3 \right\}, \emptyset\rp
& \lp\left\{ -5 \right\}, \emptyset\rp
& \lp\emptyset, \left\{ -1, -2, -3 \right\}\rp
\\ \hline
C3 & \lp\left\{ -1 \right\}, \emptyset\rp
& \lp\emptyset, \emptyset\rp
& \lp\emptyset, \left\{ -1, -2, -3 \right\}\rp
& \lp\left\{ 33 \right\}, \emptyset\rp
\\ \hline
C4 & \lp\emptyset, \emptyset\rp
& \lp\left\{ 44 \right\}, \emptyset\rp
& \lp\left\{ -4 \right\}, \emptyset\rp
& \lp\emptyset, \emptyset\rp
\\ \hline
B1 & \lp\emptyset, \emptyset\rp
& \lp\left\{ 1 \right\}, \emptyset\rp
& \lp\left\{ 1 \right\}, \emptyset\rp
& \lp\left\{ -11 \right\}, \emptyset\rp
\\ \hline
B2 & \lp\left\{ 2 \right\}, \emptyset\rp
& \lp\left\{ 2 \right\}, \emptyset\rp
& \lp\left\{ -22 \right\}, \emptyset\rp
& \lp\left\{ -5 \right\}, \emptyset\rp
\\ \hline
B3 & \lp\left\{ 3 \right\}, \emptyset\rp
& \lp\left\{ -33 \right\}, \emptyset\rp
& \lp\left\{ -5 \right\}, \emptyset\rp
& \lp\left\{ 3 \right\}, \emptyset\rp
\\ \hline
A0 & \lp\emptyset, \emptyset\rp
& \lp\left\{ 4 \right\}, \emptyset\rp
& \lp\left\{ 55 \right\}, \emptyset\rp
& \lp\emptyset, \left\{ -1, -11 \right\}\rp
\\ \hline
A1 & \lp\emptyset, \emptyset\rp
& \lp\left\{ 1 \right\}, \emptyset\rp
& \lp\left\{ 5 \right\}, \emptyset\rp
& \lp\left\{ -1 \right\}, \emptyset\rp
\\ \hline
A2 & \lp\left\{ 2 \right\}, \emptyset\rp
& \lp\left\{ 5 \right\}, \emptyset\rp
& \lp\left\{ -2 \right\}, \emptyset\rp
& \lp\left\{ -5 \right\}, \emptyset\rp
\\ \hline
A3 & \lp\left\{ 5 \right\}, \emptyset\rp
& \lp\left\{ -3 \right\}, \emptyset\rp
& \lp\left\{ -5 \right\}, \emptyset\rp
& \lp\left\{ 3 \right\}, \emptyset\rp
\\ \hline
\end{tabular}
}
\end{center}
\caption{List of unit tile rotation class representatives for the L-Shape Tiling: tile sides and labels.}
\label{tab:L-sides}
\end{table}
\begin{table}
\begin{center}
{
\renewcommand{\arraystretch}{0.9}
\begin{tabular}{|M|M|M|}
\hline
\multicolumn{3}{|c|}{\textbf{Tile Index: Transmission and Activation Signal Sets}}\\
\hline
\text{Tile} & \St & \A \\
\hline
X1 
& \{ 55_{0}^{-x} \} 
& \emptyset 
\\ \hline
X2 
& \{ 55_{+x}^{+y} \}
& \emptyset 
\\ \hline
X3 
& \{ 11_{0}^{+y}, 66_{0}^{+y} \}
& \{ 55_{-y}^{+y} \}
\\ \hline
G1 
& \{ 1_{0}^{-x}, 4_{-x}^{-y}, 5_{-y}^{-x} \}
& \emptyset
\\ \hline
G2 
& \{ 2_{0}^{-y}, 4_{-y}^{+x}, 5_{0}^{-y}, 55_{+x}^{-x} \}
& \emptyset
\\ \hline
G3 
& \{ 2_{0}^{-y}, 55_{-x}^{+y}, 55_{-y}^{-x} \}
& \{ 4_{+y}^{-x} \}
\\ \hline
G4 
& \{ 4_{0}^{+x}, 4_{-x}^{-y}, 55_{-y}^{-x}, 55_{+x}^{-y} \}
& \{ 4_{+y}^{-x} \}
\\ \hline
F1 
& \{ 1_{0}^{+y}, 4_{+y}^{-x}, 55_{-x}^{+y} \}
& \{ 4_{+x}^{+y}, 5_{+x}^{-x}, 44_{+x}^{-x} \}
\\ \hline
F2 
& \{ 2_{0}^{-x}, 4_{-x}^{-y}, 5_{0}^{-x}, 55_{-y}^{+y} \}
& \{ 4_{+y}^{-x}, 5_{+y}^{-y}, 44_{+y}^{-y} \}
\\ \hline
F3 
& \{ 3_{0}^{-y}, 5_{0}^{-y}, 55_{0}^{-x} \}
& \{ 4_{-x}^{-y}, 44_{-x}^{+x} \}
\\ \hline
E0 
& \{ 1_{0}^{-x}, 44_{0}^{-x} \}
& \{ 4_{-y}^{-x}, 66_{-y}^{-x} \}
\\ \hline
E1 
& \{ 4_{+x}^{-x}, 5_{+x}^{-x}, 44_{+x}^{-x}, 55_{-x}^{+x} \}
& \emptyset
\\ \hline
E2 
& \{ 44_{0}^{-y}, 55_{-x}^{+y}, 55_{-y}^{-x} \}
& \{ 4_{+y}^{-x} \}
\\ \hline
D1 
& \{ 5_{-y}^{+y}, 55_{-y}^{+y} \}
& \emptyset
\\ \hline
D2 
& \{ 1_{+x}^{-x}, 2_{+x}^{-x}, 3_{+x}^{-x}, 5_{+x}^{-x}, 55_{+x}^{-x} \}
& \emptyset
\\ \hline
D3
& \{ 1_{+y}^{-y}, 2_{+y}^{-y}, 3_{+y}^{-y}, 4_{0}^{+y}, 5_{+y}^{-y}, 55_{+y}^{-y} \}
& \emptyset
\\ \hline
C0 
& \{ 4_{0}^{-y}, 55_{-y}^{+x} \}
& \{ 1_{+x}^{-y}, 2_{+x}^{-y}, 4_{+x}^{-y} \}
\\ \hline
C1 
& \{ 1_{+x}^{-x}, 4_{+x}^{-x} \}
& \{ 5_{-y}^{-x}, 55_{-y}^{-x} \}
\\ \hline
C2 
& \{ 5_{+x}^{+y}, 55_{+x}^{+y} \}
& \{ 1_{+x}^{-x}, 2_{+x}^{-x}, 3_{+x}^{-x} \}
\\ \hline
C3 
& \{ 1_{+y}^{-x}, 2_{+y}^{-x}, 3_{+y}^{-x}, 5_{+y}^{-x}, 55_{+y}^{-x} \}
& \{ 1_{+y}^{-y}, 2_{+y}^{-y}, 3_{+y}^{-y} \}
\\ \hline
C4 
& \{ 4_{0}^{-y}, 5_{0}^{-y}, 55_{-y}^{+x} \}
& \emptyset
\\ \hline
B1 
& \{ 1_{+x}^{-x}, 4_{+x}^{-x} \}
& \emptyset
\\ \hline
B2 
& \{ 5_{-y}^{+y}, 55_{-y}^{+y} \}
& \emptyset
\\ \hline
B3 
& \{ 1_{+x}^{-x}, 2_{+x}^{-x}, 3_{+x}^{-x}, 5_{+x}^{-x}, 55_{+x}^{-x} \}
& \emptyset
\\ \hline
A0 
& \{ 1_{0}^{-x}, 4_{0}^{-x}, 55_{+x}^{+x} \}
& \{ 1_{-y}^{-x}, 11_{-y}^{-x} \}
\\ \hline
A1 
& \{ 1_{+x}^{-x}, 1_{-x}^{+x}, 2_{-x}^{+x}, 3_{-x}^{+x}, 4_{+x}^{-x}, 5_{-x}^{+x}, 55_{-x}^{+x} \}
& \emptyset
\\ \hline
A2
& \{ 5_{-y}^{+y}, 55_{-y}^{+y} \}
& \emptyset
\\ \hline
A3 
& \{ 1_{+x}^{-x}, 2_{+x}^{-x}, 3_{+x}^{-x}, 5_{+x}^{-x}, 55_{+x}^{-x} \}
& \emptyset
\\ \hline
\end{tabular}
}
\end{center}
\caption{List of unit tile rotation class representatives for the L-Shape Tiling: signaling.}
\label{tab:L-signals}
\end{table}
\begin{figure}[h]
\begin{center}
\includegraphics[scale=.25]{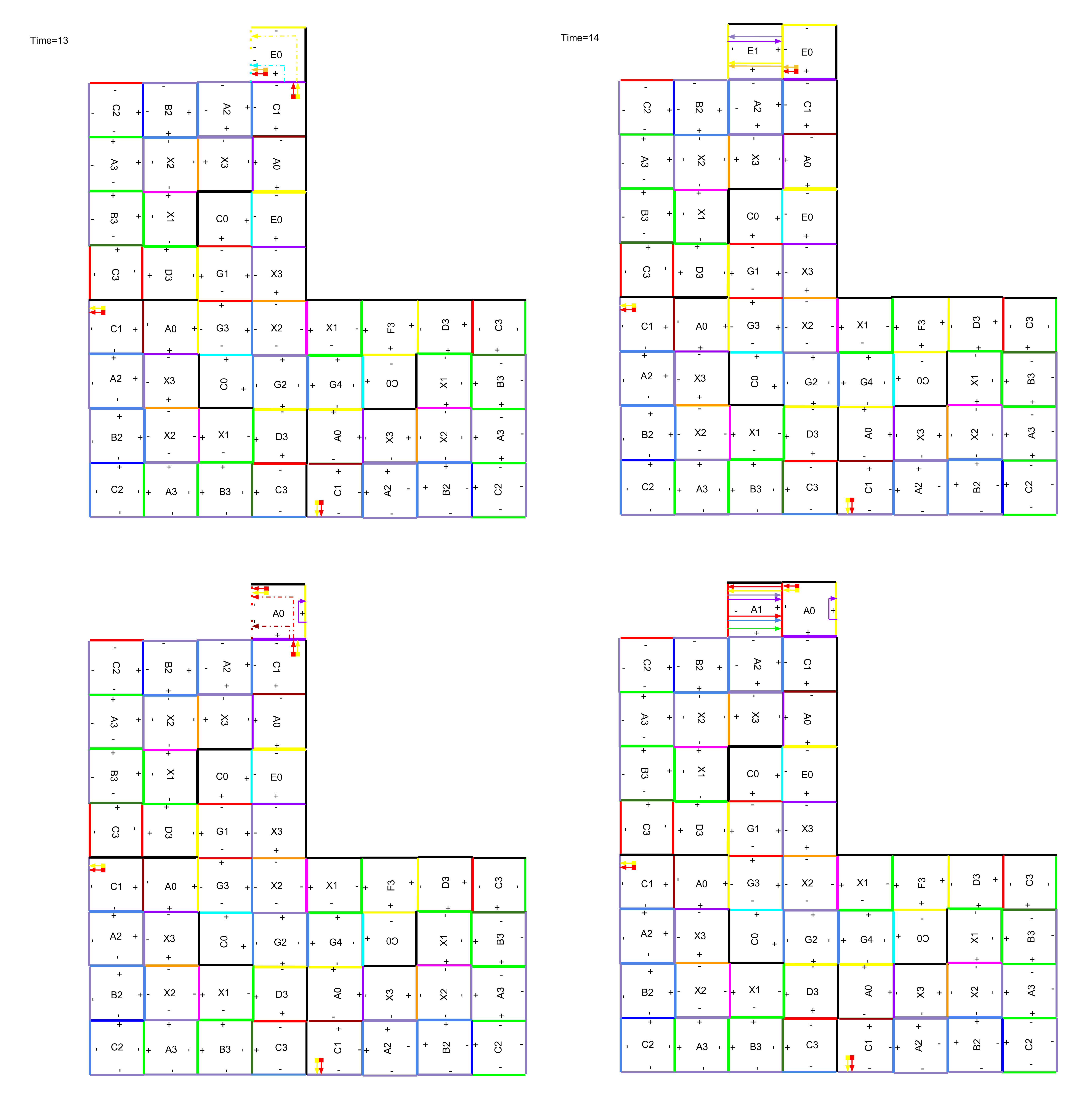}
\end{center}
\caption{Time 13-14, initiating the A0 and E0 border formation. The only two possible binding events that can occur is the level 2 L-shape binding to an A0 or an E0. Any other events are disallowed either by geometry or by complementarity and strength: observe that all of the labels on the outside of the L-shape are ``negative'' strength 1. Two ``positive'' strength 1 labels would have to match between it and another structure in order for a bond to form. The reader may verify that none of the structures which could be produced up to this point have two positive labels in the correct configuration.}
\label{fig:Time13-14}
\end{figure}
\begin{figure}[h]
\begin{center}
\includegraphics[scale=.25]{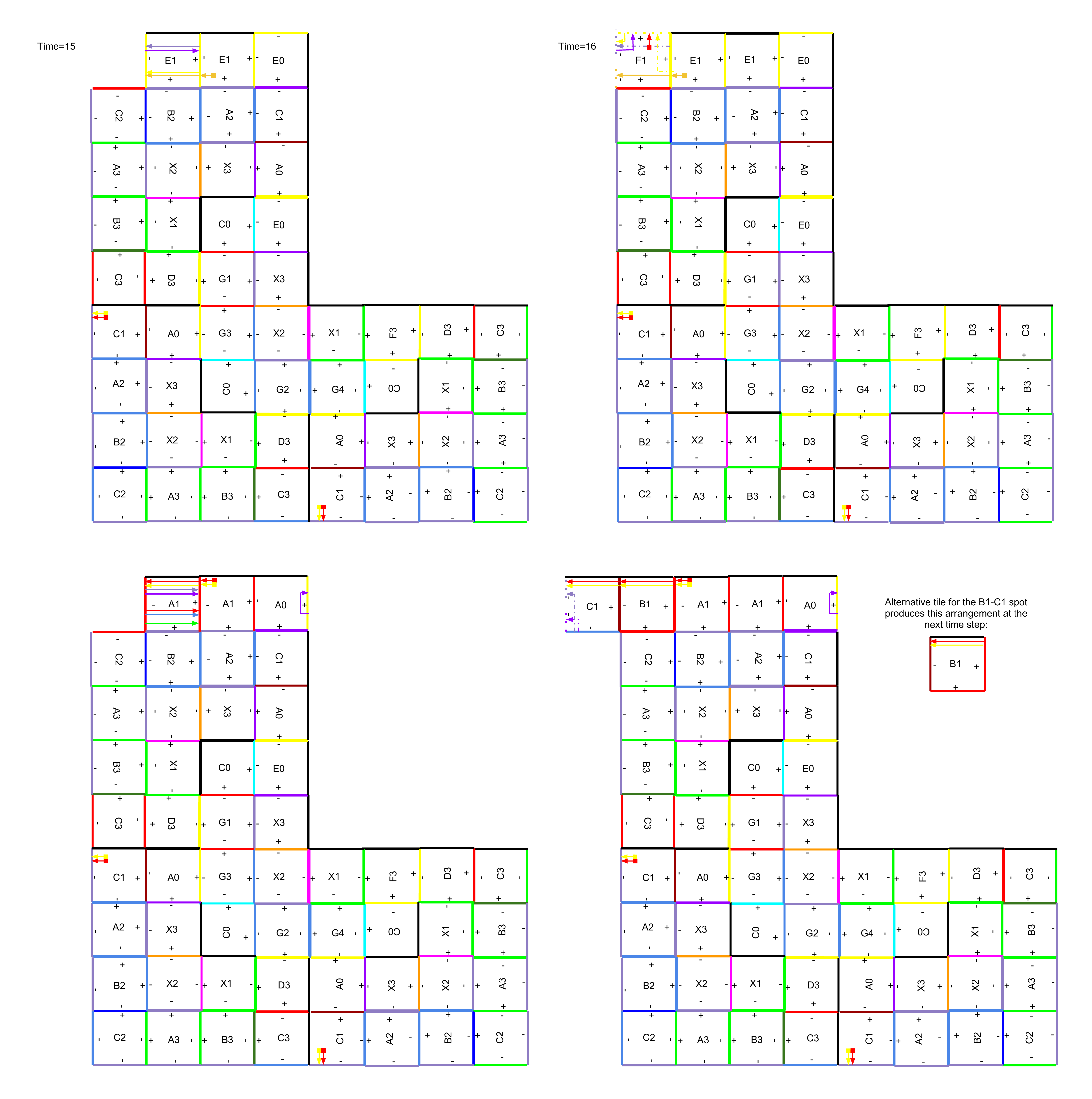}
\end{center}
\caption{Time 15-16, continuing A0-E0 border formation. The assembly of the borders continues. Since B1 may already be attached to C1, both the single and the double addition structures are produced in Time 16; but the former may produce only the latter at the next step. Note also that the positive-positive corners required to fit into the negative-negative adjacencies are only ever present on single and double tiles, never on any larger structures (most tiles have only two positive labels and they use these to attach to growing assemblies, thus no positive labels remain for binding; the ones that do not follow the pattern are G2, G3, and F1, F2, and F3, but these contain positive 4 labels and there is never more than one such label on a fixed side of the assembly, so their geometry forces them to bind only with A0-type assemblies of the same level, as seen in Fig.\ref{fig:Time37}-\ref{fig:Time42}).}
\label{fig:Time15-16}
\end{figure}
\begin{figure}[h]
\begin{center}
\includegraphics[scale=.25]{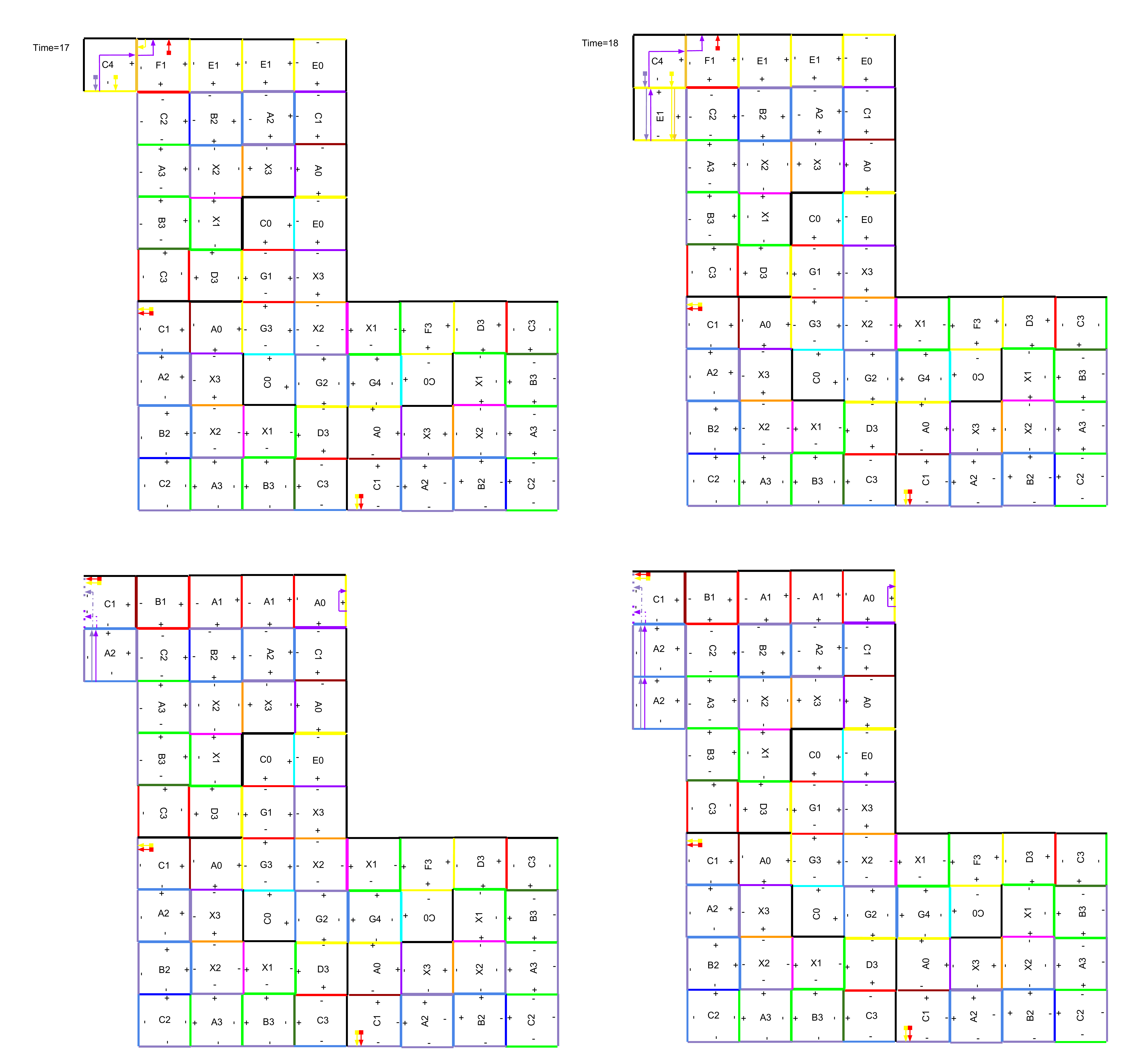}
\end{center}
\caption{Time 17-18, the assembly continues. Note that in 17, the gold label ``-44'' on F1 was activated by a signal in the previous time step, which allowed a C4 tile to attach in this step.}
\label{fig:Time17-18}
\end{figure}
\begin{figure}[h]
\begin{center}
\includegraphics[scale=.25]{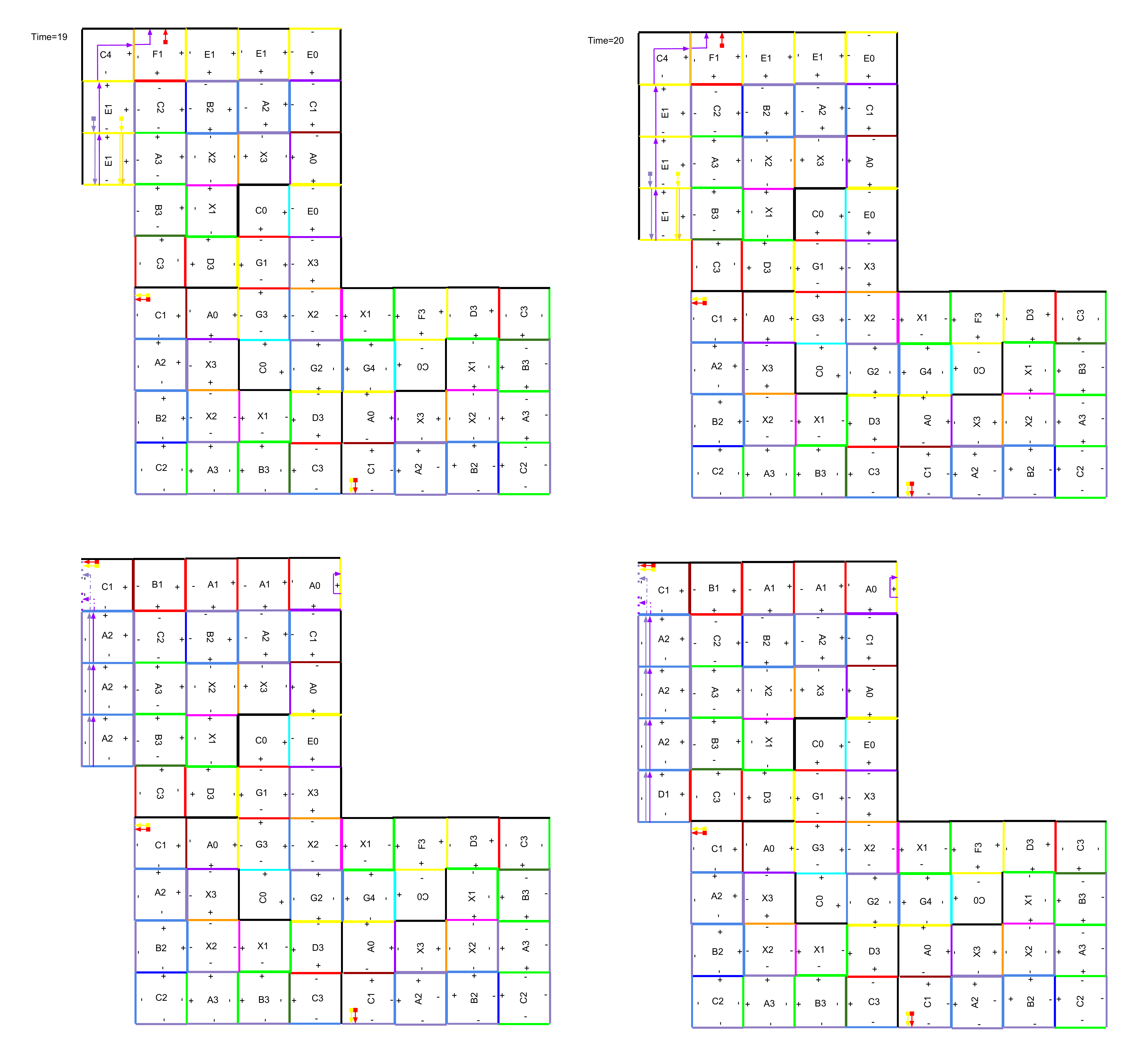}
\end{center}
\caption{Time 19-20.}
\label{fig:Time19-20}
\end{figure}
\begin{figure}[h]
\begin{center}
\includegraphics[scale=.25]{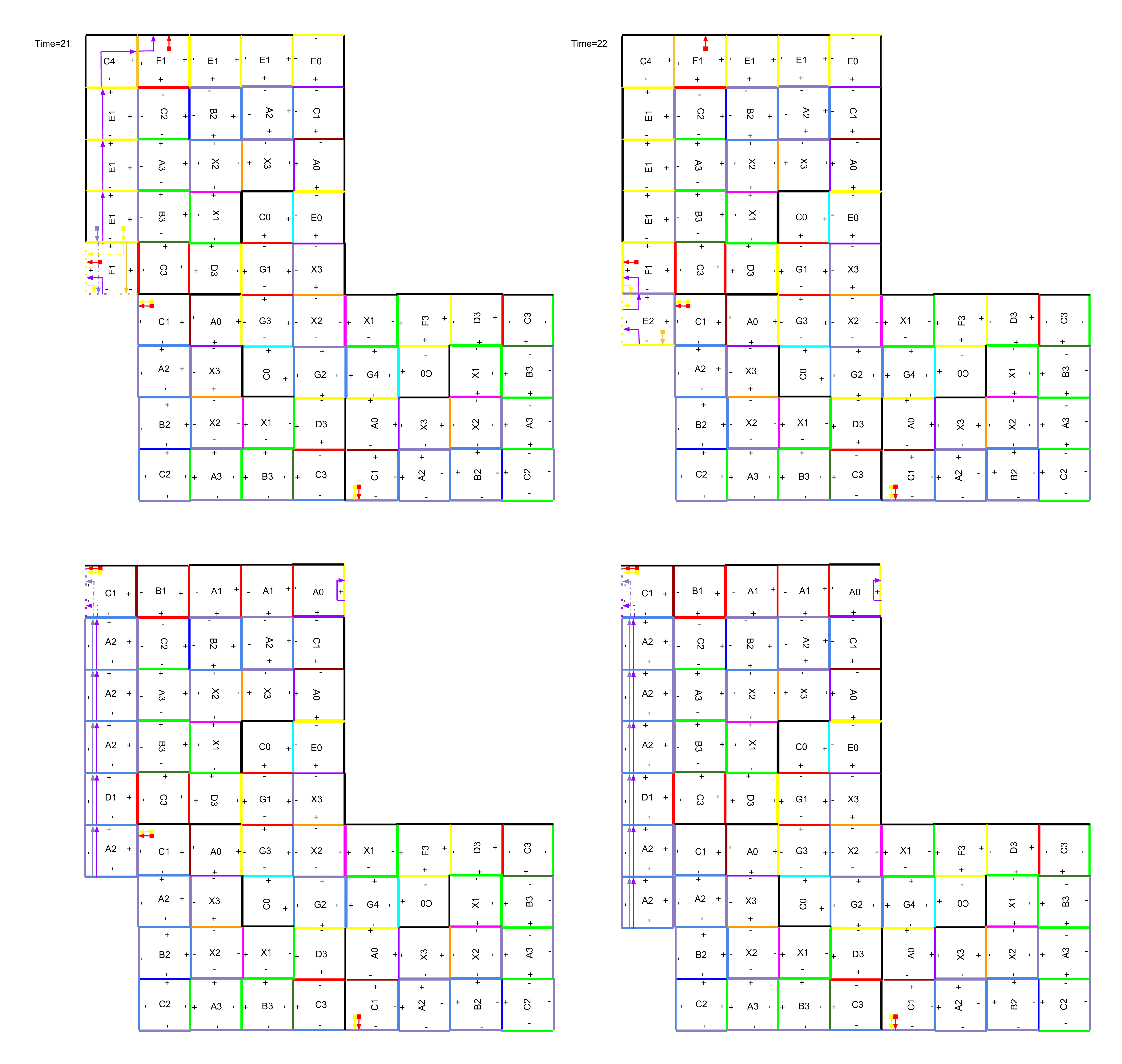}
\end{center}
\caption{When F1 attaches, it receives the signals 4 (yellow) and 5 (light purple) originally sent by C4 and transmitted by the E1's.}
\label{fig:Time21-22}
\end{figure}
\begin{figure}[h]
\begin{center}
\includegraphics[scale=.25]{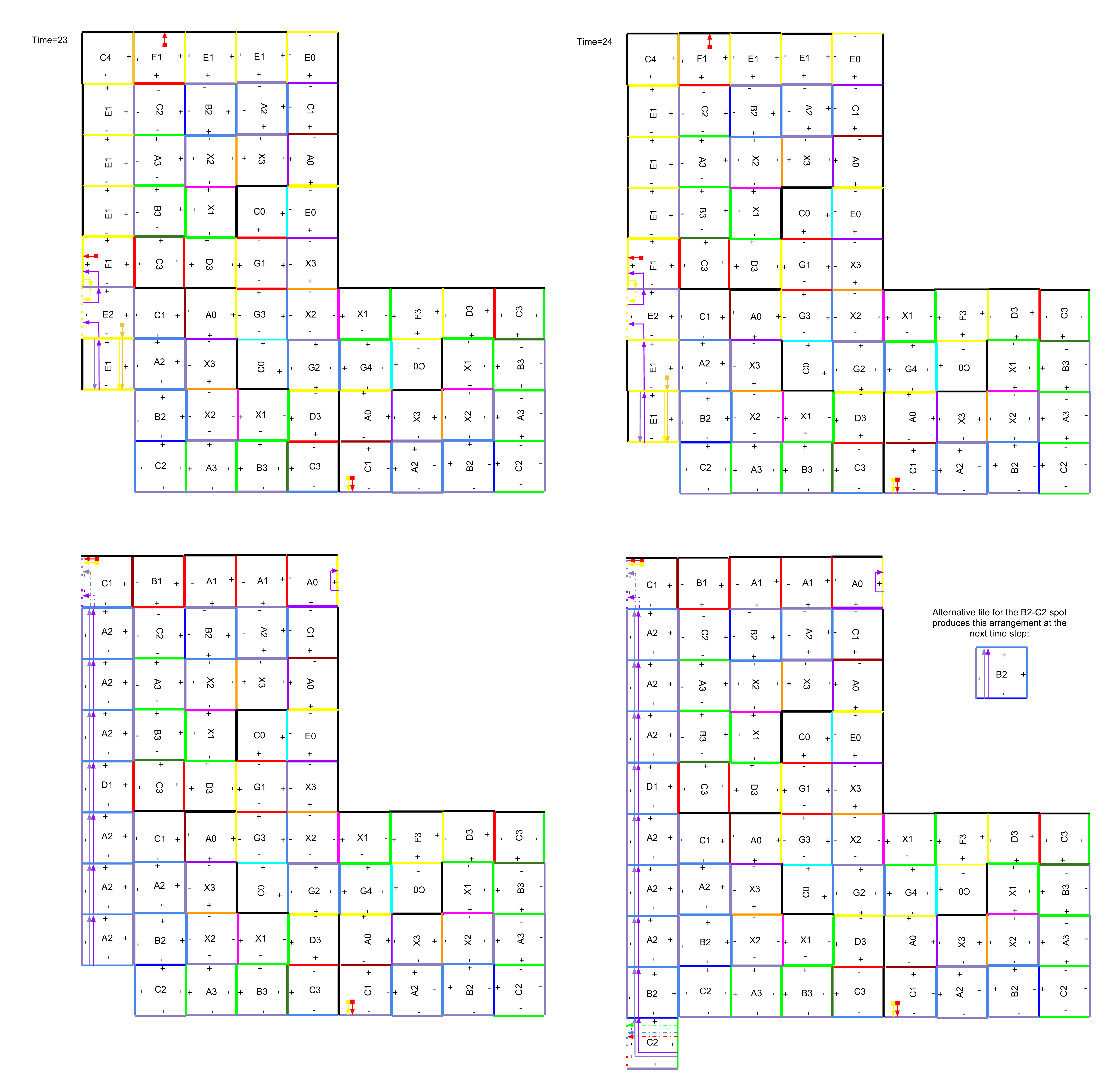}
\end{center}
\caption{Time 23-24. A similar situation arises with B2-C2 as did with B1-C1. Two assemblies are produced: one containing only B2 and one containing the B2-C2 pair; the former produces the latter at the next time step; so both evolve into the same shapes, but one does so faster.}
\label{fig:Time23-24}
\end{figure}
\begin{figure}[h]
\begin{center}
\includegraphics[scale=.25]{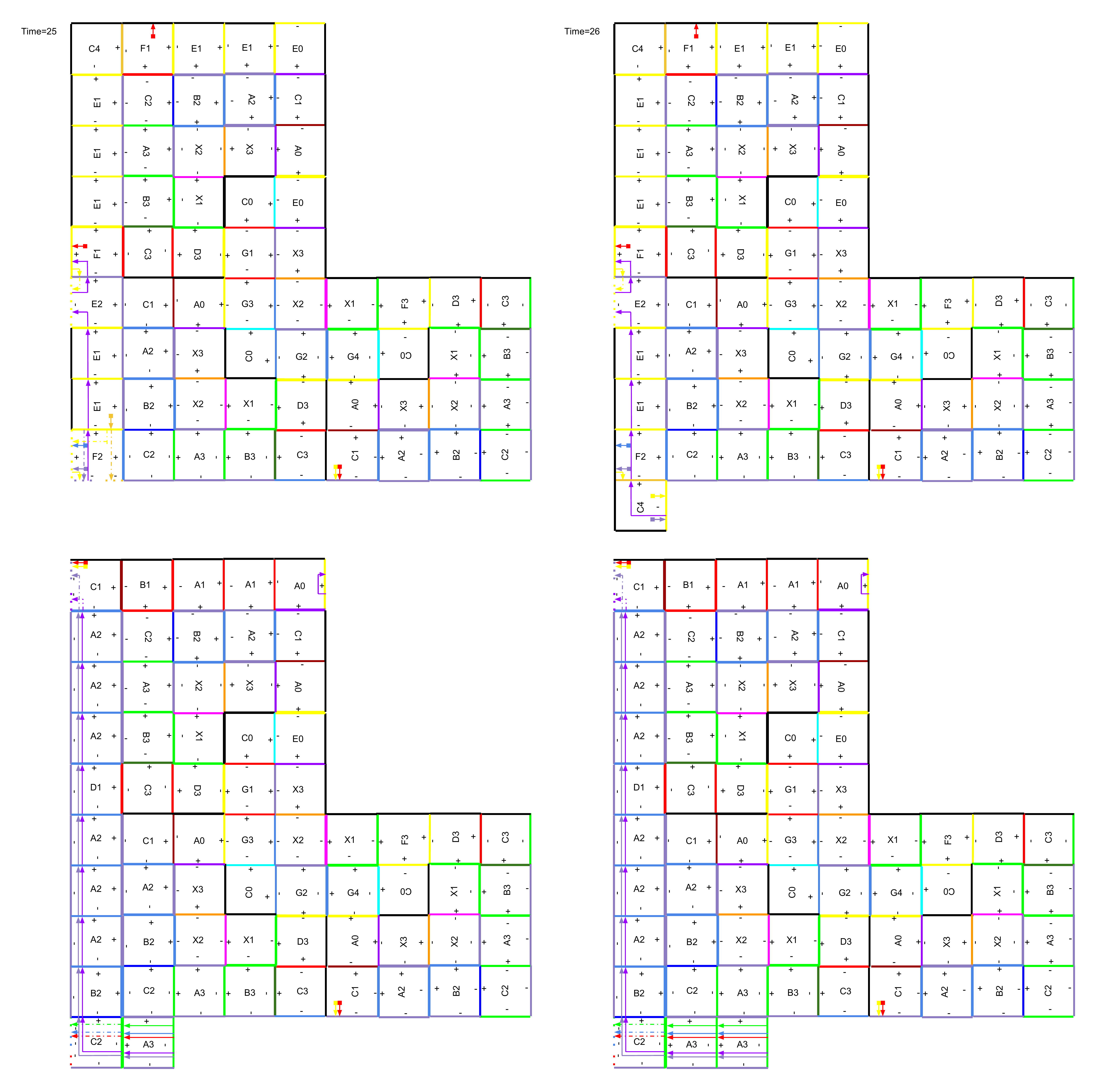}
\end{center}
\caption{Time 25-26. The signal originally sent by E2 is received by F2 and activates label -44 allowing C4 to attach.}
\label{fig:Time25-26}
\end{figure}
\begin{figure}[h]
\begin{center}
\includegraphics[scale=.25]{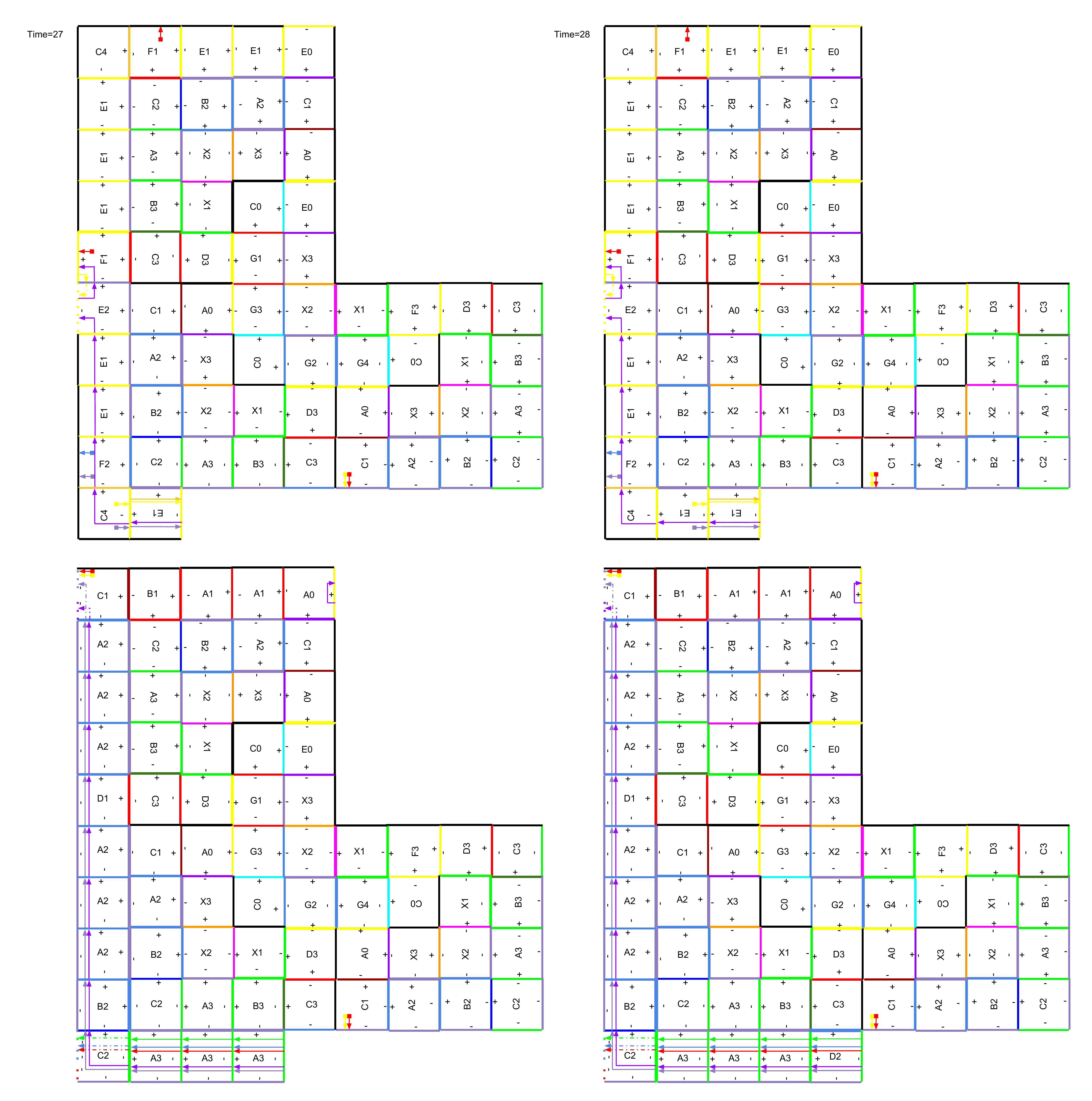}
\end{center}
\caption{Time 27-28.}
\label{fig:Time27-28}
\end{figure}
\begin{figure}[h]
\begin{center}
\includegraphics[scale=.25]{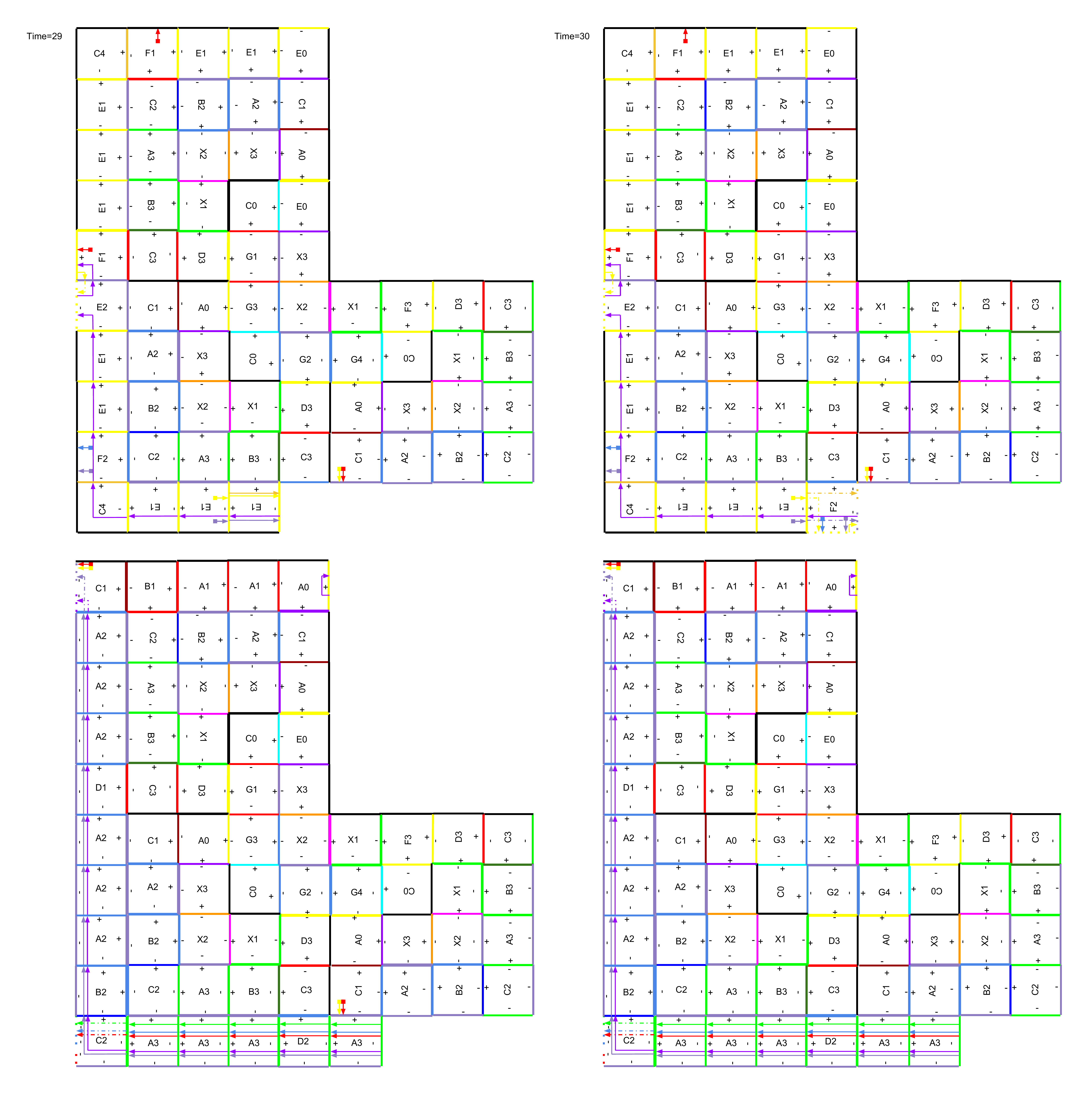}
\end{center}
\caption{Time 29-30. Notice that the second F2 receives the same signals as the second F2, activating its +4 and +5 labels and allowing E2 to attach at the next step.}
\label{fig:Time29-30}
\end{figure}
\begin{figure}[h]
\begin{center}
\includegraphics[scale=.25]{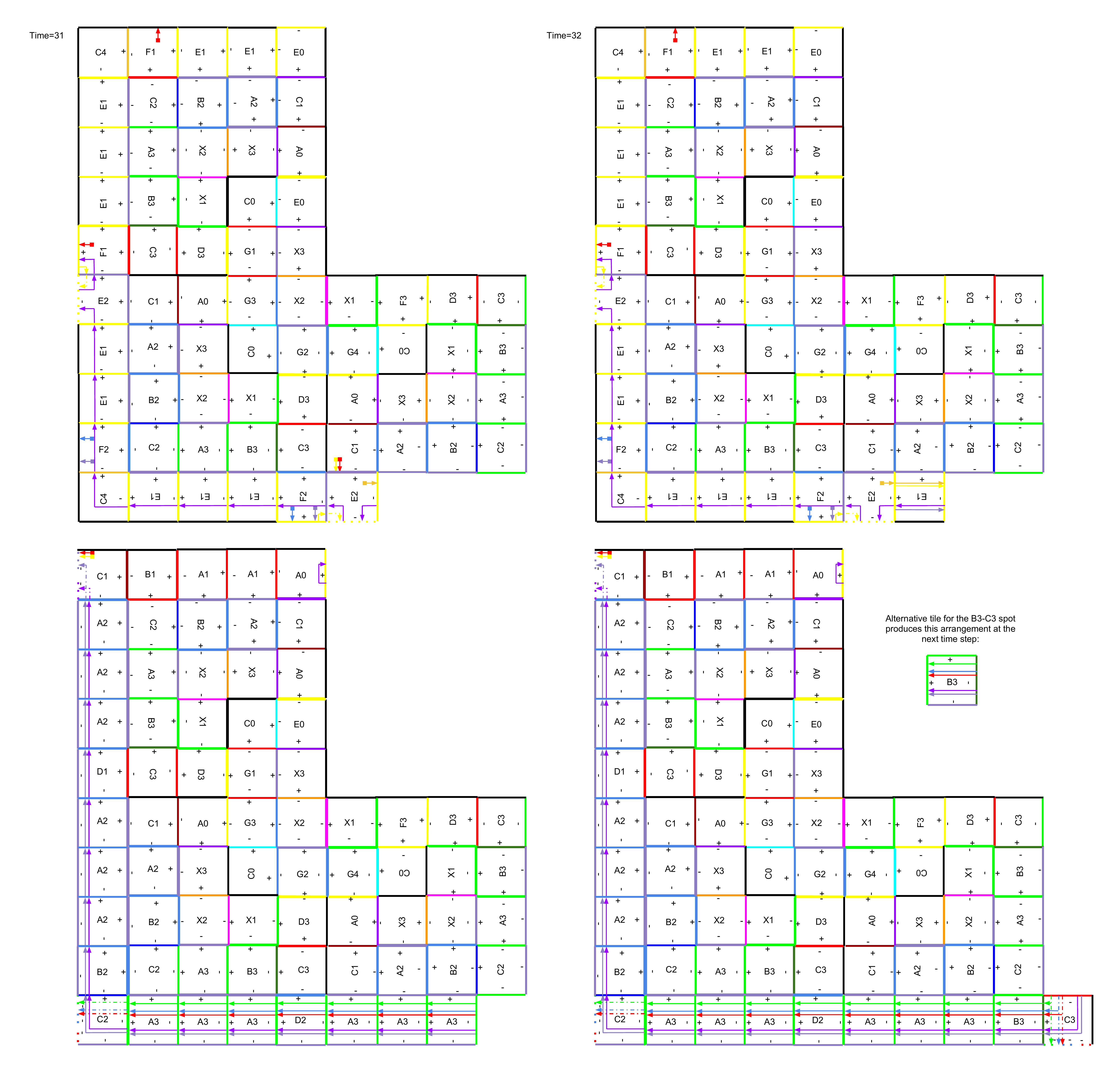}
\end{center}
\caption{Time 31-32. Just like B1-C1 and B2-C2, C3 can either be already attached to B3, or become attached immediately after.}
\label{fig:Time31-32}
\end{figure}
\begin{figure}[h]
\begin{center}
\includegraphics[scale=.23]{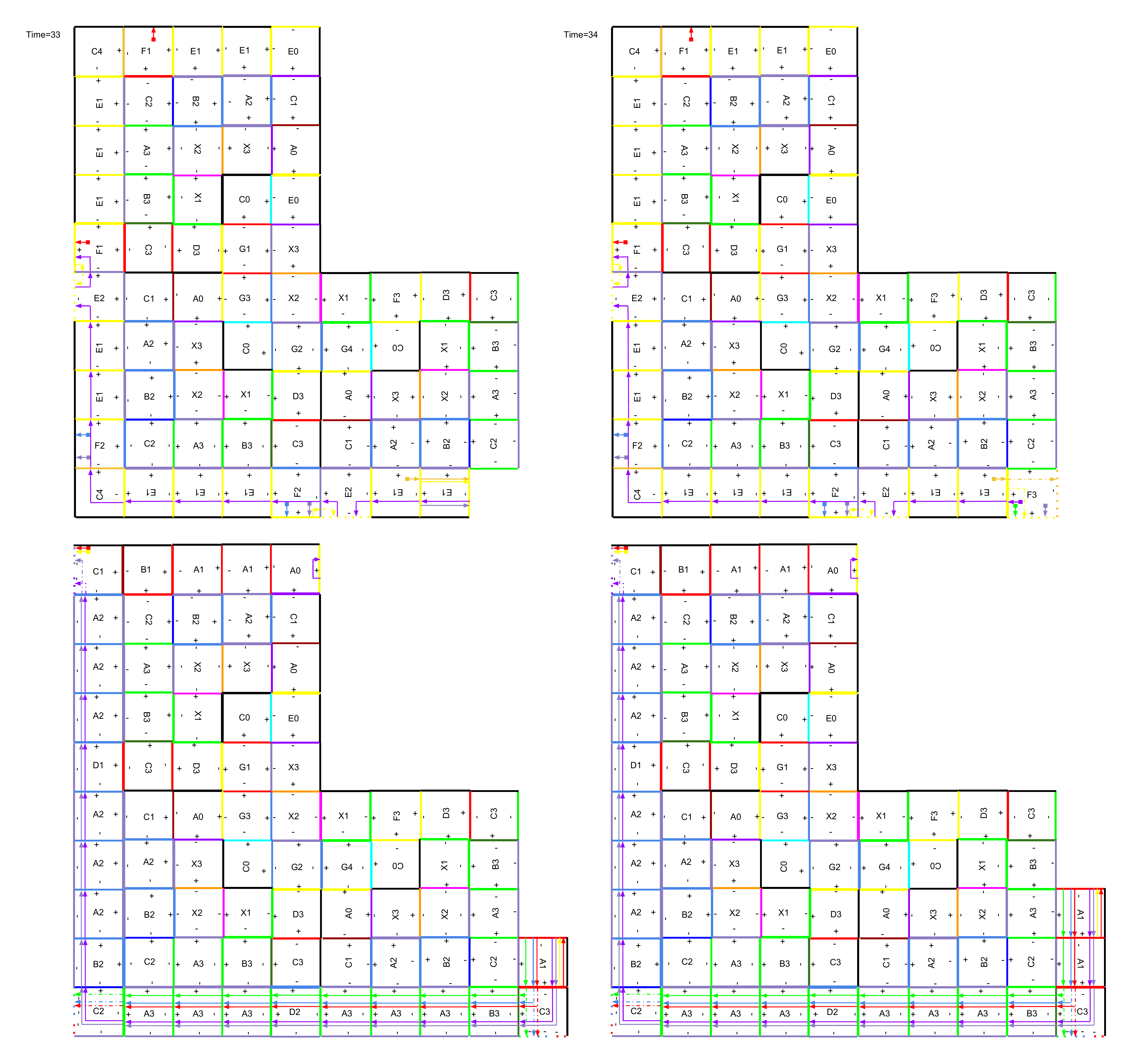}
\end{center}
\caption{Time 33-34.}
\label{fig:Time33-34}
\end{figure}
\clearpage
\begin{figure}[h]
\begin{center}
\includegraphics[scale=.2]{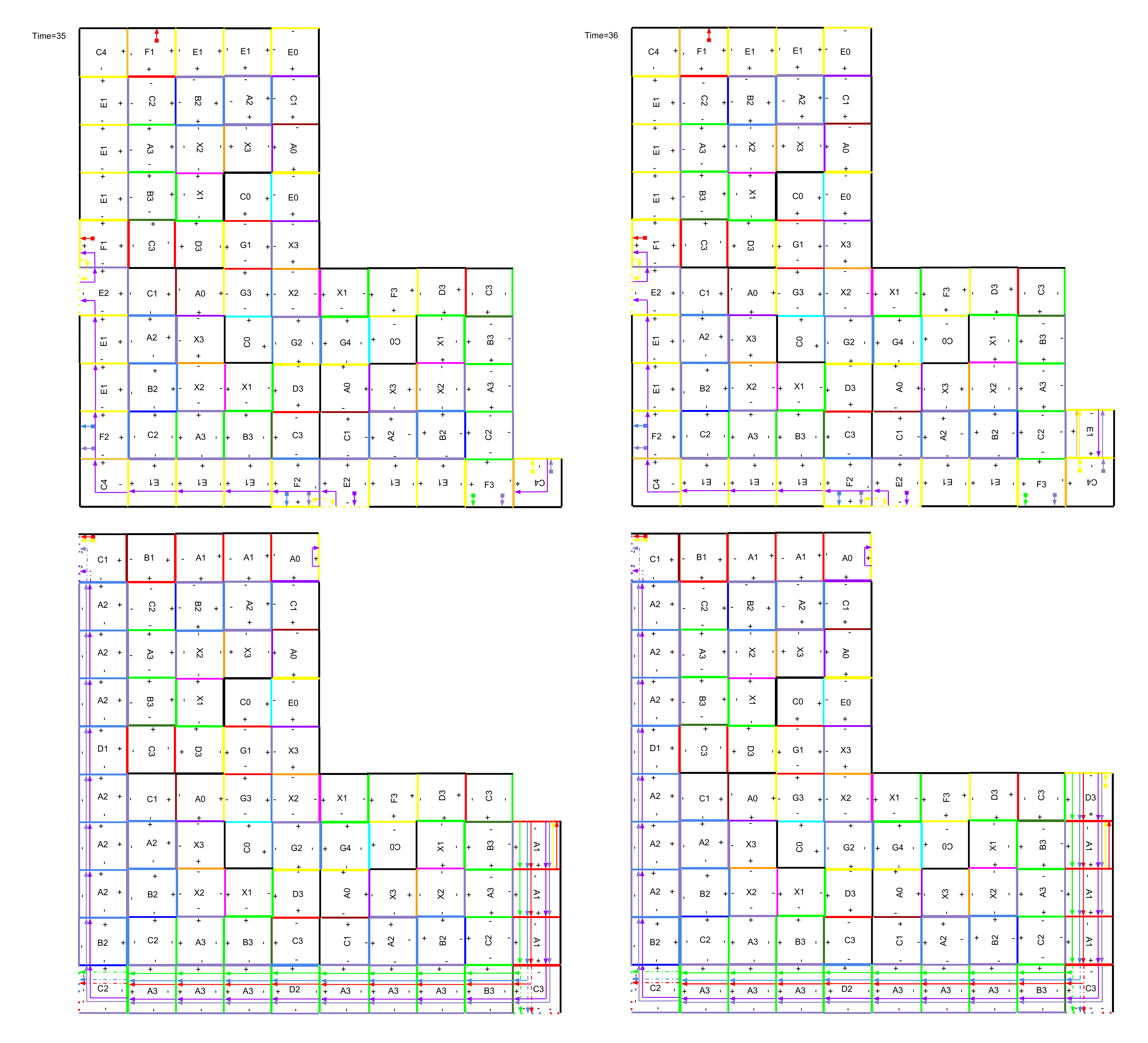}
\end{center}
\caption{Time 35-36. The A0-type border is completed at Time 36.}
\label{fig:Time35-36}
\end{figure}
\begin{figure}[h]
\begin{center}
\includegraphics[scale=.18]{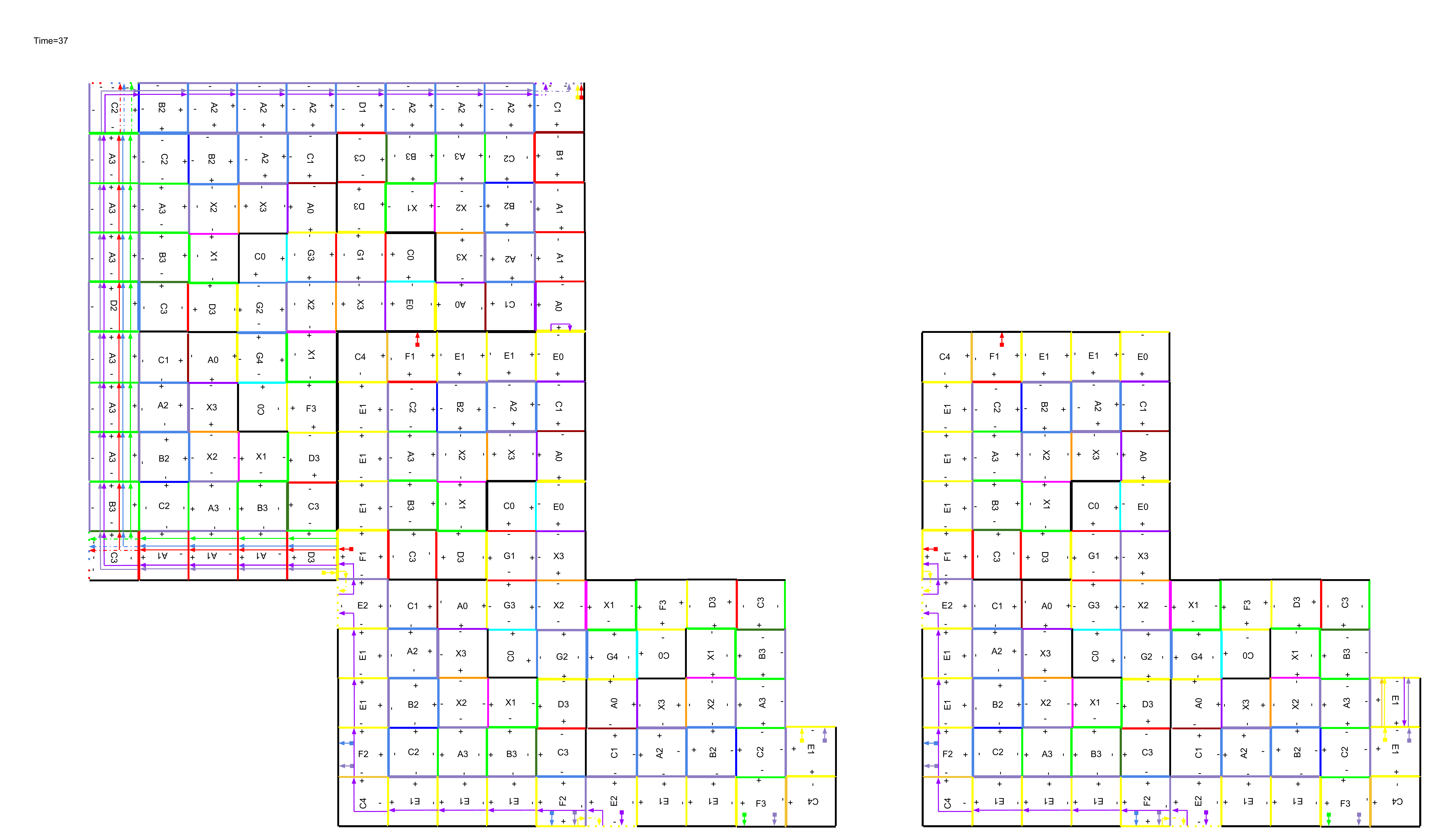}
\end{center}
\caption{Time 37. Two things can occur. Either an A0-type assembly binds to a partially complete E0-type assembly as in figure on the left, or with even less of the border complete (not shown), or an A0-type assembly may gain another border tile (right).}
\label{fig:Time37}
\end{figure}
\begin{figure}[h]
\begin{center}
\includegraphics[scale=.16]{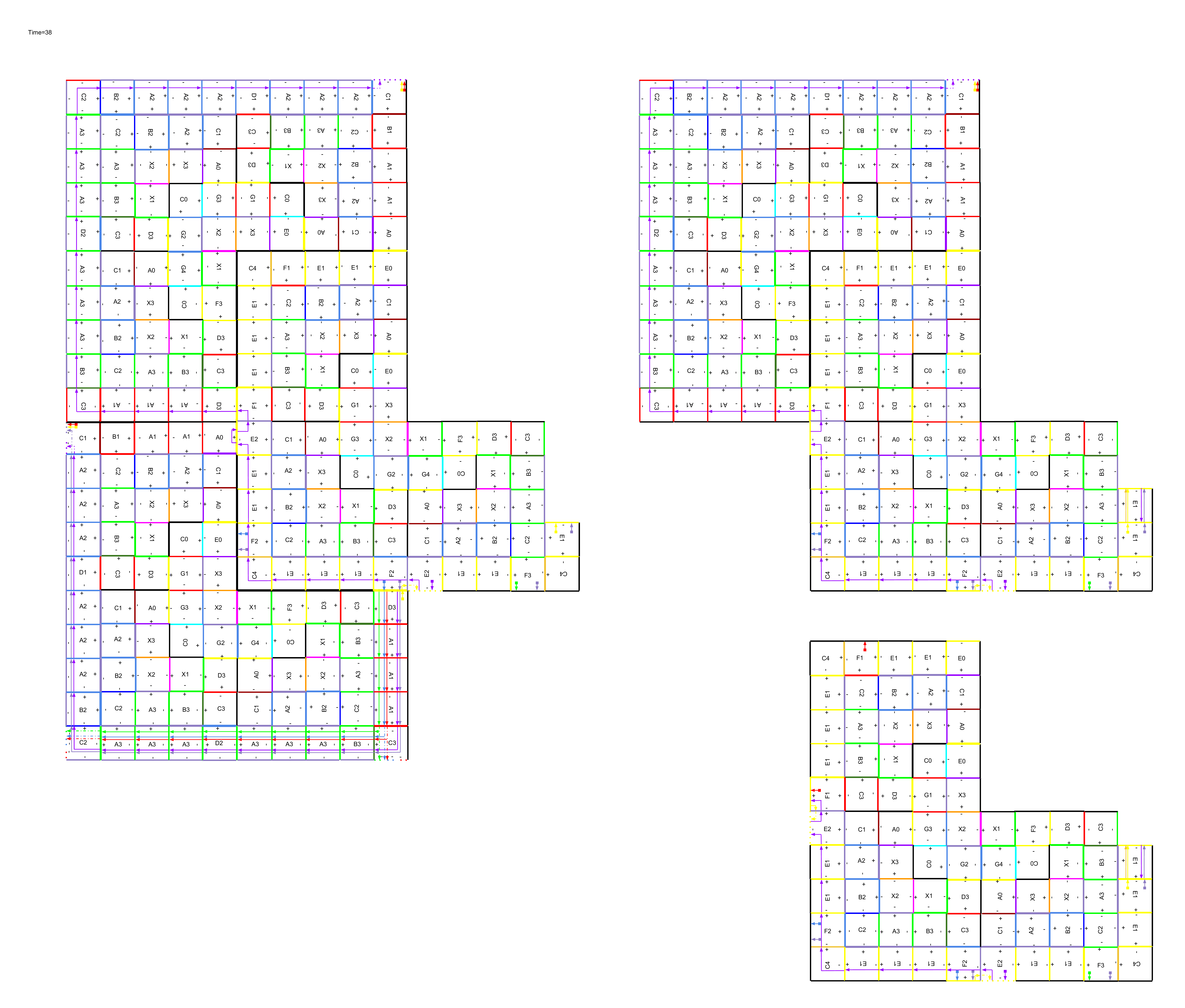}
\end{center}
\caption{Time 38. If the A0-type assembly binds, a signal from D3 is transmitted to the nearest E2, activating its -4 label and allowing it to bind to another A0-type bind if the second F2 on the E0-type border is present (left). The existing assemblies may also simply gain another border tile (right).}
\label{fig:Time38}
\end{figure}
\begin{figure}[h]
\begin{center}
\includegraphics[scale=.16]{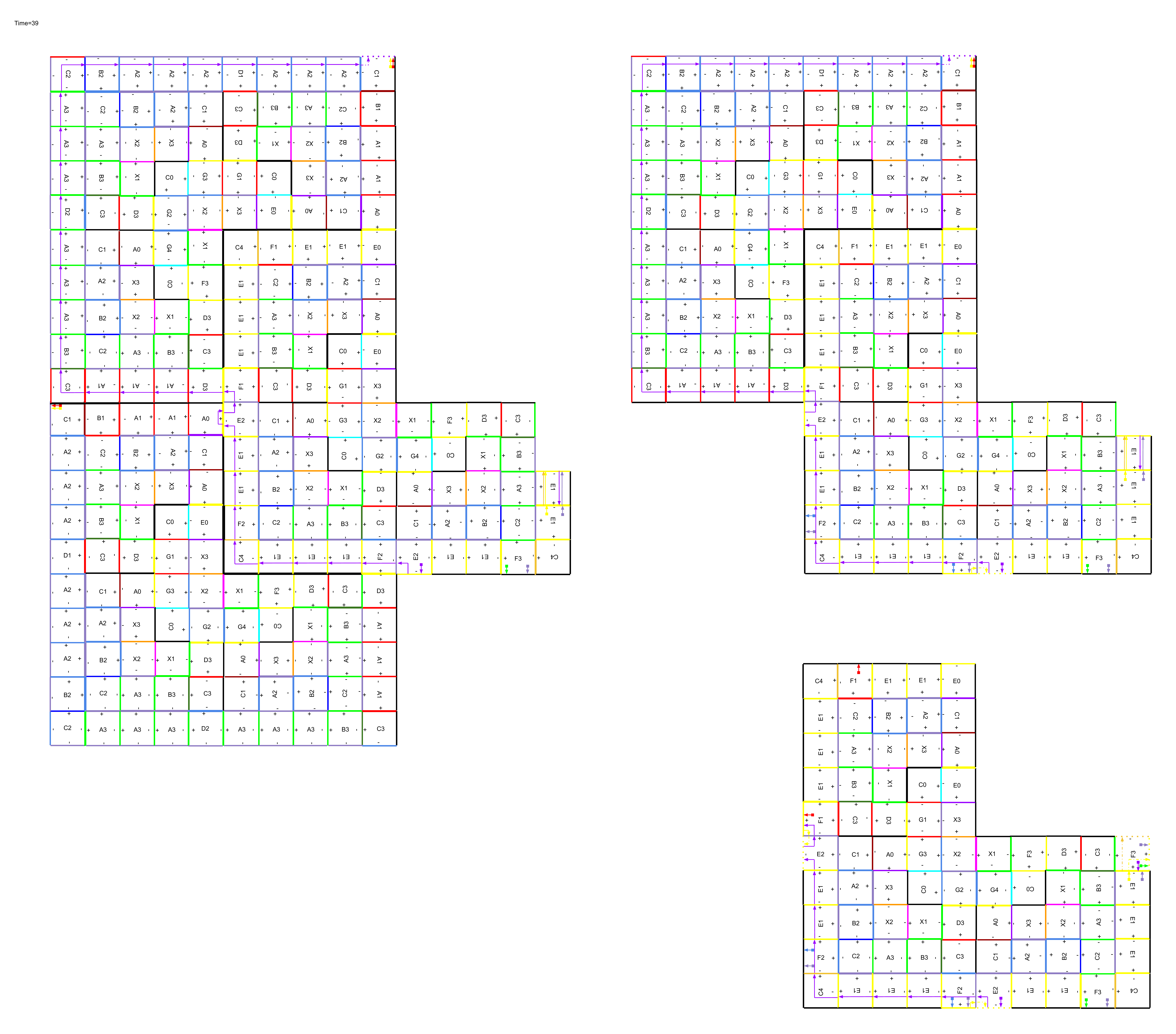}
\end{center}
\caption{Note that the final A0-type assembly cannot attach until the border on the E0-type assembly is completed, so at this step only border additions may be gained.}
\label{fig:Time39}
\end{figure}
\begin{figure}[h]
\begin{center}
\includegraphics[scale=.25]{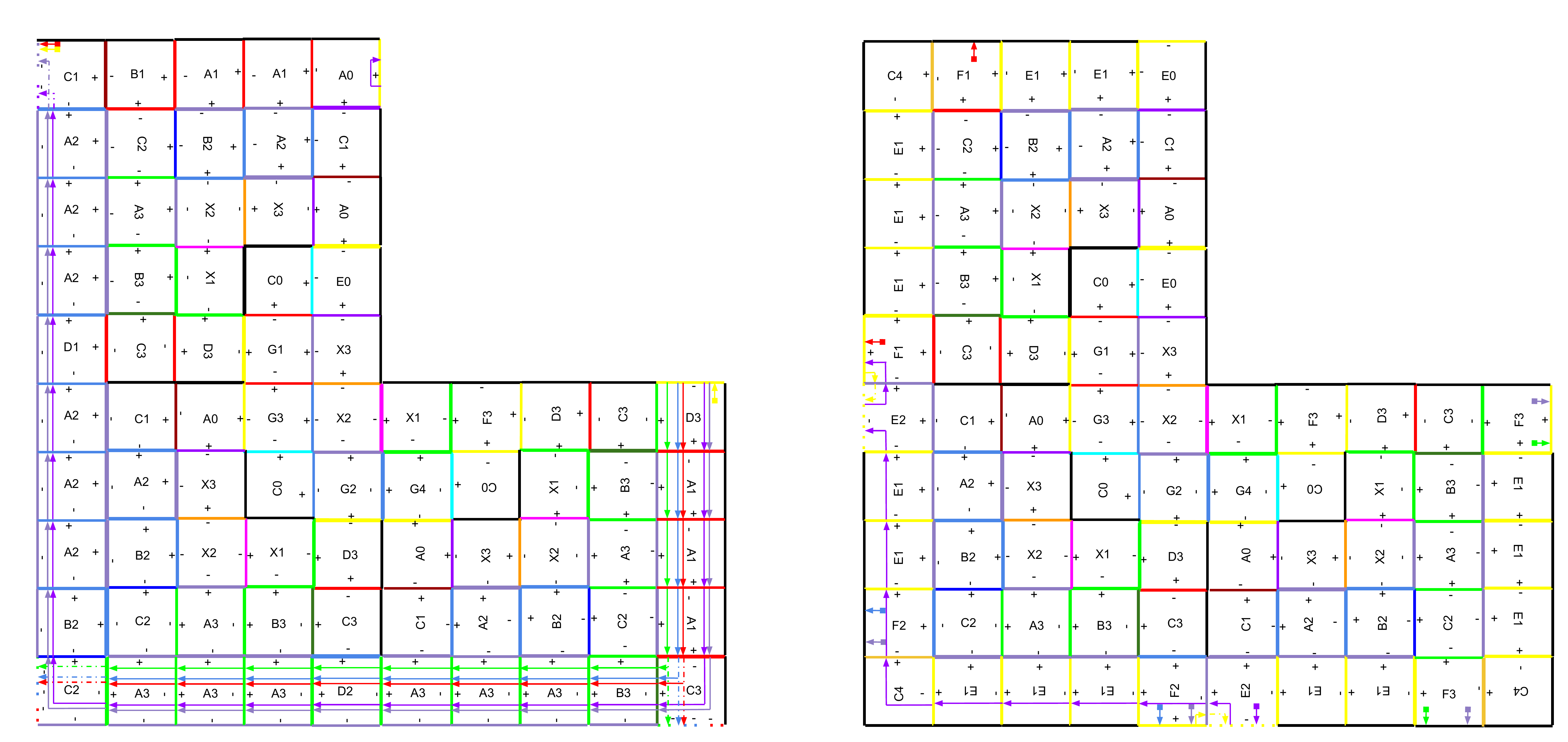}
\end{center}
\caption{$T_{2}(1)$ and $T_{3}(1)$ (left and right, respectively).}
\label{fig:Lvl2Completed}
\end{figure}

\begin{figure}[h]
\begin{center}
\includegraphics[scale=.16]{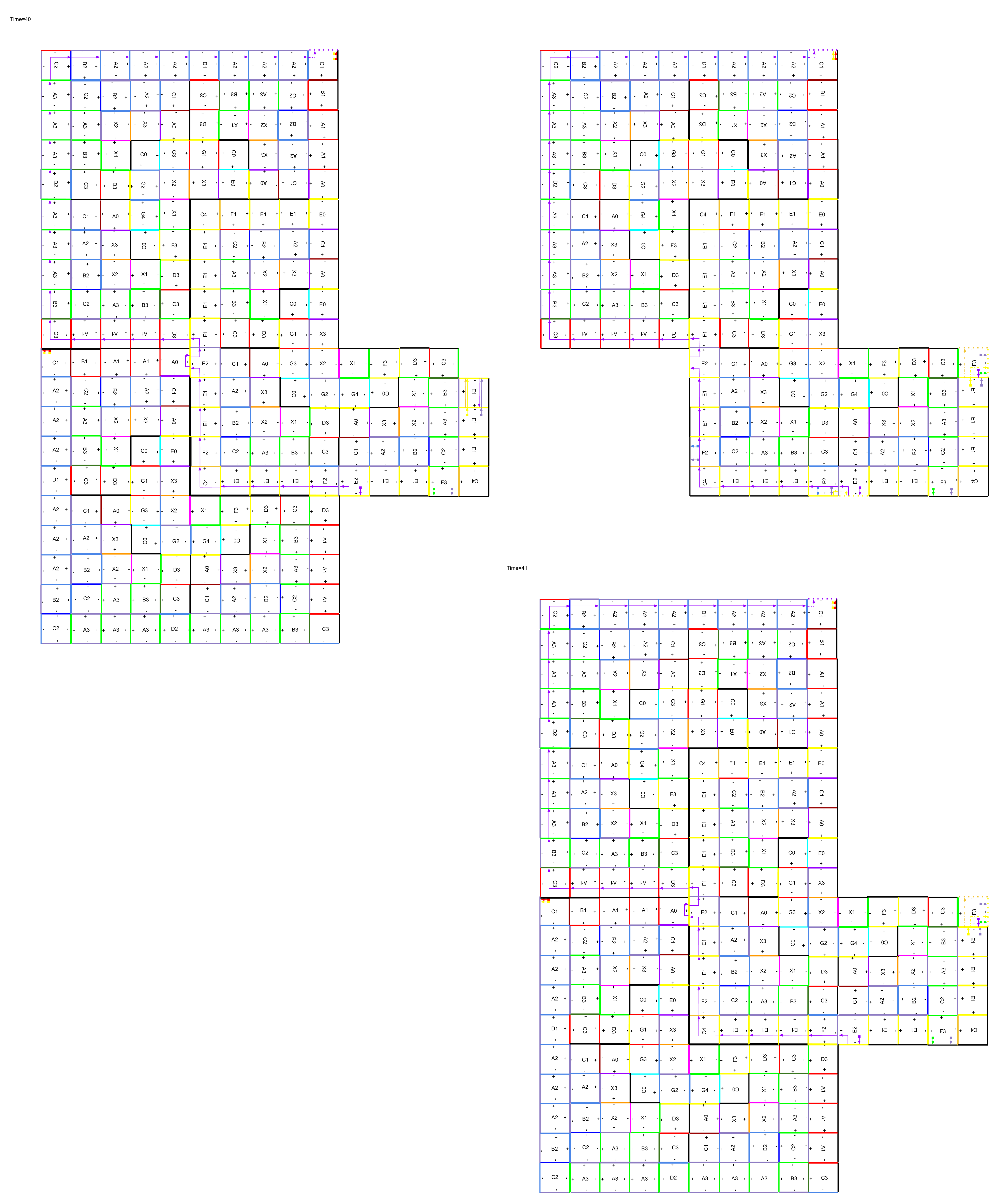}
\end{center}
\caption{Time 40 (top two assemblies) and 41 (bottom assembly). We see that in 41 only one new kind of assembly can be produced, because the variants of the partially assembled L-shapes converge to the same shape as more binding events occur.}
\label{fig:Time40-41}
\end{figure}
\begin{figure}[h]
\begin{center}
\includegraphics[scale=.16]{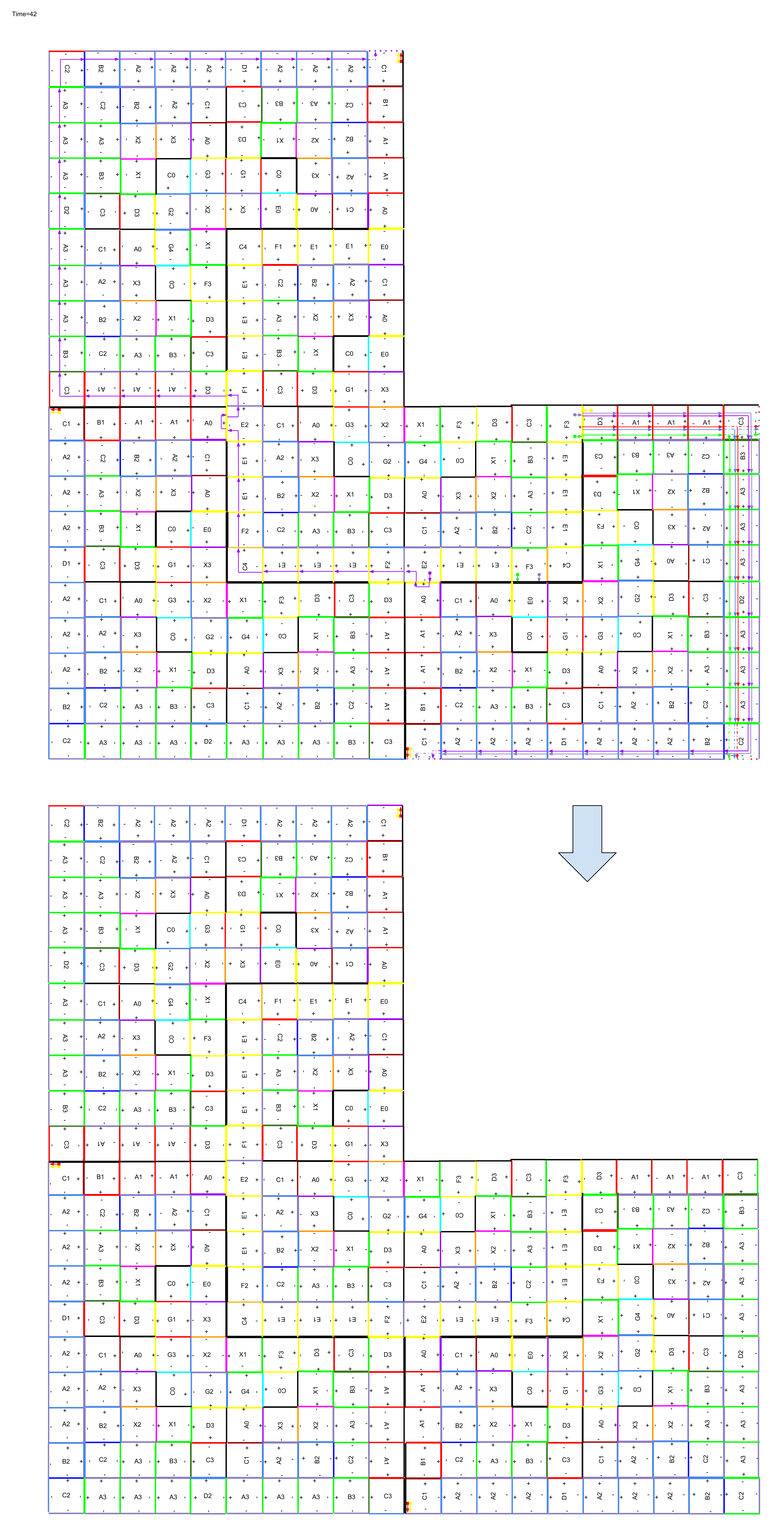}
\end{center}
\caption{Time 42. The first level 3 shape is assembled. Note the similarities between this and the level 2 shape. Both are given side by side in Fig.\ref{fig:Lvl2-3L-Shape}.}
\label{fig:Time42}
\end{figure}

\end{document}